\documentclass[preprint,nofootinbib,superscriptaddress,preprintnumbers]{revtex4}
\usepackage{tikz}
\usepackage{graphicx, hyperref, amsmath, amssymb, slashed, color, bbm}
\usepackage{hyperref}
\usepackage{breakurl}
\usepackage{array}
\usepackage{subfigure}
\usepackage{multirow}
\pdfoutput=1

\usepackage{float}
\usepackage{amsfonts}
\usepackage{morefloats}

\usepackage{slashed}
\usepackage{comment}
\usepackage{tablefootnote}
\usetikzlibrary{shapes.geometric,calc}

\usepackage{color}
\definecolor{rosso}{cmyk}{0,1,1,0.3}
\definecolor{verde}{cmyk}{0.8,0,0.6,0.25}
\definecolor{bluc}{cmyk}{1,0.4,0,0.1}
\definecolor{blucc}{cmyk}{0.8,0.3,0,0}

\usepackage{setspace}

\makeatletter
\def\simgt{\mathrel{\lower2.5pt\vbox{\lineskip=0pt\baselineskip=0pt
           \hbox{$>$}\hbox{$\sim$}}}}
\def\simlt{\mathrel{\lower2.5pt\vbox{\lineskip=0pt\baselineskip=0pt
           \hbox{$<$}\hbox{$\sim$}}}}
\makeatother
\def\simgt{\mathrel{\lower2.5pt\vbox{\lineskip=0pt\baselineskip=0pt
           \hbox{$>$}\hbox{$\sim$}}}}
\def\simlt{\mathrel{\lower2.5pt\vbox{\lineskip=0pt\baselineskip=0pt
           \hbox{$<$}\hbox{$\sim$}}}}
\makeatother

\newcommand{\be}{\begin{equation}}
\newcommand{\ee}{\end{equation}}
\newcommand{\bea}{\begin{eqnarray}}
\newcommand{\eea}{\end{eqnarray}}

\newcolumntype{M}{>{\centering\arraybackslash}m{\dimexpr.3\linewidth-2\tabcolsep}}

\def\mdm{m_{\rm DM}}

\def\GEV{\rm GeV}
\def\be{\begin{equation}}
\def\ee{\end{equation}}
\def\ba{\begin{eqnarray}}
\def\ea{\end{eqnarray}}
\def\met{\slashed{E}_T}

\makeatletter

\newcommand{\Rmnum}[1]{\expandafter\@slowromancap\romannumeral #1@}
\makeatother

\begin{document}

\title{Mono-$X$ Versus Direct Searches: Simplified Models for Dark Matter at the LHC}

\author{Seng Pei Liew}
\affiliation{Department of Physics, University of Tokyo, Bunkyo-ku, Tokyo 113-0033, Japan}
\author{Michele Papucci}
\affiliation{Theoretical Physics Group, 
  Lawrence Berkeley National Laboratory, Berkeley, CA 94720, USA}
\affiliation{Berkeley Center for Theoretical Physics, 
  University of California, Berkeley, CA 94720, USA}
\author{Alessandro Vichi}
\affiliation{Institute of Physics, \'Ecole Polytechnique F\'ed\'erale de Lausanne, CH-1015, Lausanne, Switzerland}
\author{Kathryn M. Zurek}
\affiliation{Theoretical Physics Group, 
  Lawrence Berkeley National Laboratory, Berkeley, CA 94720, USA}
\affiliation{Berkeley Center for Theoretical Physics, 
  University of California, Berkeley, CA 94720, USA}

\begin{abstract}

We consider simplified models for dark matter (DM) at the LHC, focused on mono-Higgs, -$Z$ or -$b$ produced in the final state.  Our primary purpose is to study the LHC reach of a relatively complete set of simplified models for these final states, while comparing the reach of the mono-$X$ DM search against direct searches for the mediating particle.  We find that direct searches for the mediating particle, whether in di-jets, jets+$\met$, multi-$b$+$\met$, or di-boson+$\met$, are usually stronger.  We draw attention to the cases that the mono-$X$ search is strongest, which include regions of parameter space in inelastic DM, two Higgs doublet, and squark mediated production models with a compressed spectrum.  
\end{abstract}

\maketitle

\tableofcontents


\section{Introduction}

Dark matter (DM) production at colliders is a potentially powerful complementary probe to searches for DM in direct and indirect detection experiments.  
Traditionally, searches for DM at colliders have focused on the signatures of DM candidates belonging to simple, non-singlet representation of the Standard Model (SM) weak gauge group $SU(2)\times U(1)$, motivaved by the most popular incarnations of the weakly interative massive particle (WIMP) ideas, such as the neutralino in supersymmetry (SUSY).  
More recently, however, the idea that the LHC can search for WIMP DM in more general types of theories and interactions has gained traction.  That one can look for DM via a jet, photon or $Z$-boson recoiling off missing energy has a long history~\cite{Nachtmann:1984xu,Dicus:1989gg,Brignole:1998me,Brhlik:1998uq,Birkedal:2004xn,Petriello:2008pu,Gershtein:2008bf}. 

Casting these bounds in the context of an effective field theory (EFT) allows one to compare the results from a collider in a straightforward way to direct and indirect detection constraints~\cite{Beltran:2010ww, Goodman:2010yf,Goodman:2010ku,Bai:2010hh,Fox:2011fx, Fox:2011pm, Bai:2012xg} simply by placing a bound on the scale of the EFT operator, $\Lambda$, that can be easily ported from one type of DM search experiment to the next.  Perhaps because of this ease of comparison to direct and indirect detection experiments, DM searches at the Large Hadron Collider (LHC) have gained popularity, and the EFT framework has been utilized in many LHC searches at Run I.

It is clear, however, that the typical momenta exchanged in the collision processes probed at colliders such as the LHC are often beyond the values of $\Lambda$ that can be bounded, rendering a naive EFT characterization of DM searches at colliders invalid in many cases.  Effective operators within the EFT framework are generated by integrating out heavy mediators at a scale $\Lambda$ 
in the UV-complete theory; a lower limit on $\Lambda$ can be derived self-consistently if the energy scale of the processes used to constrain the theory is smaller than $\Lambda$.  Further discussions and more detailed analyses of this issue can be found in~\cite{Friedland:2011za,Shoemaker:2011vi,Busoni:2013lha,Chang:2013oia,An:2013xka,Bai:2013iqa,Dreiner:2013vla,DiFranzo:2013vra,Buchmueller:2014yoa,Papucci:2014iwa}.  For this reason, the collider limits obtained using the EFT approach cannot be straightforwardly used, for example, to compare with limits obtained from direct detection experiments. Various prescriptions to overcome these issues can be found, {\em e.g.}, in~\cite{Busoni:2013lha,Busoni:2014sya,Busoni:2014haa,Racco:2015dxa,Abercrombie:2015wmb}. 

These statements are especially true once constraints on the mediating particle are taken into account, generally forcing one either out of the LHC reach or out of regime of validity of the EFT  ({\em e.g.}~\cite{Dreiner:2013vla,Papucci:2014iwa}).  Identifying the regions where mono-$X$ searches provide the strongest constraint is therefore important for developing a DM LHC search program. For example, di-jet searches for the particle mediating the DM production place such strong constraints on the quark-mediator coupling that, in order for the DM-mediator coupling to be perturbative but still constrained by mono-jet searches, one finds the mediator must, in most cases, be produced on-shell. 
For the purpose of DM direct detection experiments, a given scattering cross-section will map to different parameter points that may have different exclusion status between mono-jet and di-jet LHC searches, thus requiring additional assumptions.

Therefore, in order to interpret DM search results at colliders adequately, simplified models should be employed~\cite{Abercrombie:2015wmb}. Simplified models are UV-complete models that do not necessarily represent the full theory, but enable one to study the kinematics and topologies of DM production at the LHC in a precise manner. Moreover, the sensitivity comparisons between collider and direct detection limits can be performed accurately. 

Simplified models immediately suggest that other signatures, apart from looking for DM recoiling against a visible SM particle, must be considered.  Searching directly for the mediator of the SM-DM interaction may generally be more powerful for constraining the parameter space.   For example, returning to the earlier example, assuming that the mediator is coupled to both quarks and DM, where the monojet search is expected to be important, models with $t$-channel DM production (squark mediator) are constrained by jets plus missing transverse energy ($\met$) searches, while models with $s$-channel DM production ($Z'$ mediator) are constrained by di-jet searches. Various aspects of such simplified models have been studied extensively in the literature\footnote{For a comprehensive list of references, see~\cite{Abdallah:2014hon,Malik:2014ggr,Abercrombie:2015wmb}.}~\cite{An:2012va,Frandsen:2012rk,Chang:2013oia,An:2013xka,Bai:2013iqa,DiFranzo:2013vra,Alves:2013tqa,Buckley:2014fba,Papucci:2014iwa,Dreiner:2013vla,Hamaguchi:2014pja,Garny:2014waa,Buchmueller:2014yoa,Chala:2015ama,Alves:2015mua,Goncalves:2016iyg}.   
  
Simplified models for mono-$X$ searches, where here $X$ will be taken to be an object different from a jet, such as mono-Higgs~\cite{Petrov:2013nia,Carpenter:2013xra,Berlin:2014cfa,Ghorbani:2016edw}, mono-$W$~\cite{Bell:2015rdw}, -$Z$~\cite{Petriello:2008pu,Bell:2012rg}, and -$b$~\cite{Lin:2013sca,Izaguirre:2014vva}
 have received comparatively less attention. Understandably, one does not expect DM to be produced copiously while radiating from the initial-state a particle such as Higgs, Z or W at the LHC. In most cases, DM production with a jet from the initial state imposes the most stringent constraints. Even so, as dedicated searches for various mono-$X$ channels have already been performed~\cite{Aad:2014vka,Aad:2013oja,Aad:2015dva,ATLAS:2014wra,Khachatryan:2014tva,Khachatryan:2016mdm,Khachatryan:2015bbl} and will be extensively carried on in the current and future LHC runs, it is important and timely to consider a relatively exhaustive set of simplified models that give rise dominantly to such mono-$X$ signals. A systematic study considering a broad range of simplified models is still lacking in the literature.  
The present work aims to bridge this gap and propose a comprehensive set of simplified models that characterizes mono-$X$ searches. In the following, we focus on the interplay of mono-$X$ limits with other collider searches as well as their phenomenological implications.  We also provide UV completions of these DM production topologies. Table~\ref{tab:ab} shows diagrammatically the simplified models in consideration for mono-Higgs and mono-$Z$ as well as the models' constraints from other collider searches.  In general, many models which feature a mono-$Z$ signal also have a mono-$W$ signal.  For most of our analysis, we focus on singlet DM where there is only mono-$Z$ and mono-$H$ signals; the exception is the ``inelastic squark'' model, where the topology demands the presence of both mono-$Z$ and mono-$W$ signatures.  In general, however, the constraint on the production cross-section times branching fraction is weaker for mono-$W$ as compared to mono-$Z$, rendering the former less powerful, unless the latter is strongly suppressed for, {\em e.g.}, kinematic reasons.  We also do not further consider mono-$\gamma$ searches~\cite{Gershtein:2008bf,Lopez:2014qja}.  When the photon is radiated from the initial state, the constraint is generically weaker than when a jet is radiated from the initial state.  The other options are that that photon is radiated from the mediator or from the final state.  Since the final state is charge neutral, the latter does not occur at tree level.  The photon may instead be radiated from a charged non-colored mediating particle\footnote{If the mediating particle is also colored, mono-jet searches tend to provide stronger limits than the corresponding mono-photon ones.}.  In this case a charged particle must be produced in the final state as well, which must decay to additional charged SM states.  These may be lost if they are sufficiently soft, but in this case, it has been shown that mono-$X$ searches alone are not very powerful~\cite{Gori:2013ala}, although they may provide stronger limits if complemented with other signatures present in the event, such as a soft lepton or a disappearing track~\cite{Giudice:2010wb}. The only exception is if the mediating particle is present in a $t$-channel in the vector-boson-fusion (VBF) topology~\cite{Brooke:2016vlw}. We leave the study of the corresponding search of two forward jets and a single central photon~+~$\met$ to future work.
  
Among possible other mono-$X$ searches there are also those where $X$ is a bottom or top quark. Mono-$b$ searches are very effective for models where the mediator preferentially couples to the third generation, such as Higgs-like particles. The correspondence between mono-$b$ and direct searches for this type of $s$-channel model has been thoroughly investigated in~\cite{Izaguirre:2014vva}.  In this work, we will consider a simplified model with $t$-channel mediator (sbottom), which, as will be shown below, also plays a role in mono-$h$ and mono-$Z$ searches. Table~\ref{tab:ab2} shows diagrammatically the mono-$b$ topology as well as the relevant direct searches considered in this work. In the case of mono-$t$ searches the only simplified models producing sizable signals at tree level are divided in two categories depending on whether mono-$t$ is resonantly produced, as in $R$-parity violating (RPV) SUSY, or non-resonantly produced via a $t$-channel top quark~\cite{Andrea:2011ws,Wang:2011uxa,Agram:2013wda,Boucheneb:2014wza,D'Hondt:2015jbs}. Strictly speaking, the RPV SUSY scenario does not have a dark matter candidate, as the lightest neutralino is not stable on cosmological time scales. Moreover, both scenarios involve flavor-changing neutral interactions, which potentially lead to stringent flavor constraints. Furthermore, key direct searches for the RPV case involve displaced stop decays and apart from a few (very powerful) searches performed at Run I, both experiments are ramping up search efforts for long-lived particles in Run II. Given these complications, we leave the detailed study of mono-$t$ signatures elsewhere. 

In Table~\ref{tab:mono}, we summarize our main results: for each mono-$X$ search studied in this paper we list the simplified model where it has reach. We omit simplified models where a given search can only exclude parameter space already ruled out by a different analysis.

The $s$-channel $Z'$ and Higgs mediated models are briefly commented on in the next section without performing further mono-$X$ analysis as they have been studied in detail previously~\cite{Carpenter:2013xra,Berlin:2014cfa}.  Our primary purpose there is to compare the mono-$X$ analysis against other ways to look for the mediator and/or the DM particle at the LHC.  In each of the subsequent models, we compare the strength of mono-$Z$ and mono-Higgs against each other and the constraints from other searches, such as di-jet, jets+$\met$, mono-jet, di-boson+$\met$ and mono-$b$, whenever they are relevant.   These results will serve as a guideline to both theorists and experimentalists for optimizing mono-$X$ searches. For reference,  we list all relevant collider searches utilized in our analysis in Table~\ref{tab:searches}. 

For illustrating our results, we focus here on Run I searches, since a complete set of both mono-$X$ and direct searches performed with similar amounts of integrated luminosity has been performed. At the time of writing this is not yet the case for Run II analyses with approximately $13\,{\rm fb}^{-1}$. We checked and found the set of analyses released with 2015 data do not significantly increase the Run I limits. Therefore in the following, we will perform comparisons among different searches with 8~TeV data and use the available 13~TeV searches to validate the procedure we use to make our projections for the future reach, at $300\, {\rm fb}^{-1}$, as described in Appendix~\ref{app:b}\footnote{The only exception to this rule is a boosted di-jet analysis performed for the first time in Run II with $2.7\, {\rm fb}^{-1}$.  This analysis is important for improving the low mass limits, and we utilize it because with this luminosity we expect similar constraints as with $20\, {\rm fb}^{-1}$ at 8~TeV.}. The study presented here can nevertheless be updated with new Run II analyses once those are completely available.

\begin{table}[H]
\centering

\small{\begin{tabular}{>{\centering\arraybackslash} m{3cm} >{\centering\arraybackslash} m{4cm} >{\centering\arraybackslash} m{4cm} >{\centering\arraybackslash} m{5.5cm}}
\hline
Model & mono-$h$  &mono-$Z$ &direct constraints\\
\hline
Inelastic DM&\includegraphics[scale=0.9]{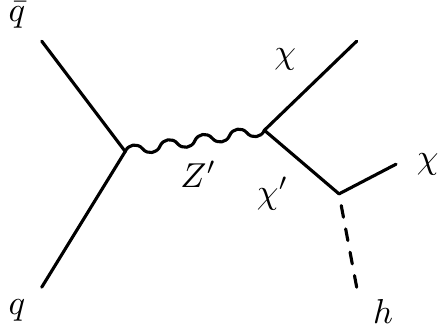}&\includegraphics[scale=0.9]{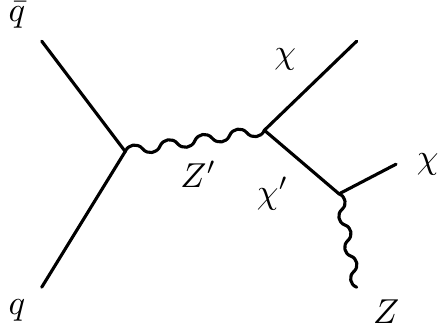}&\includegraphics[scale=0.6]{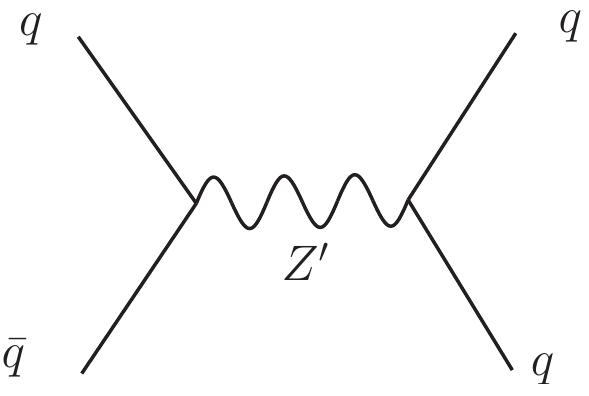} \\
\hline
2HDM&\includegraphics[scale=0.9]{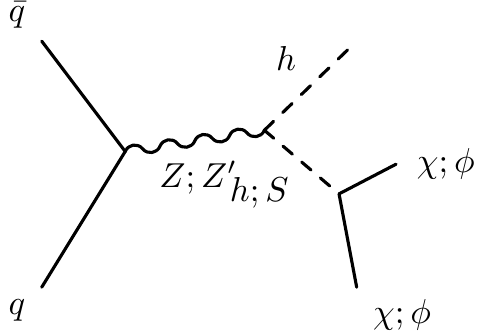}&\includegraphics[scale=0.9]{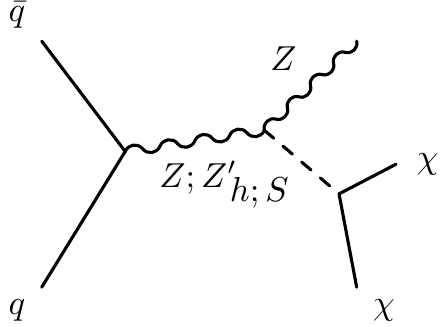}&\includegraphics[scale=0.6]{Diagrams/direct_searches/dijet.pdf}\\
\hline
Squarks/sbottoms &\includegraphics[scale=0.9]{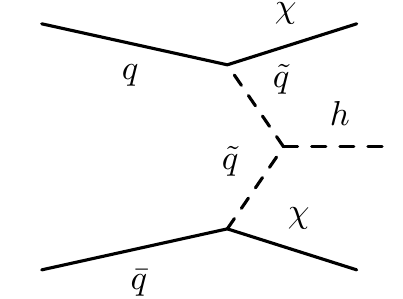}&\includegraphics[scale=0.19]{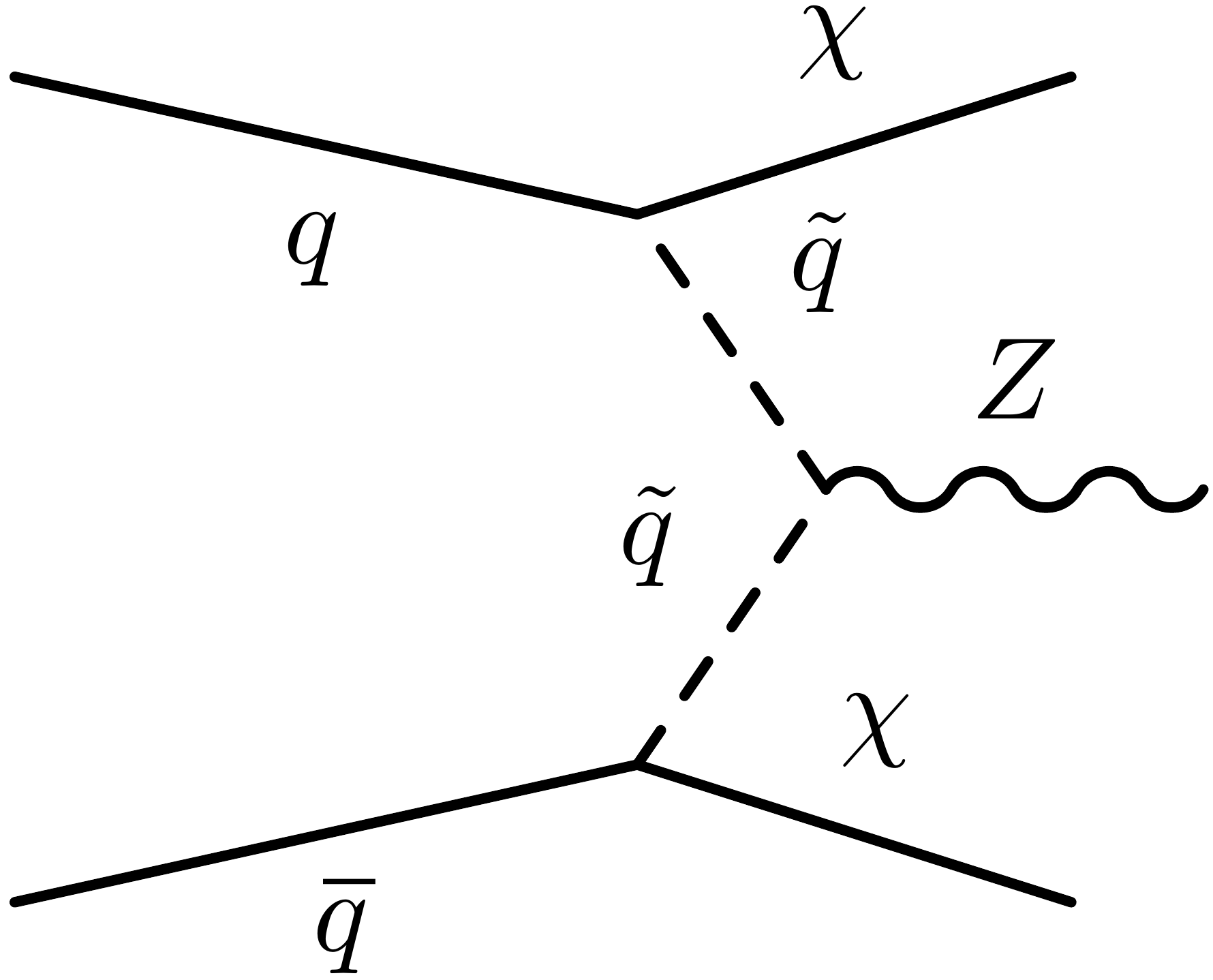}&\includegraphics[scale=0.6]{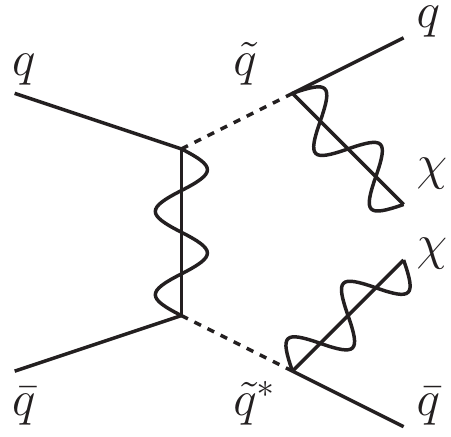}\\
\hline
$s$-channel vector &\includegraphics[scale=0.9]{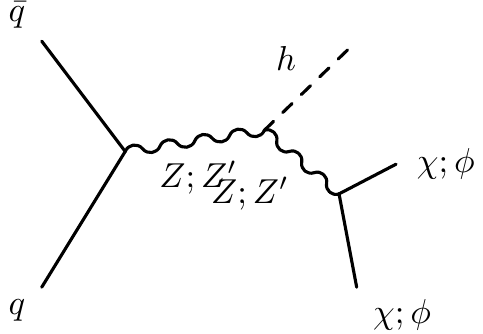}&&\shortstack{\includegraphics[scale=0.5]{Diagrams/direct_searches/dijet.pdf} \\ \includegraphics[scale=0.15]{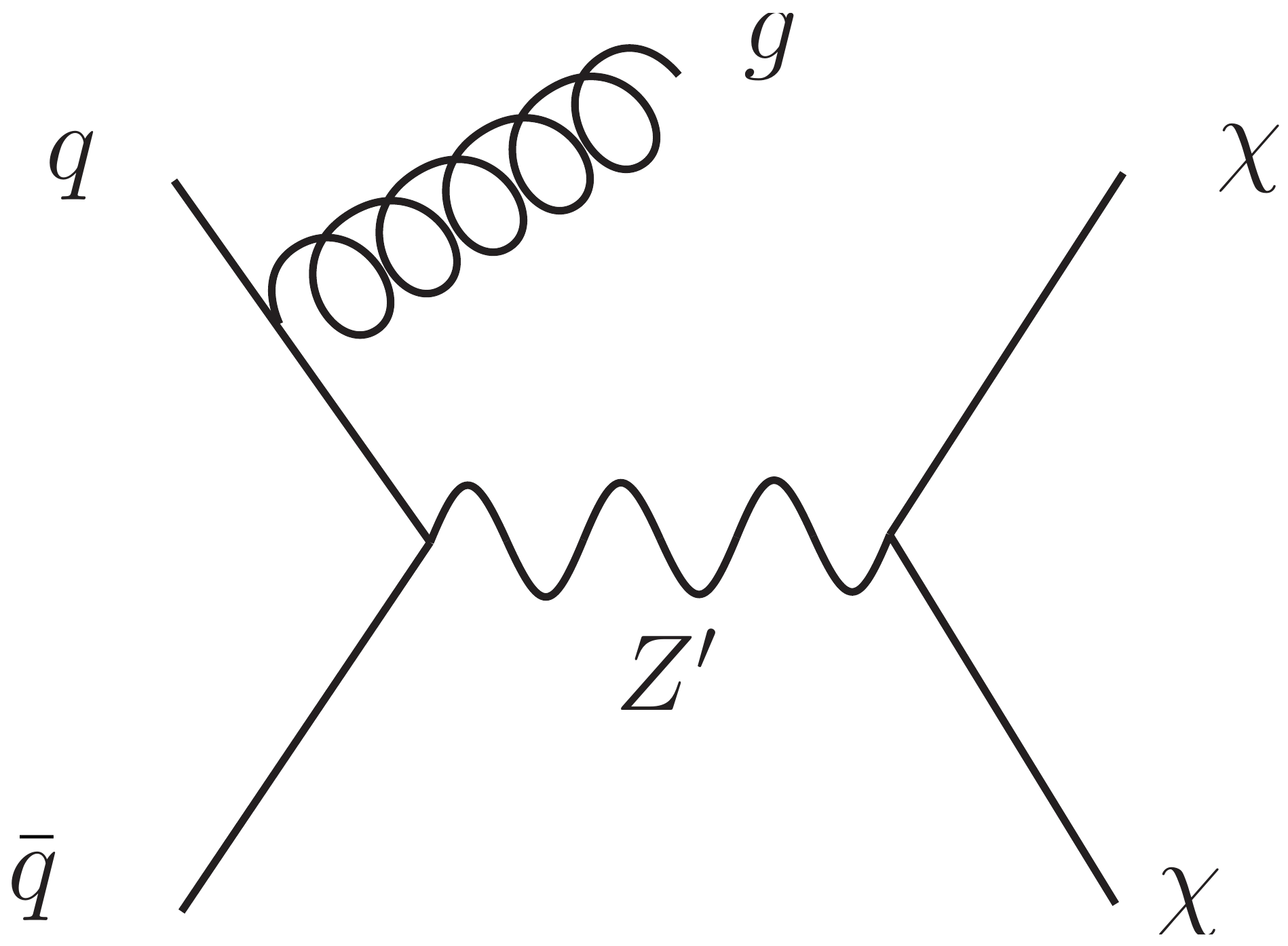}\includegraphics[scale=0.16]{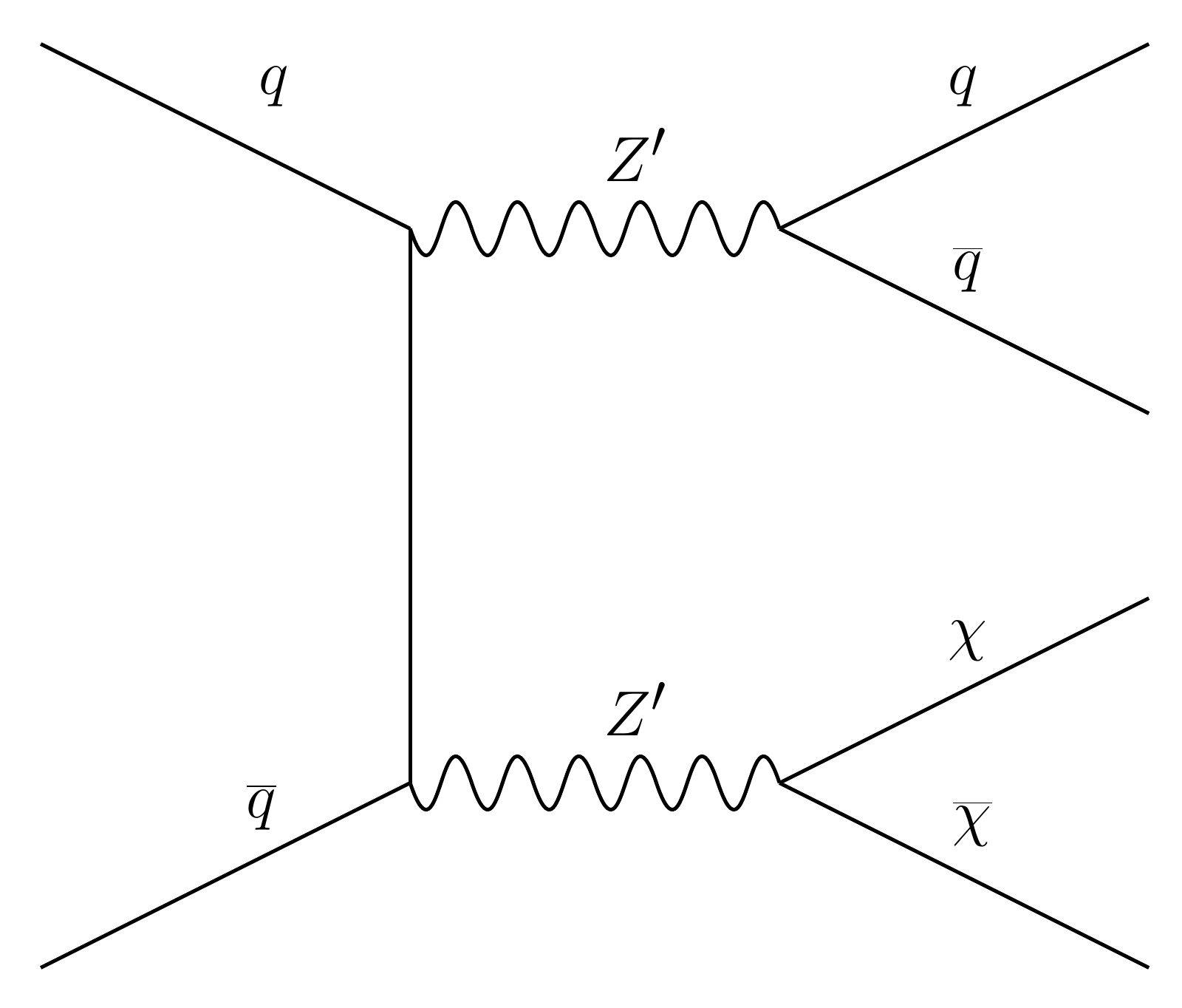}}\\
\hline
$s$-channel scalar &\includegraphics[scale=0.9]{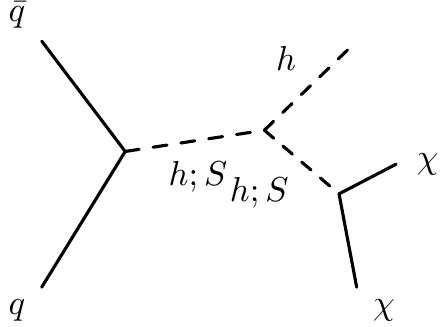}&\includegraphics[scale=0.9]{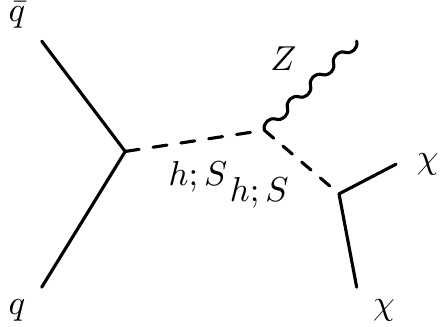}&\includegraphics[scale=0.62]{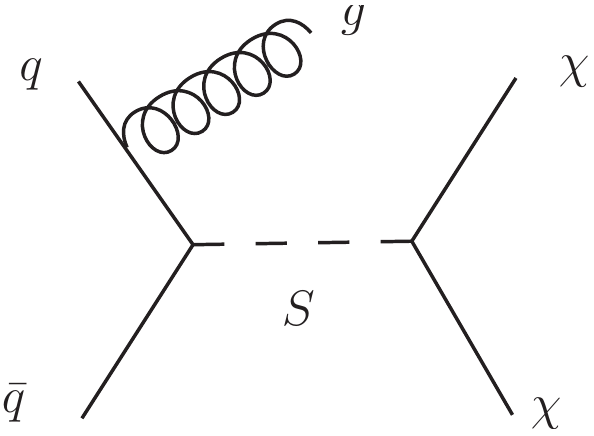}\\
\hline
Inelastic squark &\includegraphics[scale=0.9]{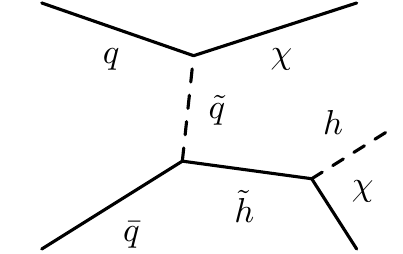}&\includegraphics[scale=0.19]{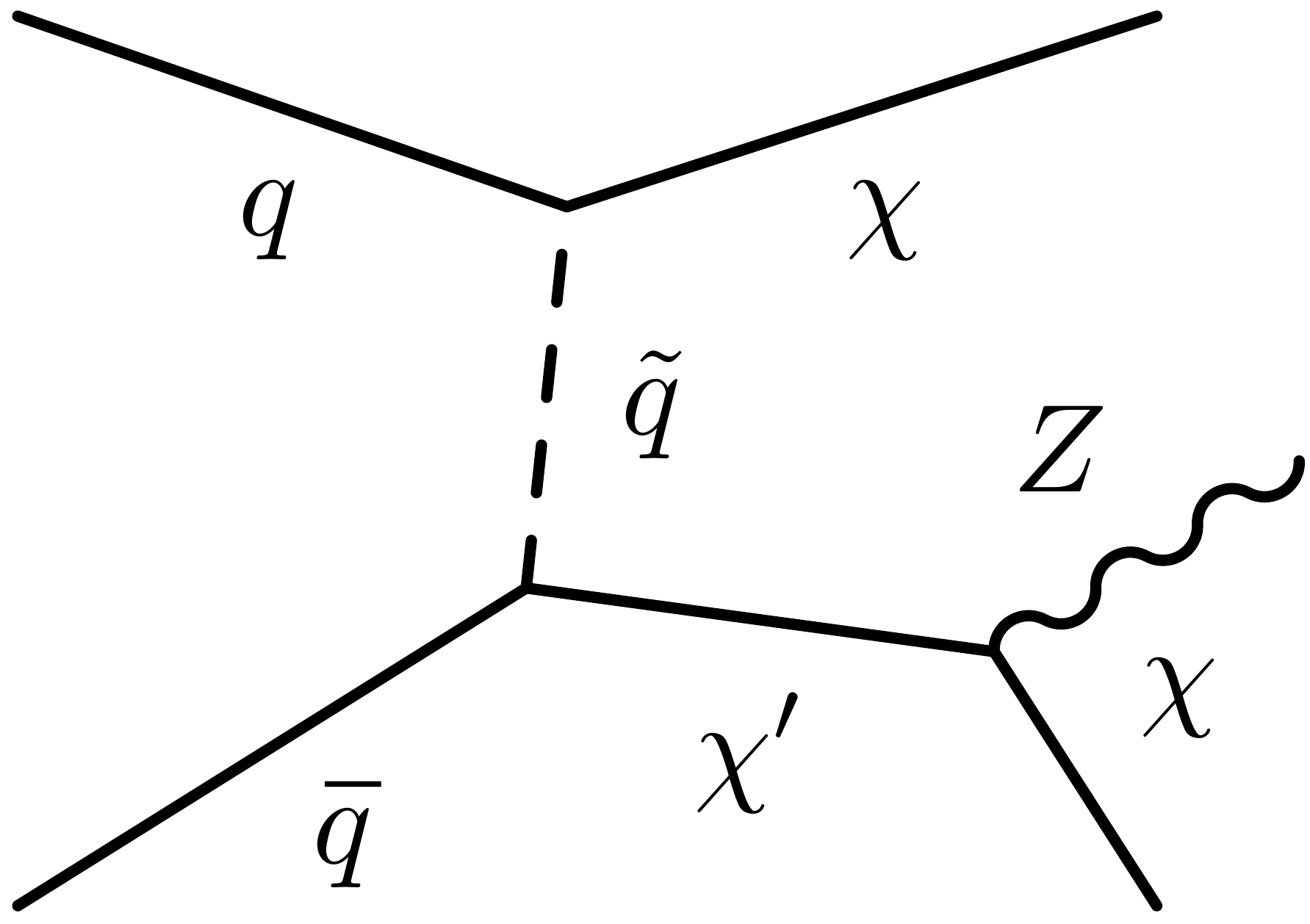}&\includegraphics[scale=0.15]{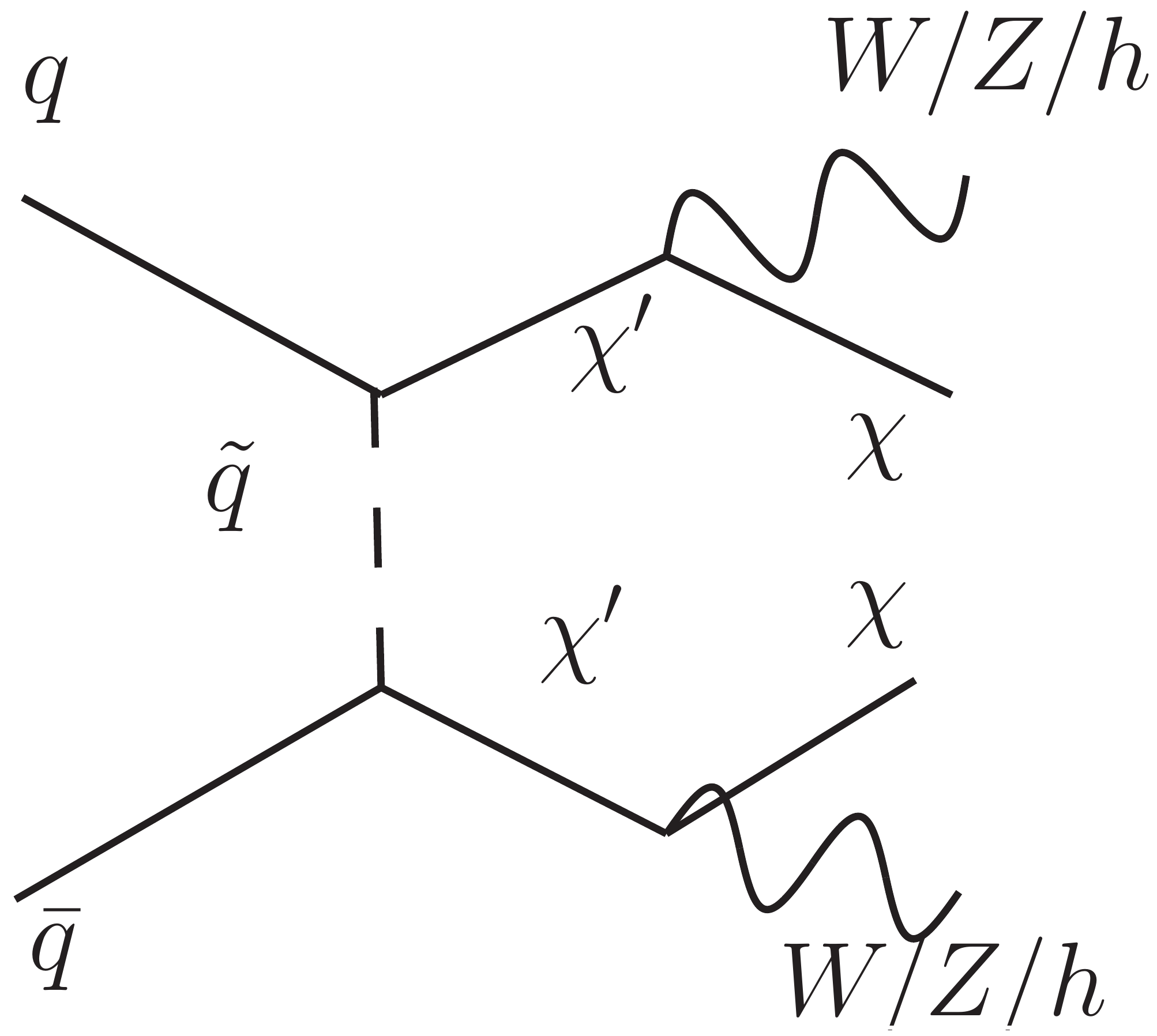}\\
\hline
\end{tabular}}
\caption{ Summary of mono-Higgs and mono-$Z$ topologies, as well as the corresponding relevant direct searches considered in this work.}
\label{tab:ab}
\end{table}   

\begin{table}[H]
\centering

\small{\begin{tabular}{ccc}
\hline
Model & mono-$b$   &direct constraints\\
Sbottoms &\includegraphics[scale=0.65]{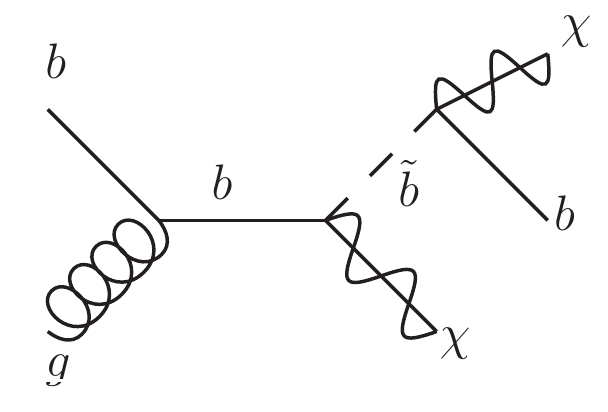} &\includegraphics[scale=0.65]{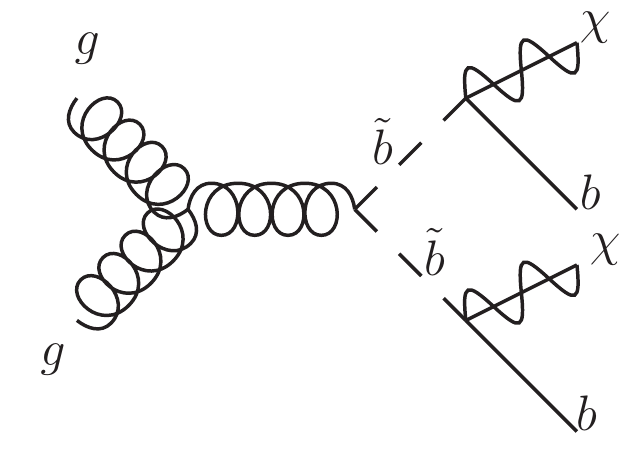}\\
\hline
\end{tabular}}
\caption{ Summary of mono-$b$ topology, as well as the corresponding relevant direct search considered in this work.}
\label{tab:ab2}
\end{table}

The outline of this paper is as follows.  In the next section we summarize the models and analyses utilized in our comparison of mono-$X$ searches against various searches for the mediating particle.  In the following subsections, we then systematically compare the constraints for each model in Table~\ref{tab:ab} and~\ref{tab:ab2} from mono-$X$ to various searches for resonances, as well as for supersymmetry.  Our goal is to highlight the classes of models where mono-$X$ constraints shed the most new light on new physics, beyond what is already constrained by more standard types of searches.  Finally, we conclude.

\begin{table}[htb] 
\centering 
\begin{tabular}{|c||c|}
\hline
Search & Model where it matters
\\
\hline
\hline
mono-$h$ & Inelastic DM, 2HDM   \\
\hline
mono-$z$ & Inelastic DM, 2HDM\\
\hline
mono-jet & Squark mediated production, compressed spectrum \\
\hline
mono-$b$ & Sbottom mediated production, compressed spectrum \\
\hline \hline
\end{tabular} 
\caption{Summary of results: for each mono-$X$ search we list the models where the analysis can exclude part of the parameter space not already ruled out by some other search. }
\label{tab:mono} 
\end{table}


\begin{table}[h] 
\centering 
\begin{tabular}{|c||c|c|}
\hline
Simplified model & searches compared & method
\\
\hline
\hline
Inelastic DM &  mono-$h$  & full recasting \\
 & mono-$z$ & full recasting\\
 \hline
2HDM &  mono-$h$  & full recasting \\
 & mono-$z$ & full recasting\\
 \hline 
 \hline
Squarks  &  mono-$z$  & full recasting \\
($u_{L,R},d_{L,R},c_{L,R},s_{L,R}$) & mono-jet & results of~\cite{Papucci:2014iwa} \\
  & multi-jet + $\met$ & results of~\cite{Papucci:2014iwa}\\
 \hline
Sbottom &  mono-$z$  & full recasting \\
 & mono-$b$ & simplified model~\cite{Aad:2014nra} \\
  & multi-$b$ jets + $\met$& simplified model~\cite{Aad:2013ija}\\
 \hline 
$s$-channel  &  mono-$h$  & results of~\cite{Carpenter:2013xra} \\
scalar mediator & mono-jet & full recasting\\
 \hline
$s$-channel  &  mono-$h$  & results of~\cite{Carpenter:2013xra} \\
vector mediator &mono-jet  & full recasting \\
& multi-jet + $\met$ & full recasting \\
 \hline 
Inelastic Squarks &  mono-$h$  & full recasting \\
 & mono-$z$ & full recasting\\
 & diboson + $\met$ & simplified model~\cite{Aad:2014vma,Khachatryan:2014mma}\\
 & bosons + jets + $\met$ & simplified model~\cite{Aad:2015mia}
\\ \hline
\hline
\end{tabular} 
\caption{Summary of simplified models and analyses considered in this work. The last column indicates whether we perform a full reinterpretation, use the results published by the experimental collaborations, or utilize previous work in the literature.}
\label{tab:searches} 
\end{table}

\section{Simplified models for mono-$X$}
\label{sc:fr}

Before describing the details of each simplified model, we discuss the general properties and assumptions made on the models considered here.  We require that:
\begin{itemize}
\item the DM is a fermionic singlet under the SM gauge group;
\item the mono-$X$ signatures are produced by tree-level topologies,
\item the model have the smallest number of mediating particles for each mono-$X$ topology we consider.
\end{itemize}
We only consider pair production of DM at colliders given that DM is stable on timescales the order of the lifetime of the Universe.  An $s$-channel vector (scalar) mediator is denoted as $Z'$ ($S$). We also use the notation of SUSY whenever a SUSY analogue is applicable to our simplified models. For example, $\tilde q$ denotes the $t$-channel colored mediator that couples to a quark ($q$) and DM ($\chi$). Other auxiliary particles may be needed for constructing our simplified models. They are defined accordingly in the respective subsection describing the details of the simplified model.      
 
Given this set of rules, one can find the list of all the possible topologies and embed each of them in the minimal incarnation of a simplified model as defined above.  We relax the requirement of singlet DM only for the case of the inelastic squark model, where the topology we consider requires the DM to take on SM quantum numbers. 
These requirements are also easy to understand: focusing on singlet DM stems from the fact that searches for DM belonging to weak doublets or triplets are more mature due to the extensive program for SUSY searches. Restricting our focus to tree level topologies and keeping the number of mediators to a minimum instead originates from the attempt to maximize the reach potential of mono-$X$ searches in comparison to direct searches for the mediators.  


For the purpose of illustrating the strength of mono-$X$ and direct searches relevant to these simplified models, we either perform Monte Carlo event simulation, or make use of results of previous works in the literature and as presented by the experimental collaborations. We do not perform full scans in the parameter space of each model, but rather focus on slices of parameter space we believe are highlighting the main qualitative features of the comparisons between mono-$X$ and other searches. A full parameter scan can in principle be performed but it is beyond the scope of this work. We summarize the methods and analyses employed for the simplified models in Table~\ref{tab:searches}. The details of the experimental analyses and our simulations, as well as our method of obtaining 14~TeV projections, are elaborated in Appendices~\ref{app:a},~\ref{app:b}.

\subsection{``Inelastic'' Dark Matter}
We begin by considering a Higgs or $Z$ radiated in the final state through the process $\chi' \rightarrow \chi h$ or $\chi' \rightarrow \chi Z$, where $\chi'$ and $\chi$ are produced via a resonant $Z'$. Here, $\chi'$ is an ``excited" state of DM $\chi$ that decays to DM along with a Higgs or $Z$. These processes arise from interaction Lagrangians of the form $Z'_\mu \chi'\gamma^\mu \chi$, $Z_\mu \chi'\gamma^\mu \chi$ and $h\chi\chi'$.   In order for mono-$h$ or mono-$Z$ to be dominant, production of $\chi' \chi'$ (which leads to di-boson signatures) and $\chi \chi$ (which will be dominated by mono-jet) must be suppressed relative to $\chi \chi'$.  We discuss a concrete model where the mono-boson signature dominates.

For concreteness, we focus on the case where only the right-handed up-quarks (all three generations) are charged under a new gauge symmetry.~\footnote{This model requires the introduction of extra (spectator) fermions to achieve anomaly cancellation. The upper limit of the masses the spectator fermions are $M_{\rm spectator} < (64\pi^2/g_{qqZ'}^3)M_{Z'}$, where $g_{qqZ'}$ is the coupling between SM quarks and the new gauge boson $Z'$, and $M_{Z'}$ is the mass of $Z'$~\cite{Preskill:1990fr,An:2012va}. To focus on the more generic collider signatures of the model, we consider these spectator fermions to be sufficiently heavy (achievable by saturating the aforementioned mass upper limit), such that LHC constraints on them are avoided.} In addition, a new SM singlet Dirac fermion $\psi$ charged under $U(1)_{Z'}$ is introduced as a doublet of DM. Moreover, we introduce a SM singlet scalar $S$ that is charged under $U(1)_{Z'}$. It plays the roles of giving the $Z'$ a mass and acting as a ``portal" to the Higgs.
We also assume that some of the SM quarks are charged under it, to allow a $q\bar q Z'$ coupling. In the following we will allow only the right-handed up quark to carry $U(1)_{Z'}$ charge which we fix to $1/2$.
We take the DM ($\psi = (\eta \ \bar{\xi} )$) Lagrangian to be: 
\ba
{\cal L}_{ \rm DM} = \bar \psi (i\slashed{D} - m_\psi) \psi - (\lambda_1\ S^* \eta \eta +\lambda_2\ S  \xi \xi+ h.c.),
\ea
where $D_\mu = \partial_\mu +ig_Xq_{\psi}\hat{X}_\mu$.  We define new bases $\chi_1, \chi_2= 1/\sqrt{2}(\eta\mp \xi)$ and new couplings $\lambda_{\pm}=\lambda_1\pm \lambda_2$ such that, after the $U(1)_{Z'}$ symmetry is spontaneously broken, the fermion bilinear terms are written as:
\be
 \label{lowWf}
{\cal L}^{\rm bi}_{ \rm DM} =  -\frac{1}{2} (-m_{\psi}\chi_1^2+m_\psi \chi_2^2 +\lambda_{+}\langle S\rangle \chi_1^2+\lambda_{+}\langle S\rangle \chi_2^2)-\lambda_-\langle S\rangle\chi_1\chi_2+{\rm h.c.}.
\ee

\begin{figure}
\centering
\subfigure[]{\includegraphics[scale=0.64]{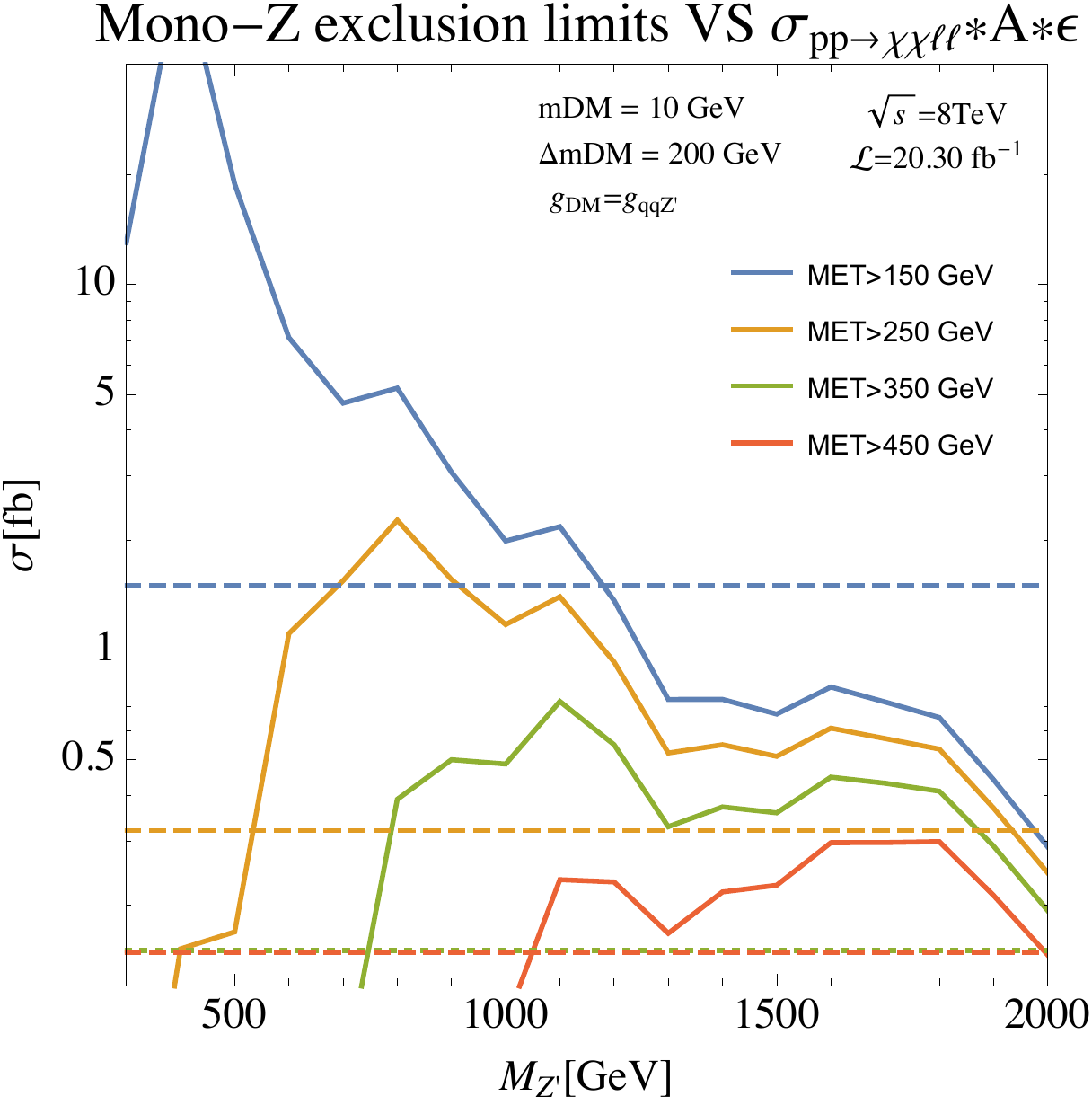}}
\subfigure[]{\includegraphics[scale=0.64]{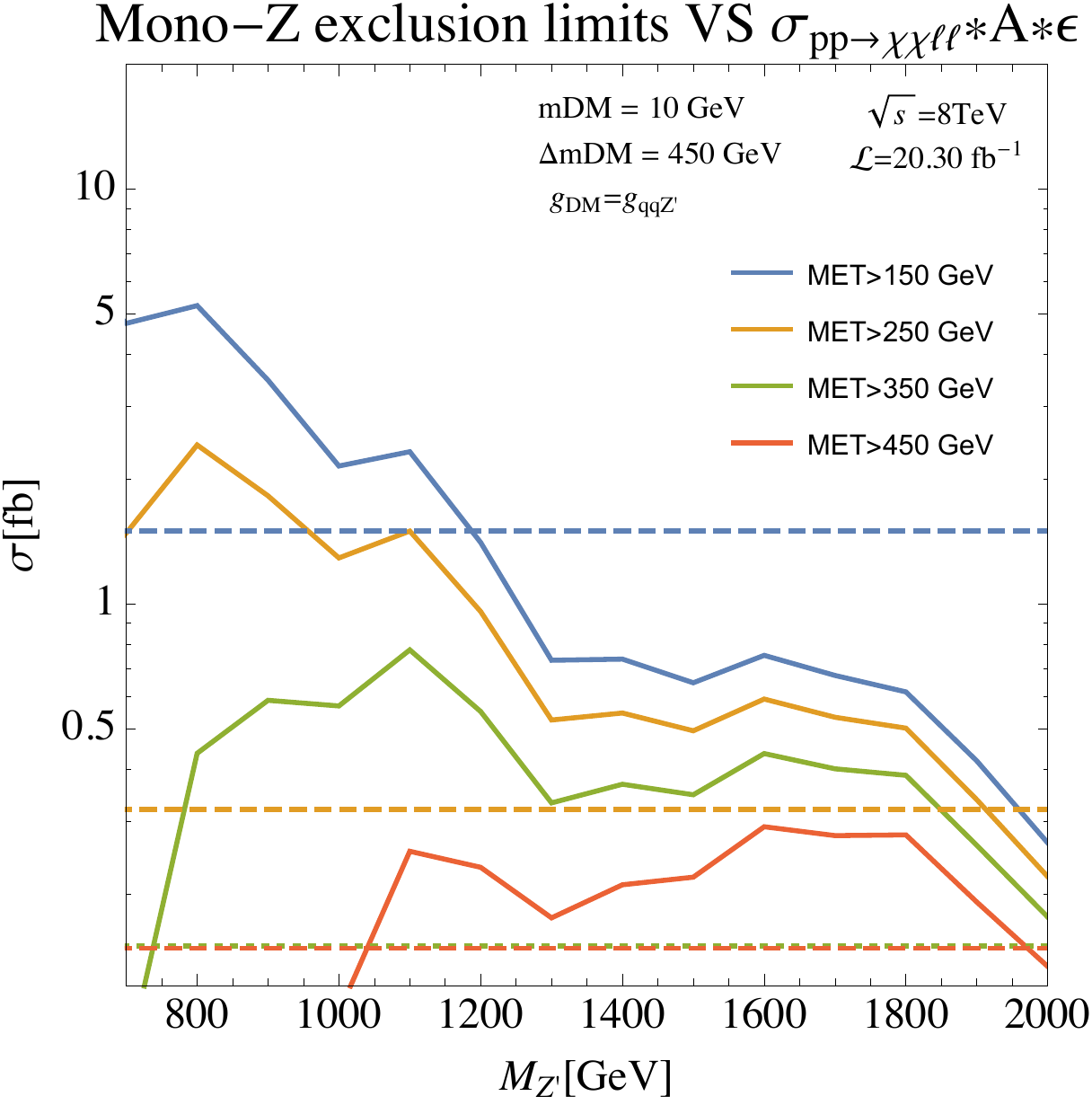}}
\subfigure[]{\includegraphics[scale=0.64]{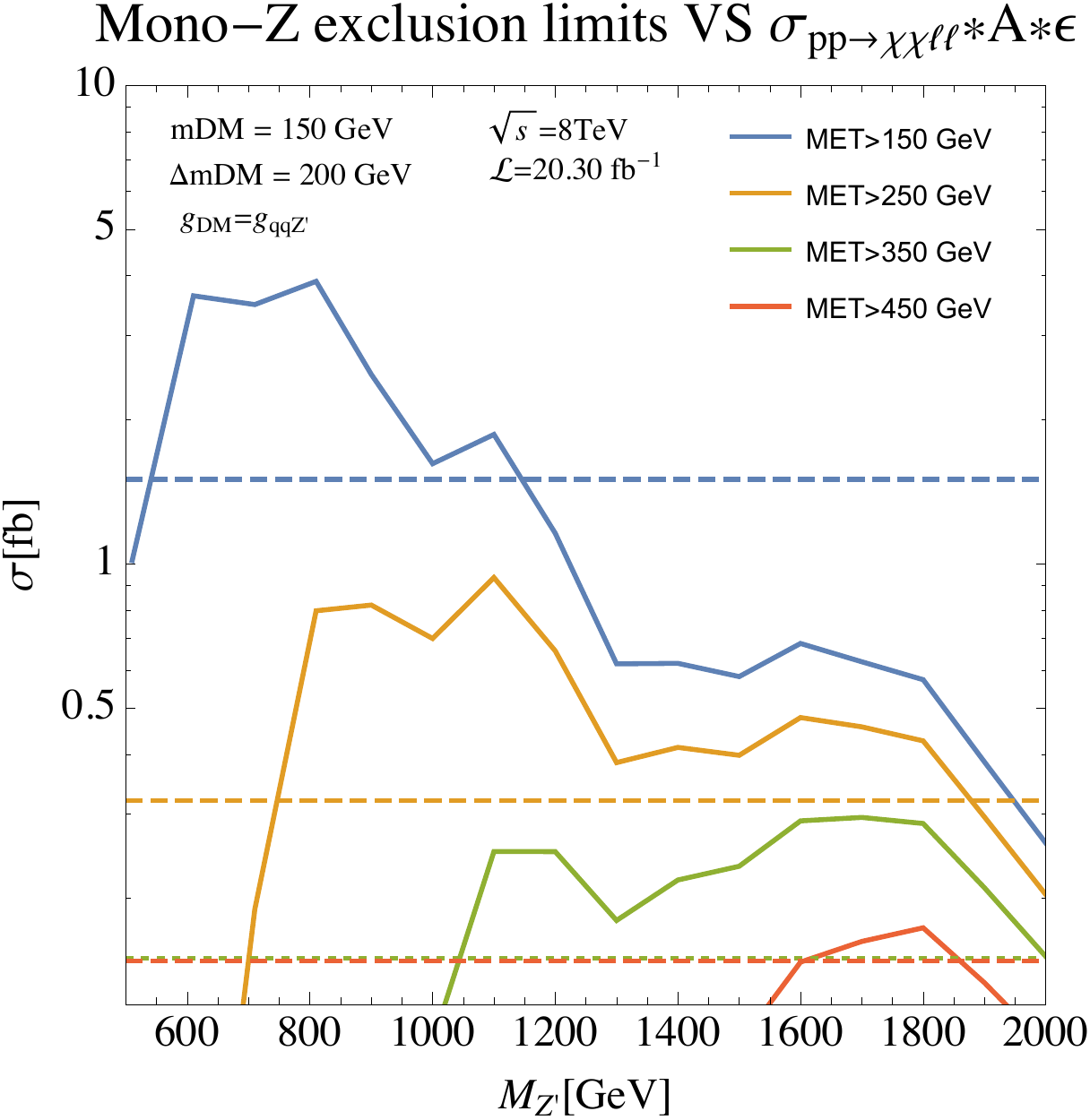}}
\subfigure[]{\includegraphics[scale=0.64]{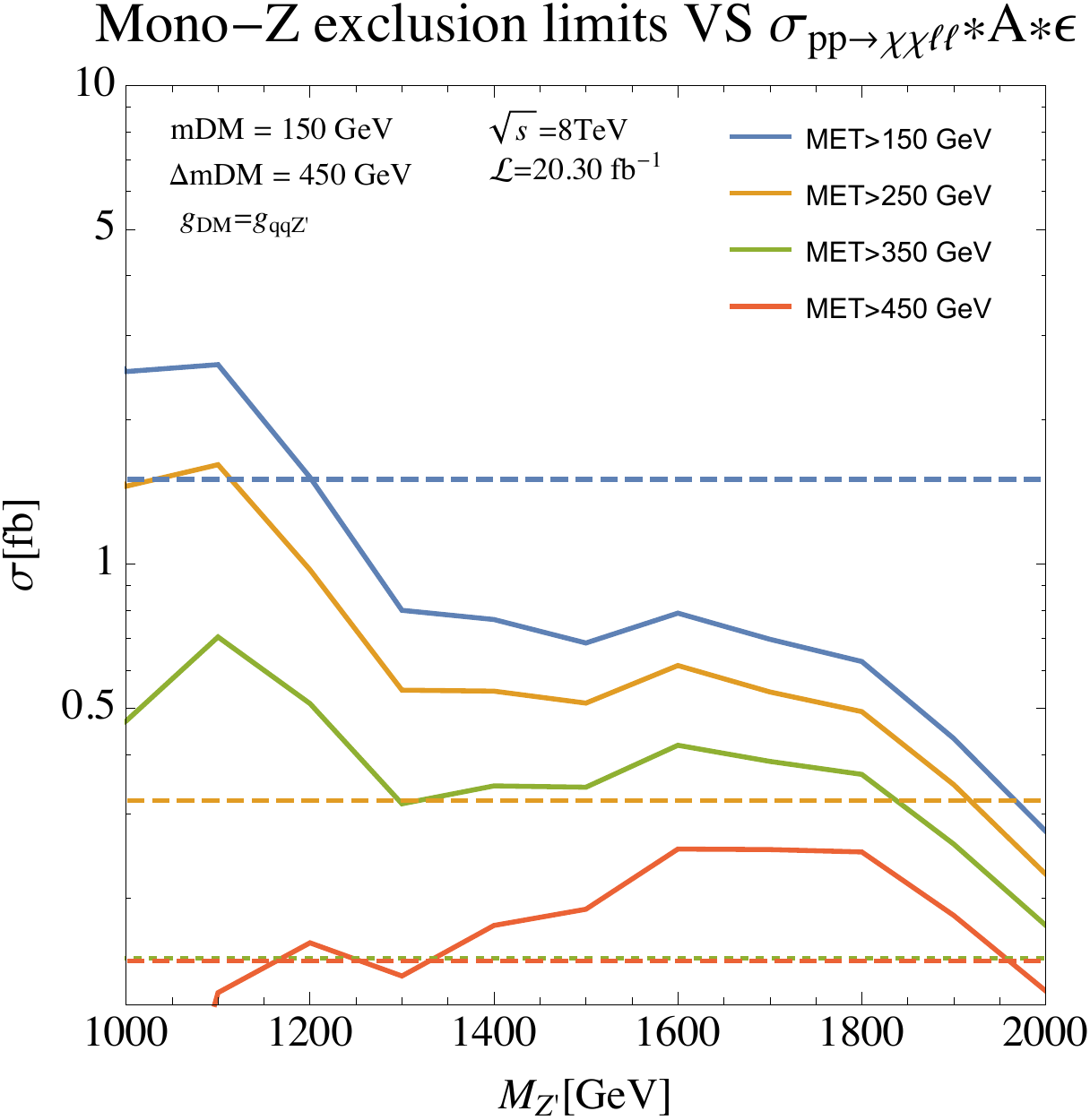}}
 \caption{Inelastic DM model: mono-$Z$ exclusion cross-section at 95\% C.L., shown as dashed lines, from 8~TeV data. $\chi'$ is assumed to have a 100\% decay to $Z\chi$. The solid lines correspond to the prediction of the model when the coupling of $Z'$ to the quarks $g_{qq Z'}$ is chosen to be equal to the the upper limit consistent with di-jet constraints at a given $Z'$ mass (see Fig.~\ref{fig:dijet1}). Panels (a)-(d) correspond to the choice of the mass parameters $(\mdm,\Delta \mdm)= (10,200), (10,450), (150,200), (150,450)$ in GeV, respectively, where $\Delta \mdm$ is the $\chi' \chi$ mass splitting. The four colors represent the four different $\met$ choices in the mono-$Z$ analysis (150, 250, 350 and 450 GeV). } 
 \label{fig:InelasticMonoZ8TeV} 
\end{figure}

\begin{figure}
\centering
\subfigure[]{\includegraphics[scale=0.64]{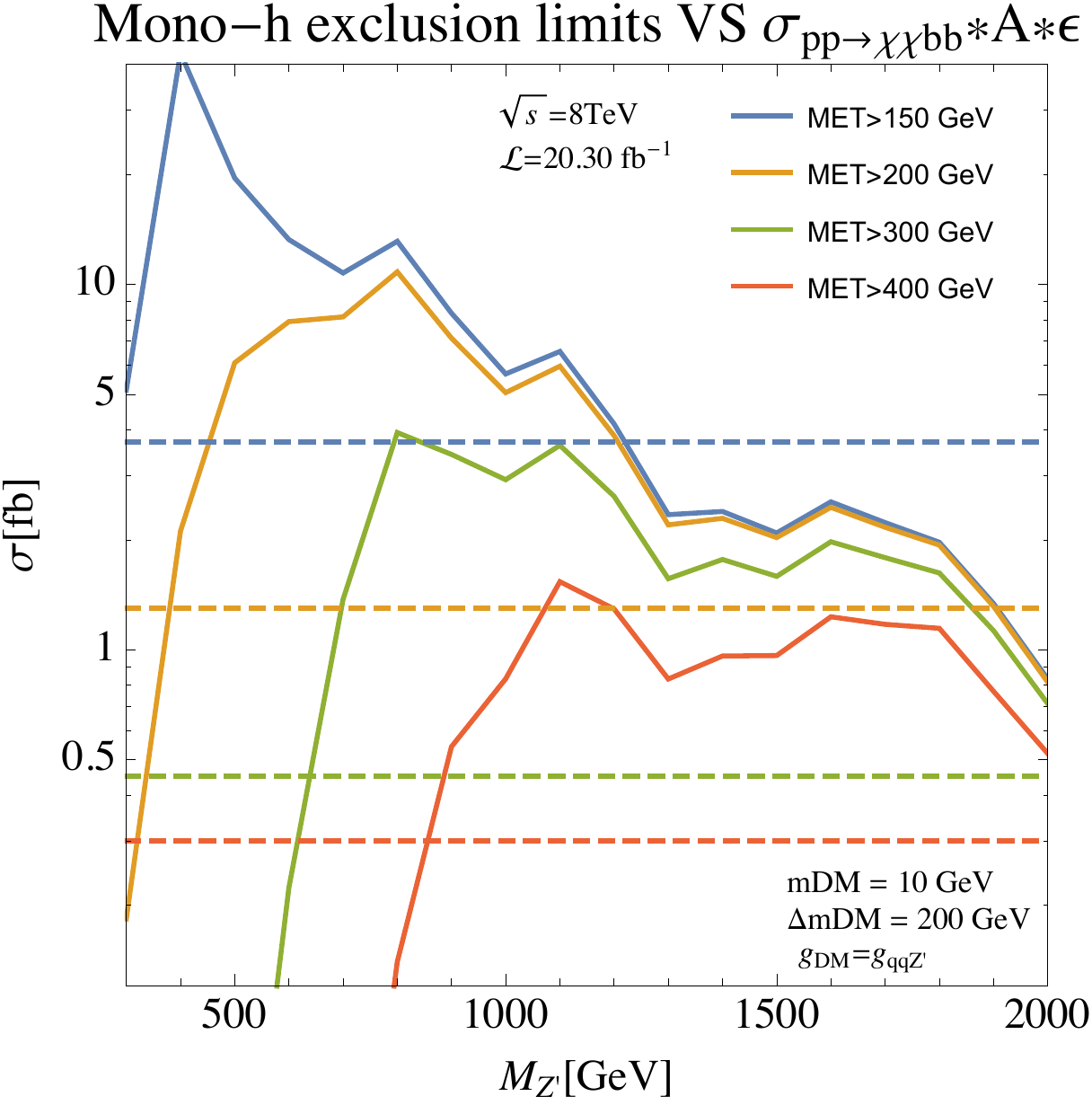}}
\subfigure[]{\includegraphics[scale=0.64]{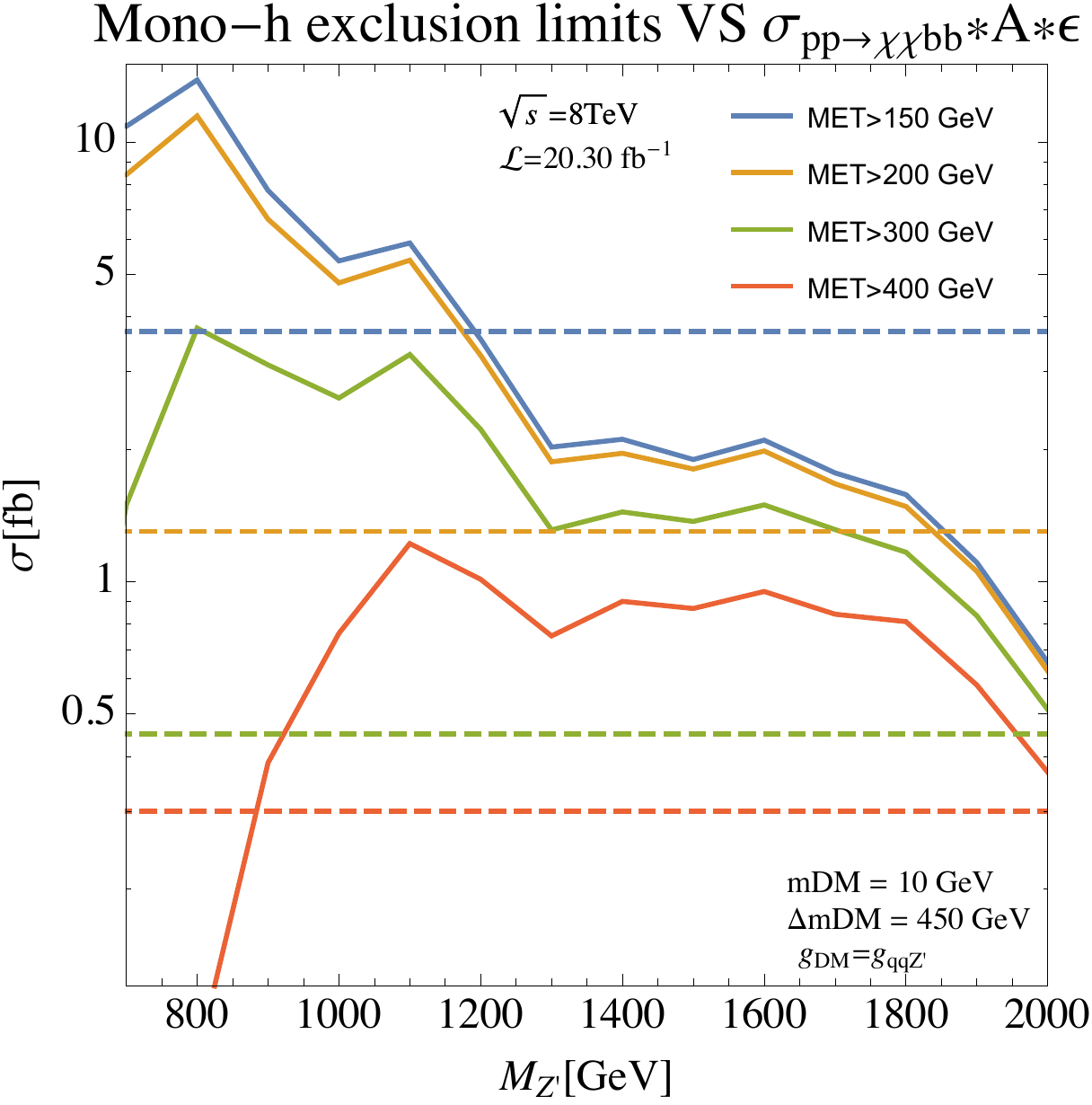}}
\subfigure[]{\includegraphics[scale=0.64]{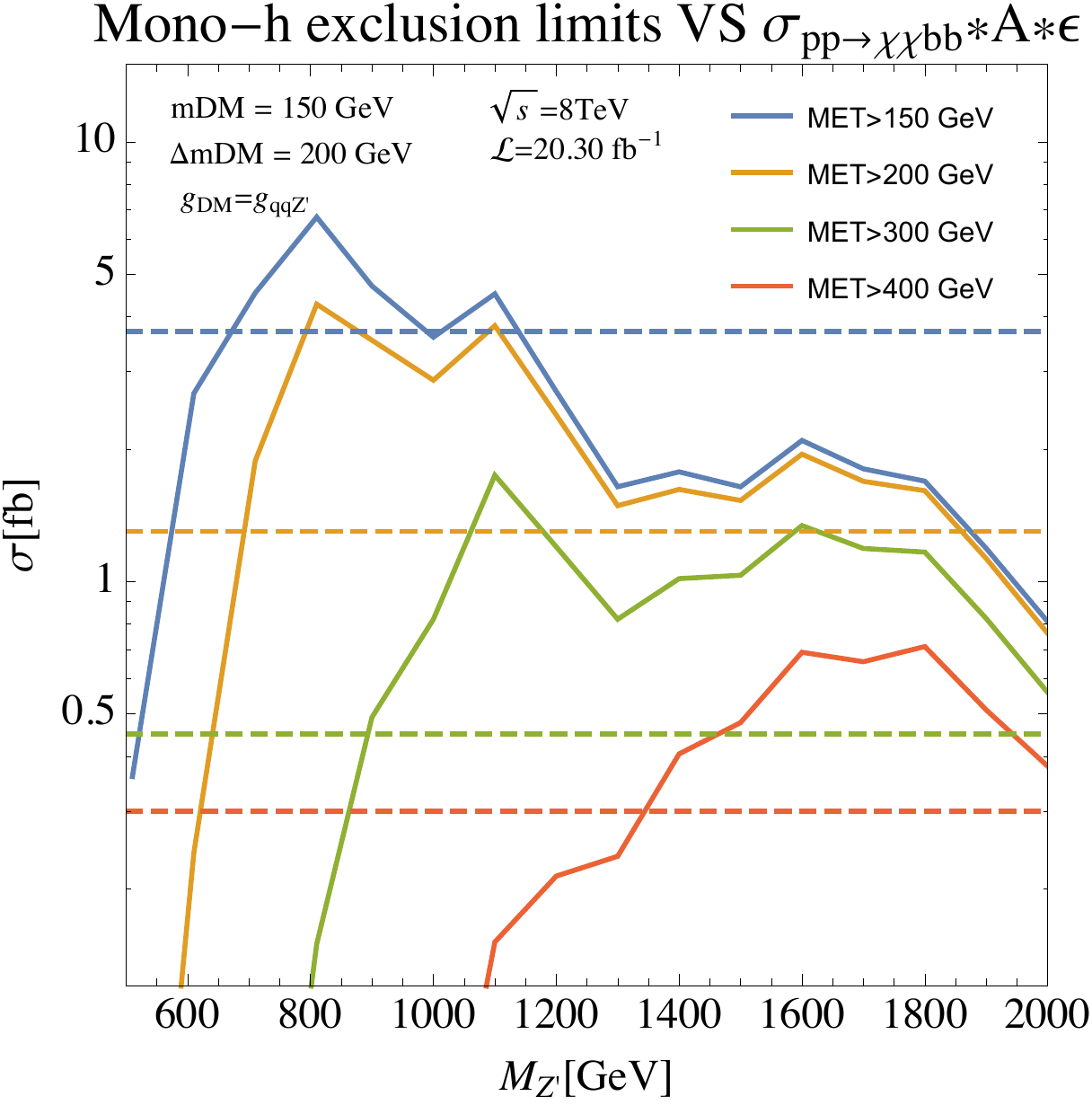}}
\subfigure[]{\includegraphics[scale=0.64]{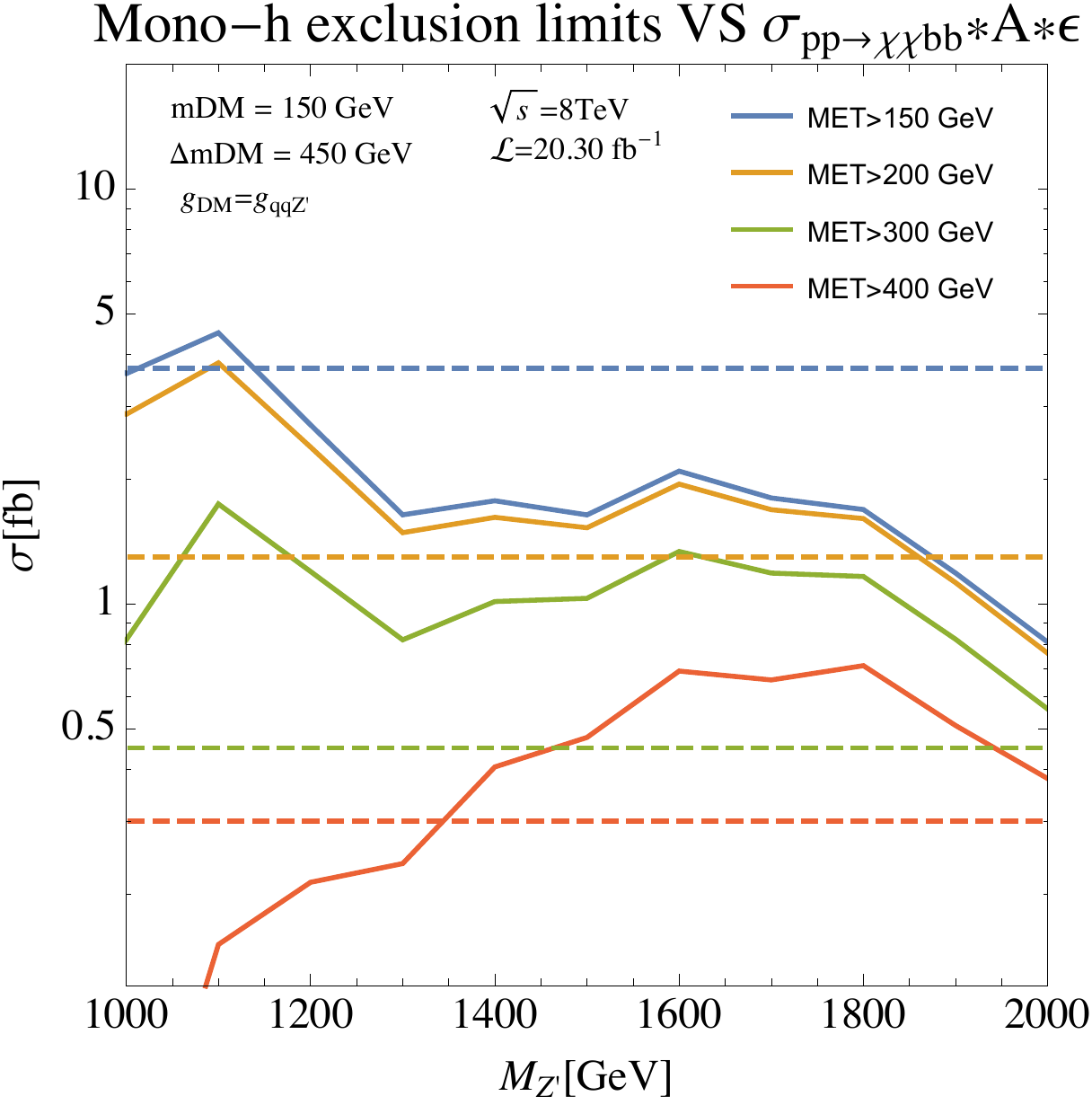}}
 \caption{Inelastic DM model: mono-$h$ exclusion cross-section at 95\% C.L., shown as dashed lines, from 8~TeV data. $\chi'$ is assumed to have a 100\% decay to $h\chi$. The solid lines correspond to the prediction of the model when the coupling of $Z'$ to the quarks $g_{qq Z'}$ is chosen to be equal to the the upper limit consistent with di-jet constraints at a given $Z'$ mass (see Fig.~\ref{fig:dijet1}). Panels (a)-(d) correspond to the choice of the mass parameters $(\mdm,\Delta \mdm)= (10,200), (10,450), (150,200), (150,450)$ in GeV, respectively, where $\Delta \mdm$ is the $\chi' \chi$ mass splitting.  The four colors represent the four different $\met$ choices in the mono-$h$ analysis (150, 200, 300 and 400 GeV).} 
 \label{fig:InelasticMonoH8TeV}
\end{figure}

Introducing the mass eigenstates (with abbreviations $c_{\chi}\equiv\cos{\theta_{\chi}}$, $s_{\chi}\equiv\sin{\theta_{\chi}}$),
\begin{displaymath} \left( \begin{array}{c} \chi_1 \\ \chi_2
\end{array} \right) = \left( \begin{array}{cc} c_{\chi} & s_{\chi} \\ -s_{\chi} &
c_{\chi} \end{array} \right) \left( \begin{array}{c} \chi \\ \chi' \end{array}
\right),
\end{displaymath}
the mixing angle and mass eigenvalues are given by 
\begin{eqnarray}
\tan{2\theta_{\chi}} &=& \frac{\lambda_-\langle S \rangle}{m_{\psi}}\, ,
\
\\ M_{\chi,\chi'}^2 =  \lambda_+\langle S \rangle
            &\mp& \sqrt{ m_{\psi}^2+\lambda_{-}^2\langle S \rangle^2} \, .
\end{eqnarray}
In the new basis, $\chi$ is the DM candidate while $\chi'$ is the ``excited" state of DM. 

$S$ and $Z'$ mix with the SM Higgs and $Z$ respectively, and facilitate the mono-$X$ processes $\chi'\to\chi h$ and $\chi'\to\chi Z$. The interaction of the scalar field $\delta S \equiv \sqrt{2}(S-\langle S \rangle)$ with the DM doublet is as follows:
\ba
{\cal L}^{\rm sc.int}_{ \rm DM} = \frac{\lambda_+}{\sqrt{2}} \delta S (\chi'^2 +\chi^2)  + \frac{\lambda_-}{\sqrt{2}} (c_{\chi}^2-s_{\chi}^2)\delta S \chi \chi' + {\rm h.c.},
\ea
while the interaction of $Z'$ with the DM doublet is:
\ba
\bar{\eta}\slashed{\hat{X}}\eta-\bar{\xi}\slashed{\hat{X}}\xi=2g_Xq_{\psi} s_{\chi}c_{\chi}(\bar{\chi'}\slashed{\hat{X}}\chi'-\bar{\chi}\slashed{\hat{X}}\chi)-g_Xq_{\psi}(c_{\chi}^2-s_{\chi}^2)(\bar{\chi}\slashed{\hat{X}}\chi'+\bar{\chi'}\slashed{\hat{X}}\chi).
\ea
 Let us note that at the limit where the mixing angle $\theta_{\chi}\to 0$, the couplings of $Z'$ to $\chi\chi$ and $\chi'\chi'$ (leading to di-boson signature) vanish, and the $\chi\chi'$ production (leading to mono-boson signature) becomes dominant.

The $Z'-Z$ mixing originates from the radiative corrections that lead to kinetic mixing between the $U(1)$ gauge bosons:
\ba
{\cal L}^{KE}_V = -\frac{1}{4} \hat{X}_{\mu\nu} \hat{X}^{\mu\nu} + \frac{\epsilon}{2} \hat{X}_{\mu\nu} \hat{B}^{\mu\nu} \ ,
\ea
where $\epsilon$ is expected to have the size $\epsilon \sim g_Xg'/16\pi^2 \lesssim 10^{-3}$ from fermion loops. The Higgs sector Lagrangian of the model is written as:
\ba 
{\cal L}_{H} &=& |D_\mu H_{SM}|^2
+ |D_\mu S |^2 
 + m^2_{S}|S|^2 + m^2_{H}|H_{SM}|^2 \nonumber\\
& & - \lambda|H_{SM}|^4 - \rho|S|^4 - \kappa
|H_{SM}|^2|S|^2. \label{Lphi.EQ} 
\ea
 $U(1)_X$ is broken spontaneously by $\left< S \right>$,
and electroweak symmetry is broken spontaneously as usual by
$\left< H_{SM}\right> = (0,v/\sqrt{2})$. The two physical Higgs bosons $h$ and $S$ mix with each other after spontaneous symmetry breaking. Whether $\chi'$ decays to $h$ or $Z$ mainly depends on the value of $\epsilon$ and $\kappa$, which are in principle free parameters. We also note that the elastic scattering of DM off nucleons via $Z'$ is suppressed as long as $\theta_{\chi}$ is small.  

\begin{figure}[H]
\centering
\subfigure[]{\includegraphics[scale=0.64]{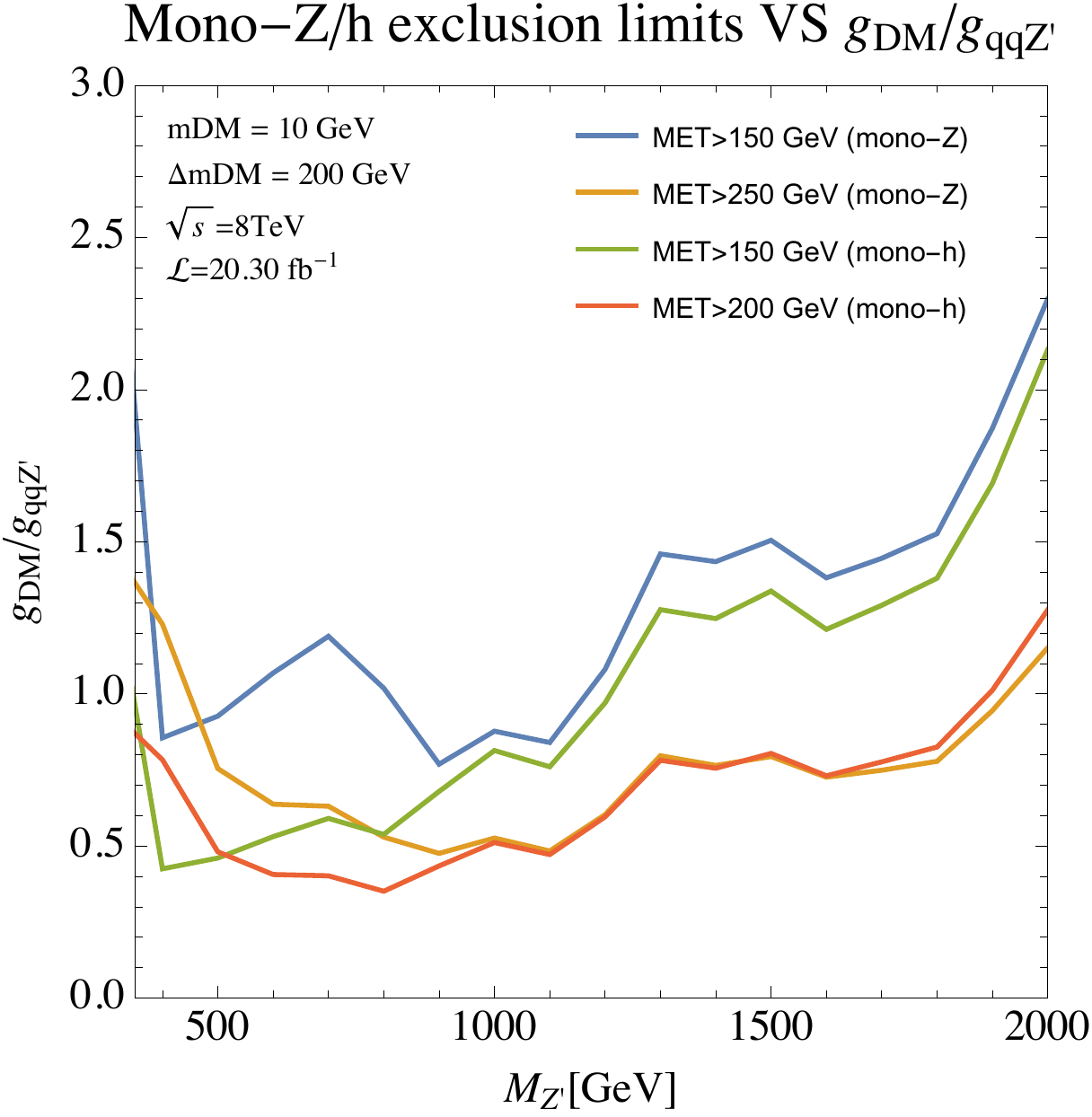}}
\subfigure[]{\includegraphics[scale=0.64]{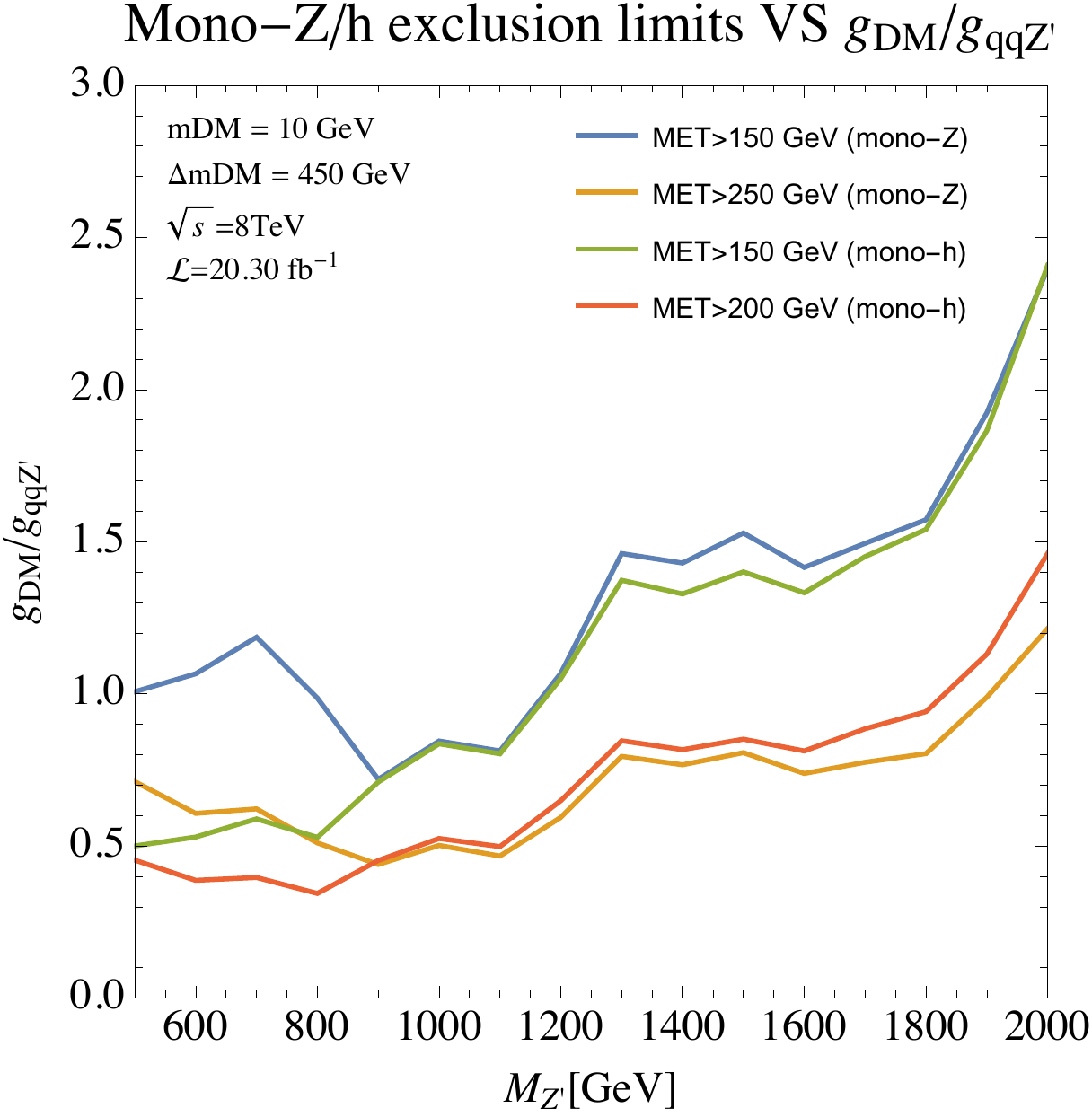}}
\subfigure[]{\includegraphics[scale=0.64]{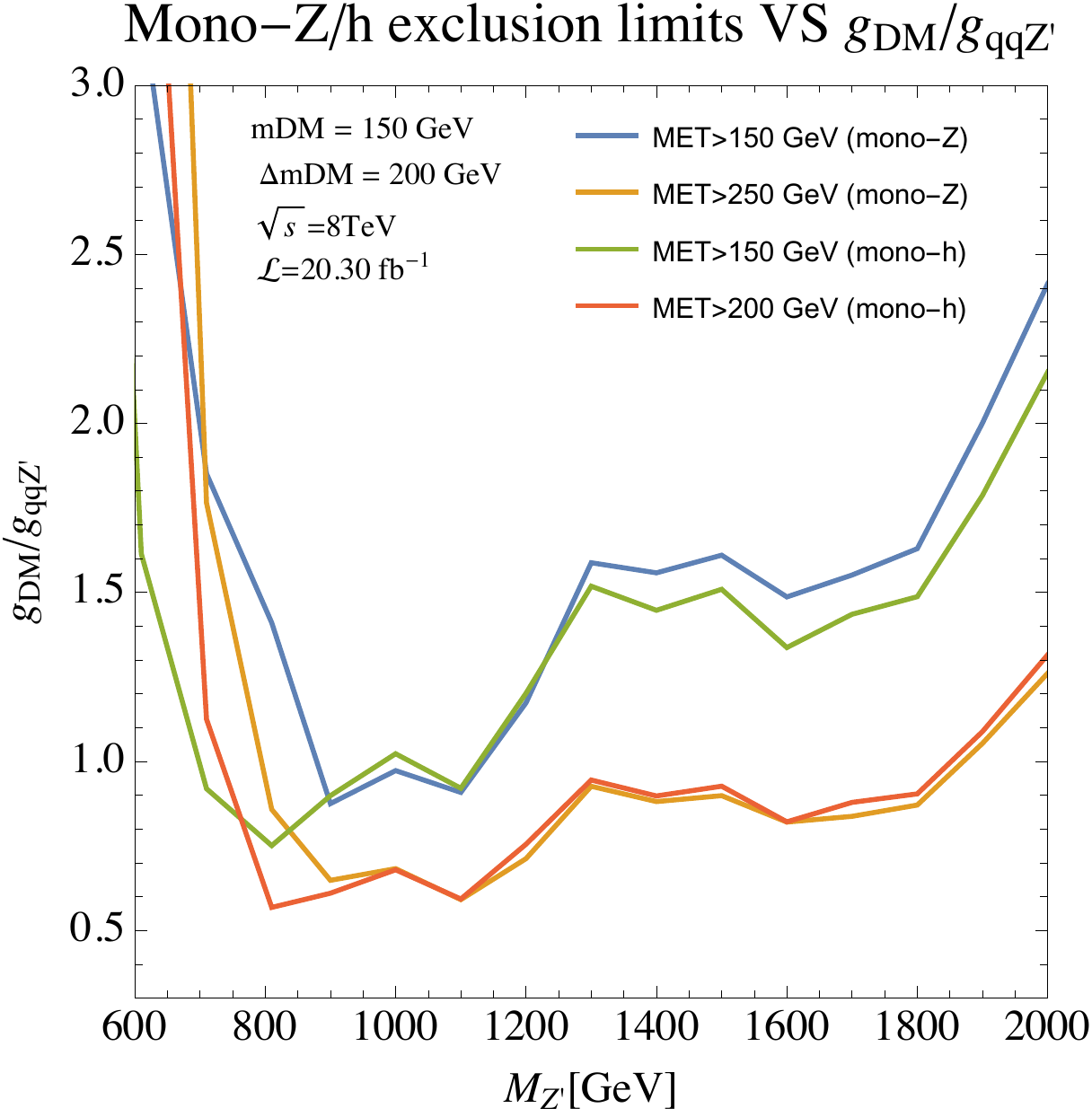}}
\subfigure[]{\includegraphics[scale=0.64]{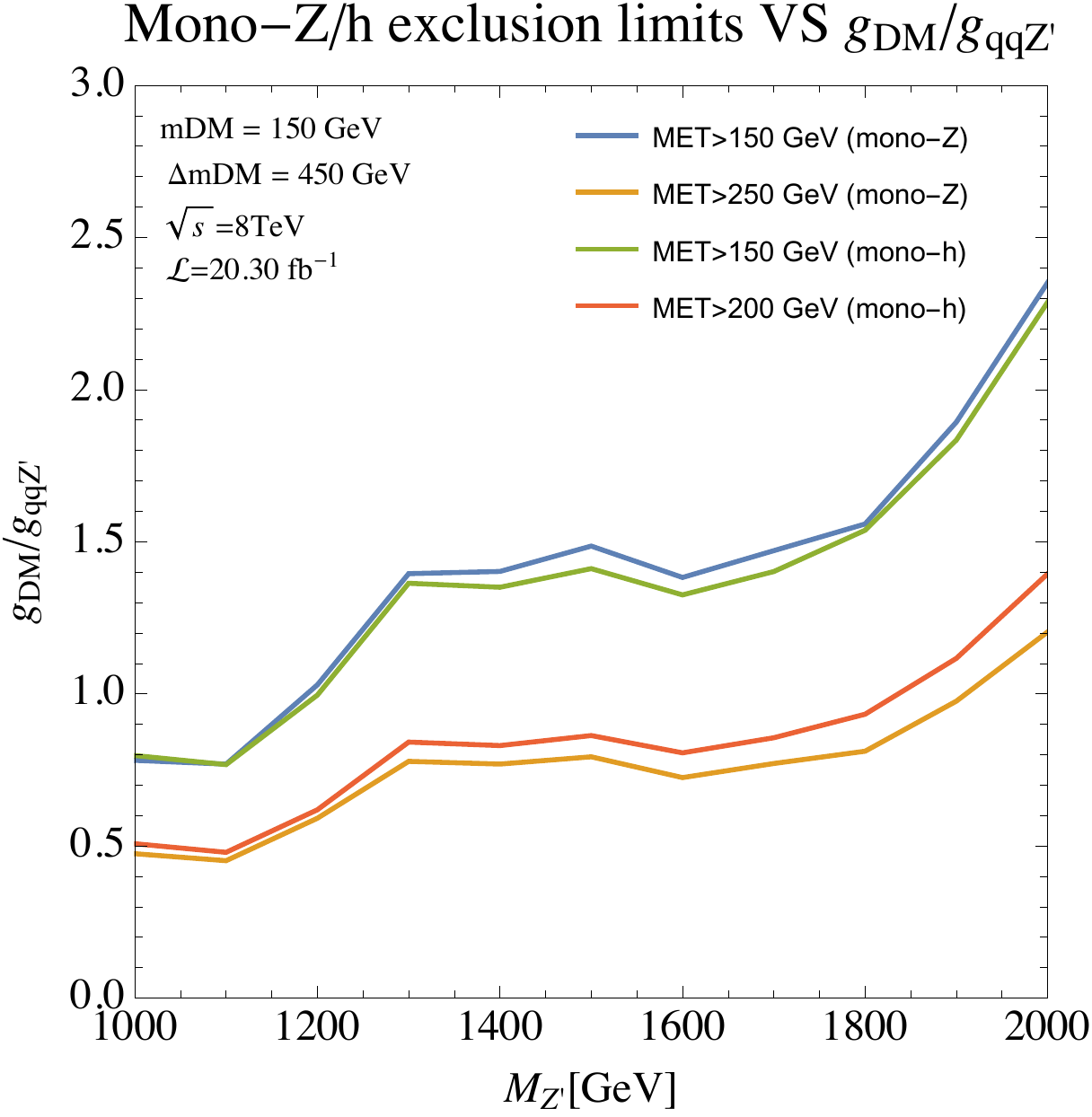}}
\caption{Inelastic DM model: constraints on the ratio of the couplings ($g_{\rm DM}/g_{qqZ'}$) as a function of the $Z'$ mass from mono-$Z$ and mono-$h$ searches at 8~TeV. The coupling of $Z'$ to the quarks $g_{qq Z'}$ is chosen to be equal to the the upper limit consistent with di-jet constraints at a given $Z'$ mass (see Fig.~\ref{fig:dijet1}). Panels (a)-(d) correspond to the choice of the mass parameters $(\mdm,\Delta \mdm)= (10,200), (10,450), (150,200), (150,450)$ in GeV, respectively.} 
\label{fig:InelasticMonoW8TeV}
\end{figure}

We compare the constraints from mono-$h$ and mono-$Z$ analyses on the cross-section times branching fraction in Figs.~\ref{fig:InelasticMonoZ8TeV},~\ref{fig:InelasticMonoH8TeV}. We investigate four benchmark points which have different combinations of DM mass $\mdm$ (10 GeV and 150 GeV) and $\Delta \mdm\equiv m_{\chi'}-m_{\chi}$ (200 GeV and 450 GeV).  We can see in these two figures that both final states can be constraining, though the mono-$Z$ search with 250 GeV $\met$ cut (mono-$h$ search with 300 GeV $\met$ cut) is typically strongest.  In all of the figures, we have chosen the coupling to quarks $g_{qq Z'}$ to saturate the di-jet resonance search constraints at a given $Z'$ mass (see Fig.~\ref{fig:dijet1}). In addition, we vary the DM-$Z'$ coupling ($g_{\rm DM}$) and show in Fig.~\ref{fig:InelasticMonoW8TeV} the 95\% C.L. upper limit on the ratio $g_{\rm DM}/g_{qqZ'}$ from the mono-$Z$ and mono-$h$ searches. We further compare future projections for 14~TeV  mono-$h$ and mono-$Z$ analyses in Fig.~\ref{fig:InelasticMonoZ14TeV}, taking a $\met$ cut of 400 GeV, as described in Appendix~\ref{app:b}. It is observed that the bounds on the production cross-section for large mediator mass is vastly improved at increased center of mass energy.

\begin{figure}[H]
\centering
\subfigure[]{\includegraphics[scale=0.64]{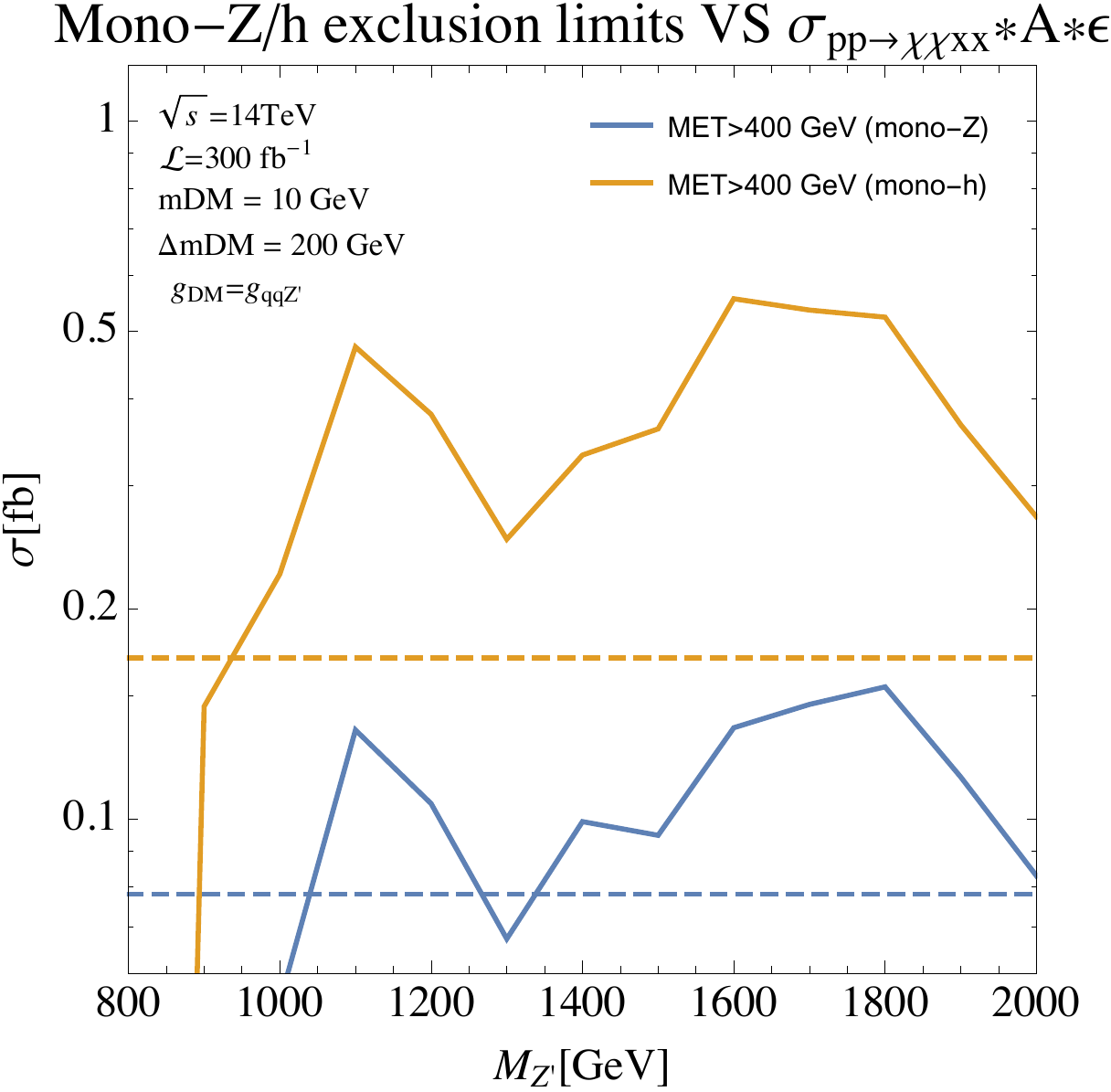}}
\subfigure[]{\includegraphics[scale=0.64]{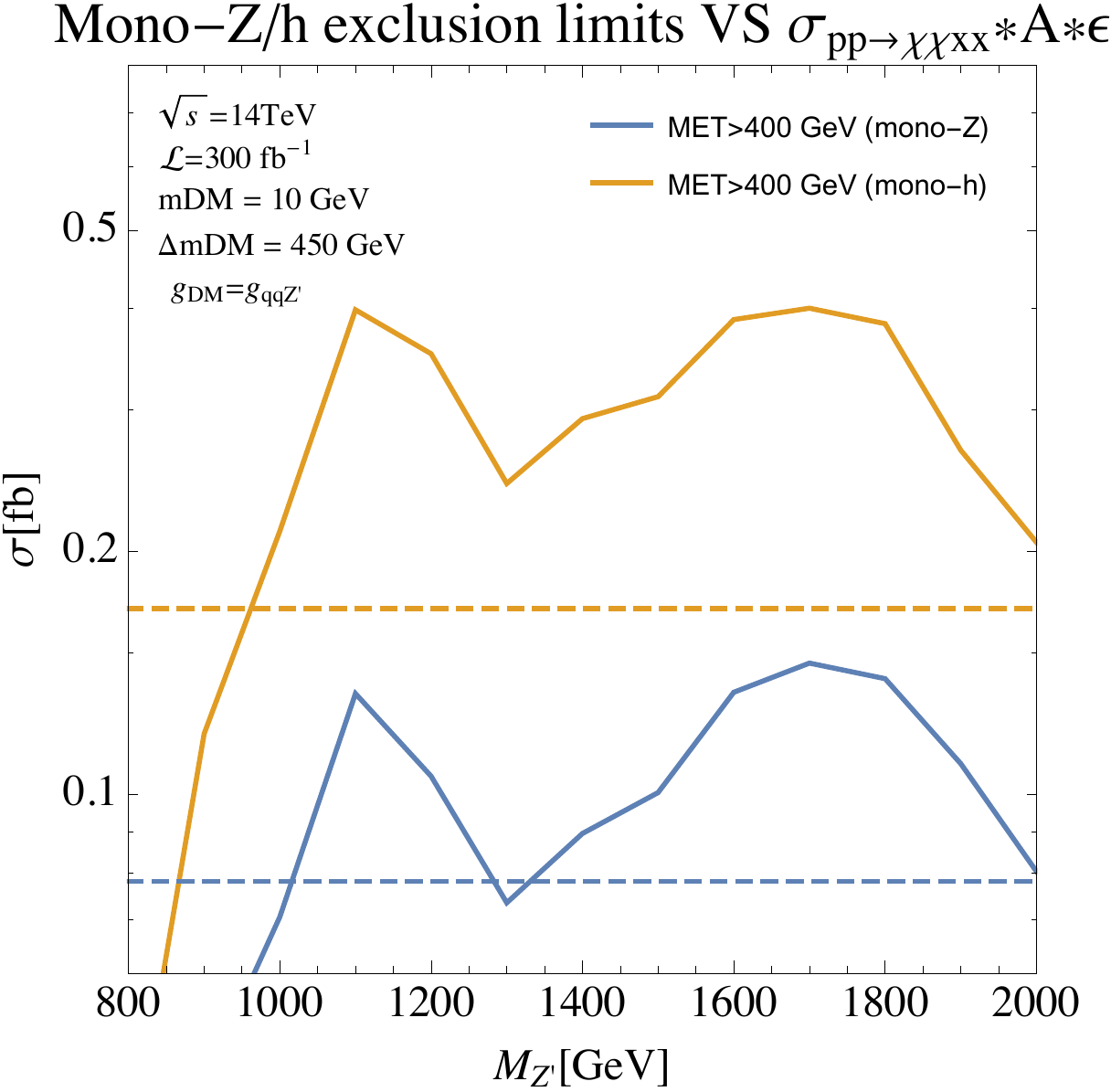}}
\subfigure[]{\includegraphics[scale=0.64]{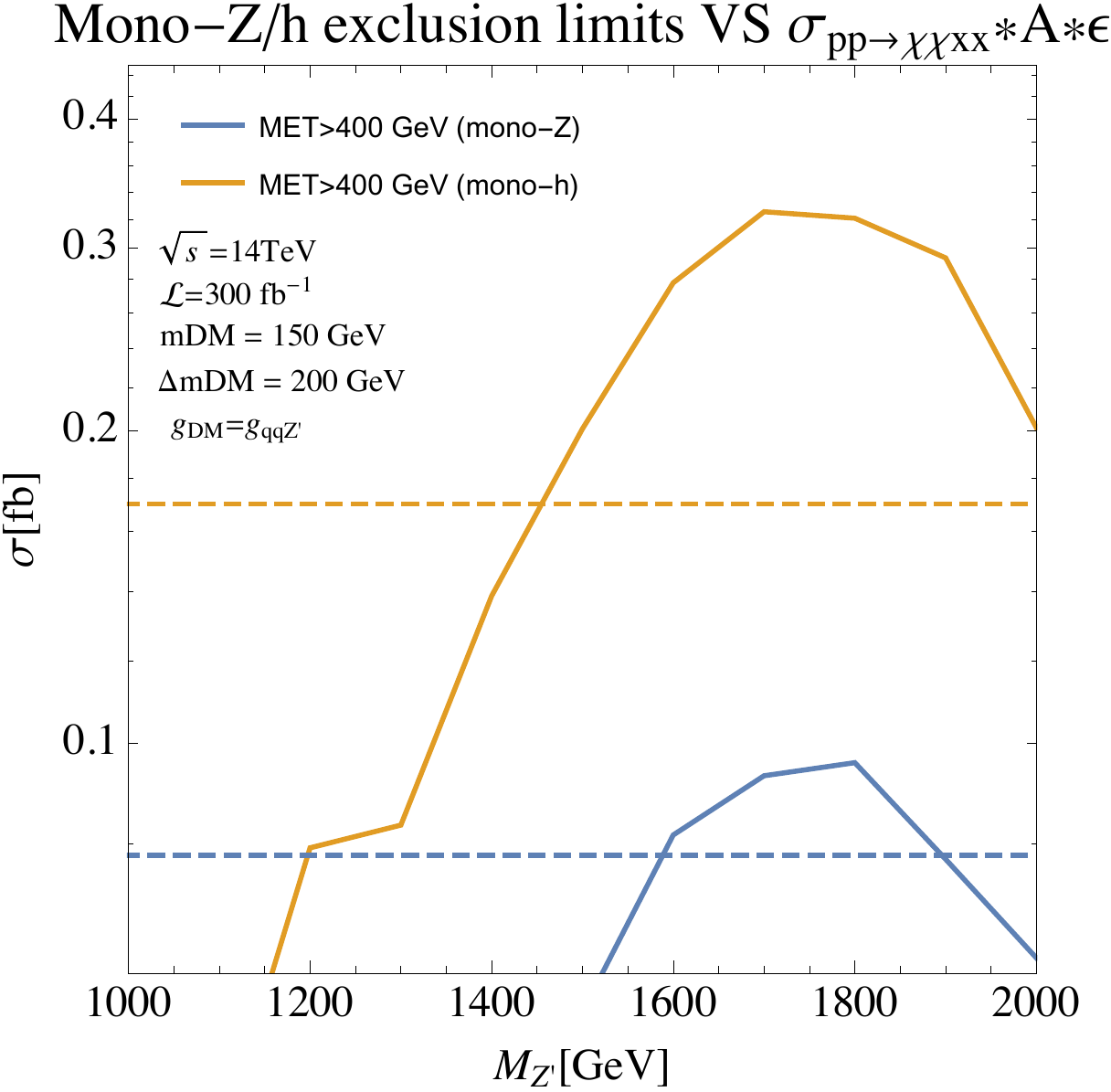}}
\subfigure[]{\includegraphics[scale=0.64]{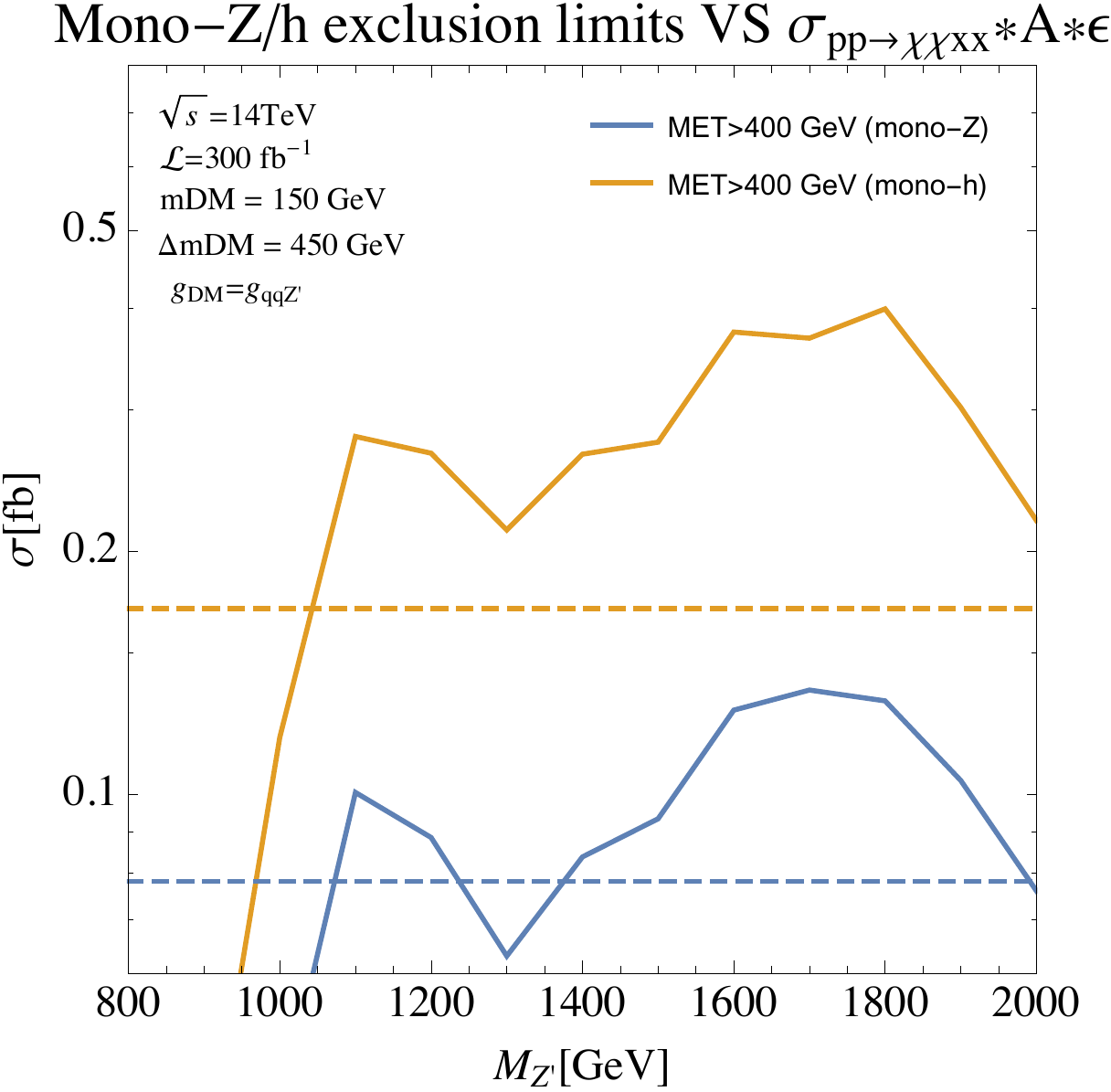}}
 \caption{Inelastic DM model: 95\% C.L. mono-$Z$ and mono-$h$ exclusion cross-section (dashed lines), projected at 14~TeV with a total integrated luminosity of 300 fb$^{-1}$, assuming $\met > 400~\GEV$. The predictions of the inelastic DM model, when the coupling of the $Z'$ to the quarks $g_{qq Z'}$ are chosen to be equal to the the upper limit consistent with di-jet constraints at a given $Z'$ mass (see Fig.~\ref{fig:dijet1}), is shown as solid lines. Panels (a)-(d) correspond to the choice of the mass parameters $(\mdm,\Delta \mdm)= (10,200), (10,450), (150,200), (150,450)$ in GeV respectively.}
 \label{fig:InelasticMonoZ14TeV} 
\end{figure}

\subsection{Two Higgs Doublet Model}

We consider the resonant production of a new heavy gauge boson $Z'$ which decays to Higgs ($Z$) and a CP-odd (CP-even) scalar $A^0$ ($H$), as considered in~\cite{Berlin:2014cfa}. The CP-odd (CP-even) scalar then is taken to exclusively decay into a pair of DM particles.  The dominant mono-$X$ signal is therefore mono-Higgs or mono-$Z$. In general, the simplified model Lagrangian of this topology can be written as:  
\bea
\mathcal L &\supset&  g_{q} Z'_{\mu} \sum_{i=1,2} \left(\bar{Q}_L^i\gamma^\mu  Q^i_L +  \bar{u}^i_R \gamma^\mu u^i_R +  \bar{d}^i_R\gamma^\mu d^i_R \right) \nonumber \\
&& + \frac12 m_{Z'}^2 Z'_{\mu}Z^{'\mu} +i \mu_A \partial_\mu A^0 Z'^\mu h +\mu_H Z'_\mu Z^\mu H.
\label{eq:2hdm}
\eea
Let us consider a UV completion of this DM production topology in order to make concrete comparisons of collider constraints as well as precision electroweak constraints. Our model and analysis follow Ref.~\cite{Berlin:2014cfa} closely, though here we perform the mono-$Z$ analysis for the first time and update the mono-$h$ constraints with newer di-jet limits. We introduce a two Higgs doublet model (2HDM) with Type-II Yukawa structure ($H_u, H_d$), {\em i.e.} $H_u$ couples with $u$-type quarks while $H_d$ couples with $d$-type quarks and charged leptons. Following Ref.~\cite{Berlin:2014cfa}, we assume that only $H_u$ and $u_R$ are charged under the new gauge symmetry $U(1)_{Z'}$ (the charge for both $H_u$ and $u_R$ is assumed to be 1/2). The $U(1)_{Z'}$ gauge symmetry is assumed to be broken spontaneously above the electroweak scale due to a new SM singlet scalar.

The physical Higgs bosons can be parametrized as follows:
\ba
H_u=\frac{1}{\sqrt{2}}\left( \begin{array}{c} -{\rm sin}~\beta~H^+ \\ v_u + {\rm cos}~\alpha~h + {\rm sin}~\alpha~H + i~{\rm cos}~\beta~A^0
\end{array} \right),  \\
H_d=\frac{1}{\sqrt{2}}\left( \begin{array}{c} {\rm cos}~\beta~H^+ \\ v_u - {\rm sin}~\alpha~h + {\rm cos}~\alpha~H - i~{\rm sin}~\beta~A^0
\end{array} \right).
\ea

We take the decoupling limit (${\rm sin}~(\beta-\alpha) = 1$) so that the lighter CP-even Higgs is SM-like. Spontaneous symmetry breaking in the Higgs sector induces mixing between $Z'$ and the SM $Z$ boson proportional to $\tan~\beta$. The mixing is constrained by the precision electroweak measurement of the deviation of  $\rho \equiv m_W^2/m_Z^2{\rm cos}~\theta_W$ from unity~\cite{Berlin:2014cfa}:

\begin{figure}[H]
\centering
\subfigure[]{\includegraphics[scale=0.55]{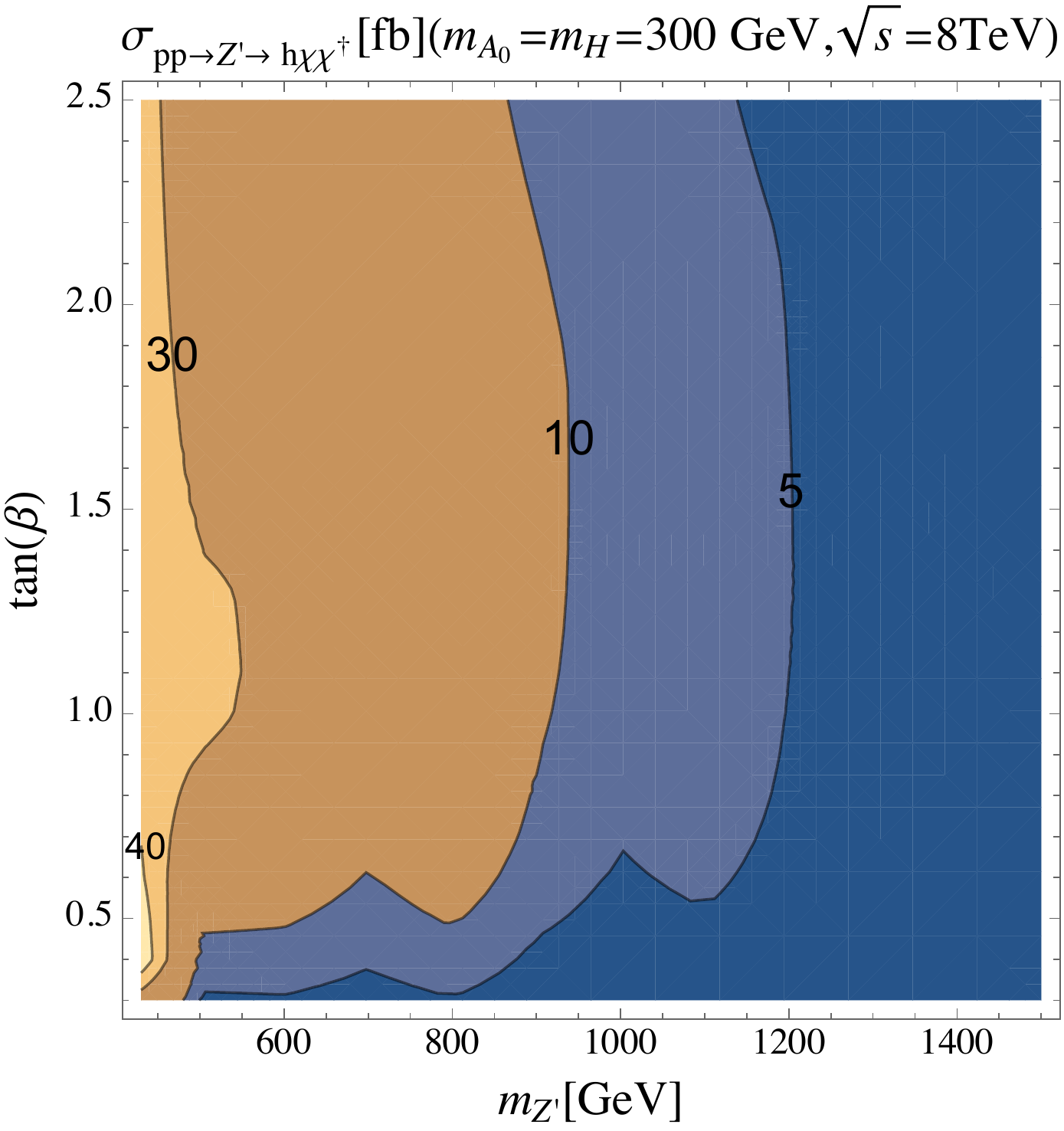}}
\subfigure[]{\includegraphics[scale=0.56]{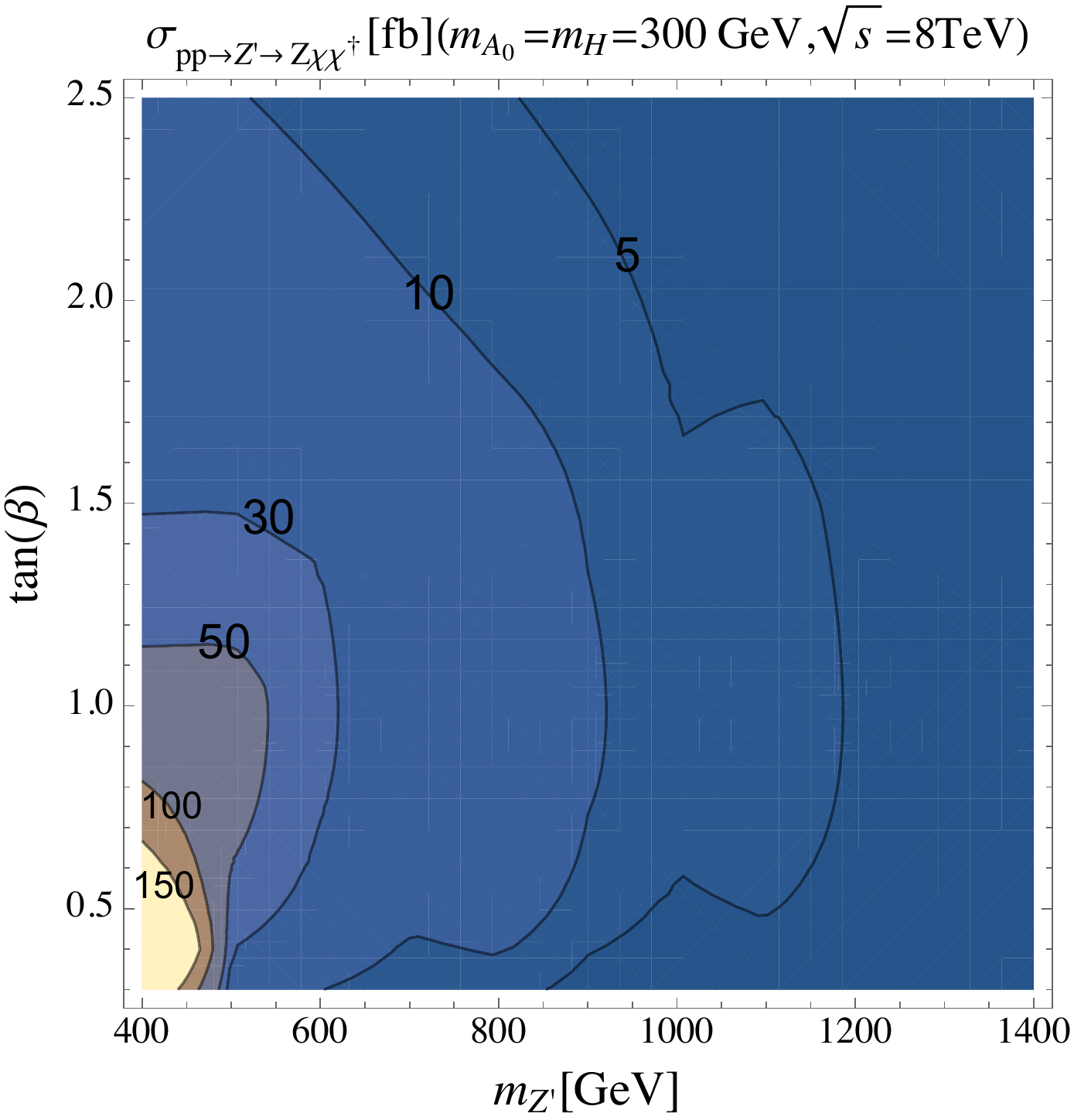}}
\subfigure[]{\includegraphics[scale=0.55]{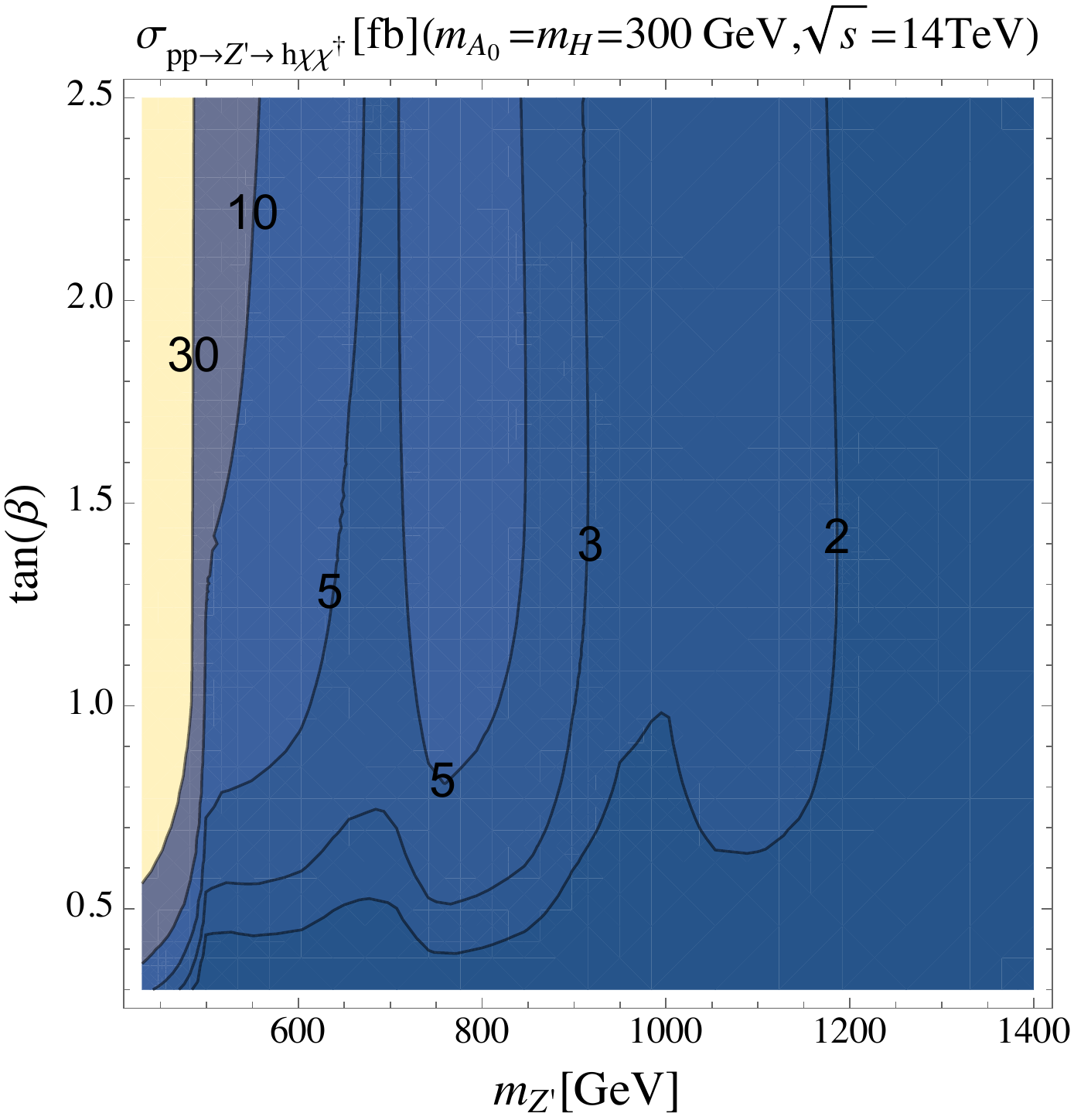}}
\subfigure[]{\includegraphics[scale=0.55]{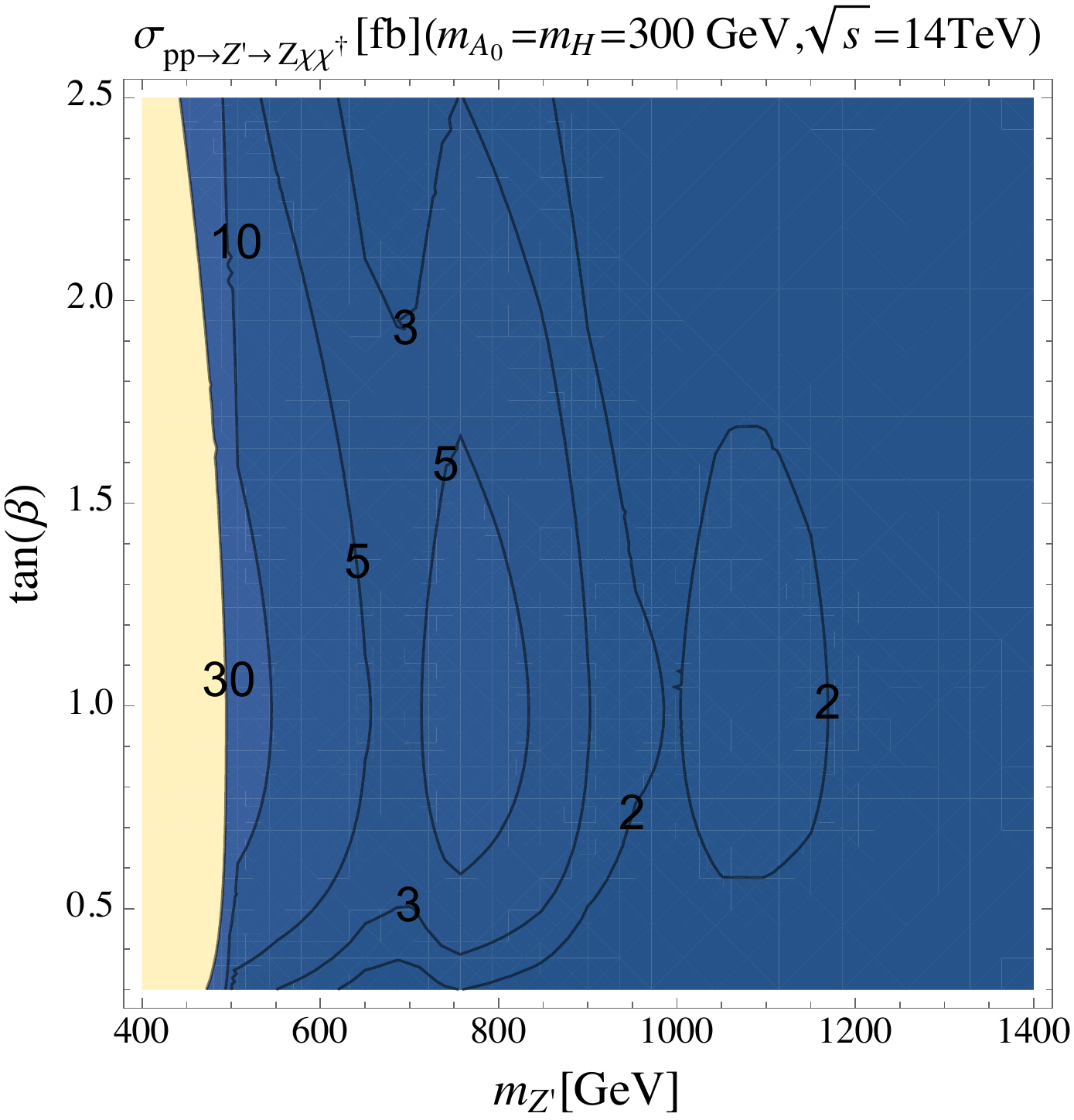}}
 \caption{2HDM model: cross-sections for the $Z'$ mediated production of $h+\met$ (a,c) and $Z+\met$ (b,d) at $\sqrt{s}=8$ TeV (top) and  $\sqrt{s}=14$ TeV (bottom). The left-hand figures include the contribution from $hA_0$ together with $hZ$. We assume a 100\% branching ratio to invisible decay $A_0\rightarrow \chi\chi^\dagger$ in (a,c) or  $H\rightarrow \chi\chi^\dagger$  in (b,d).} 
 \label{fig:2hdm_xsection}
\end{figure}

\begin{figure}[H]
\centering
\subfigure[]{\includegraphics[scale=0.55]{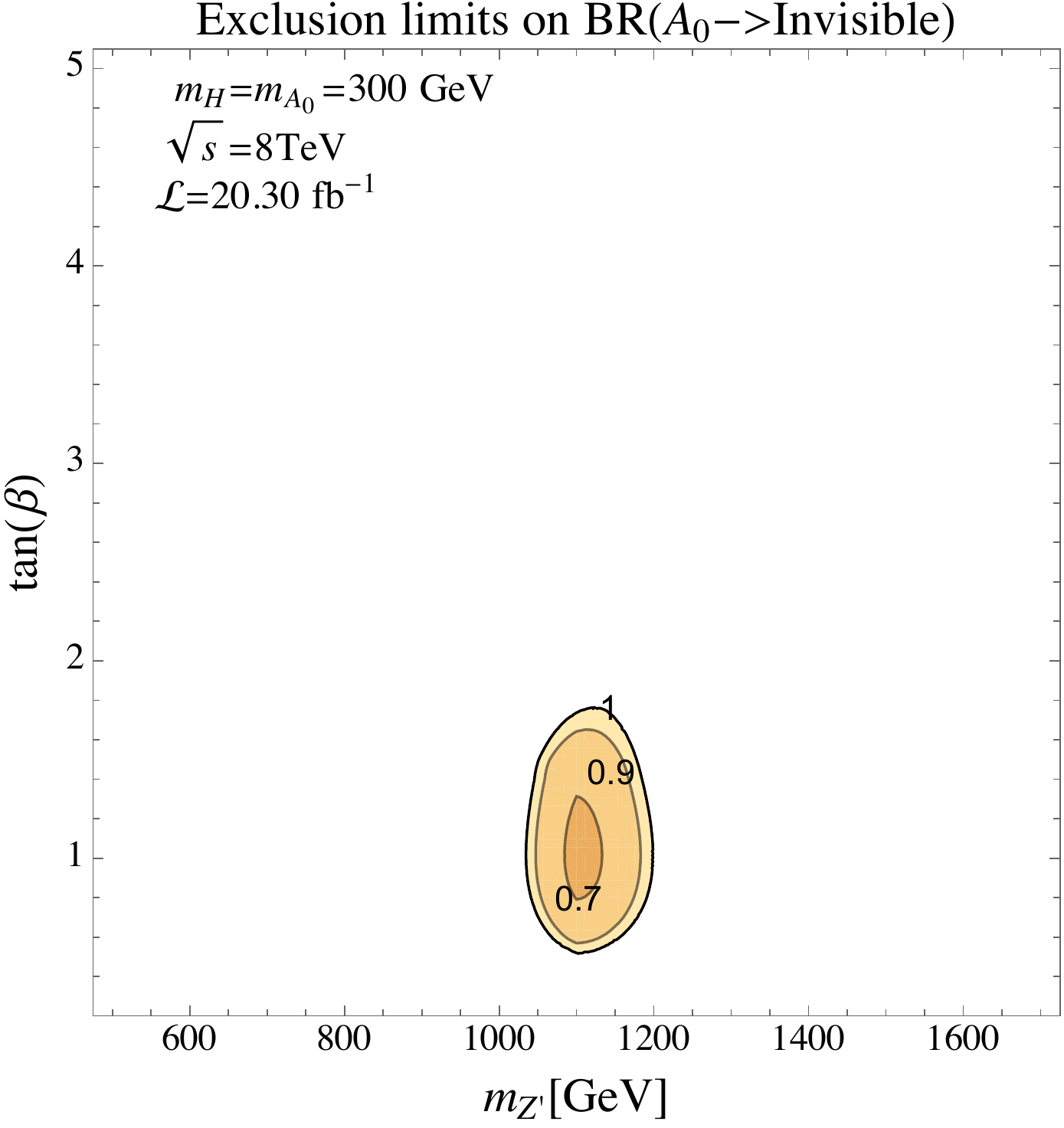}}
\subfigure[]{\includegraphics[scale=0.55]{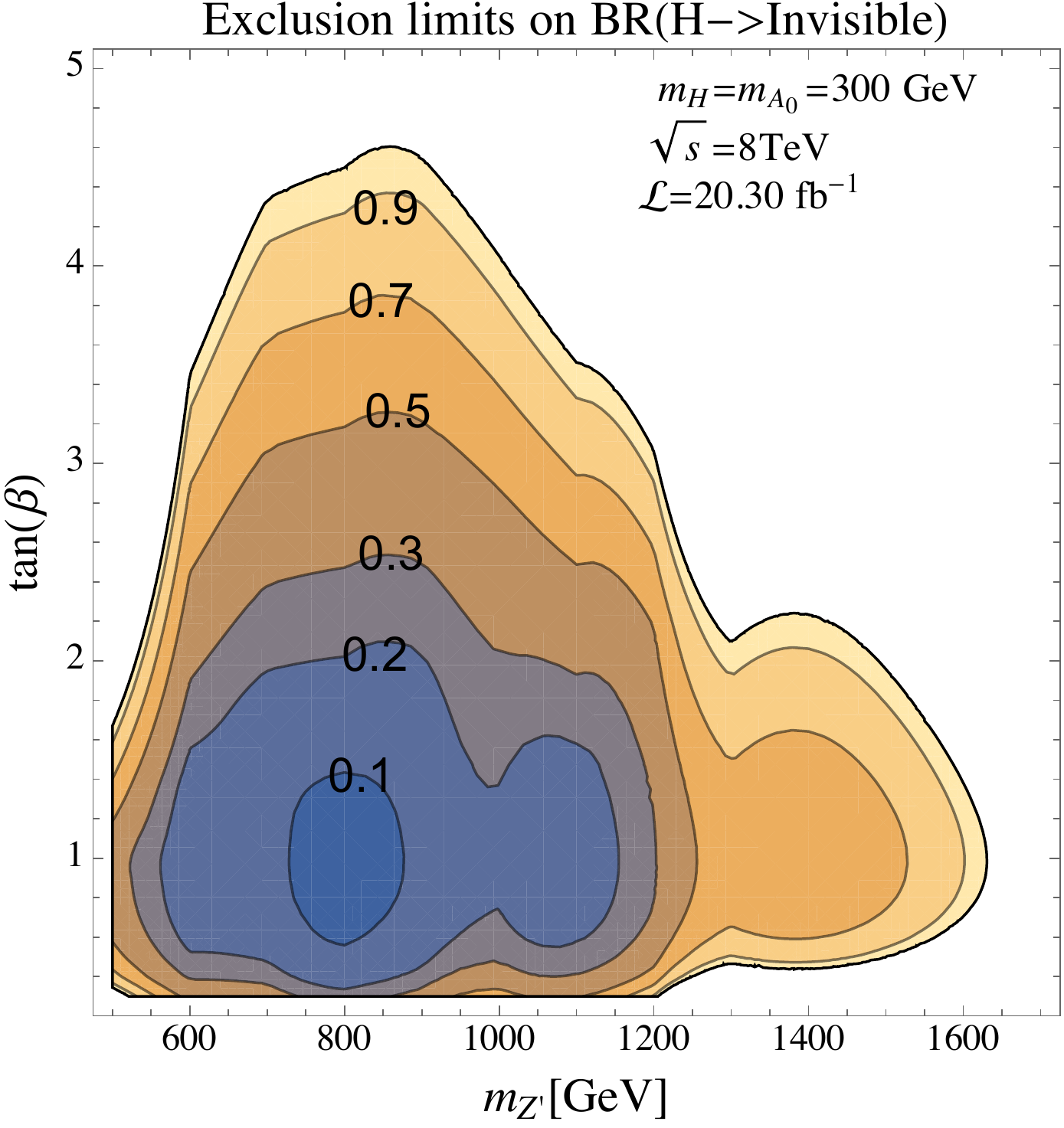}}
\subfigure[]{\includegraphics[scale=0.55]{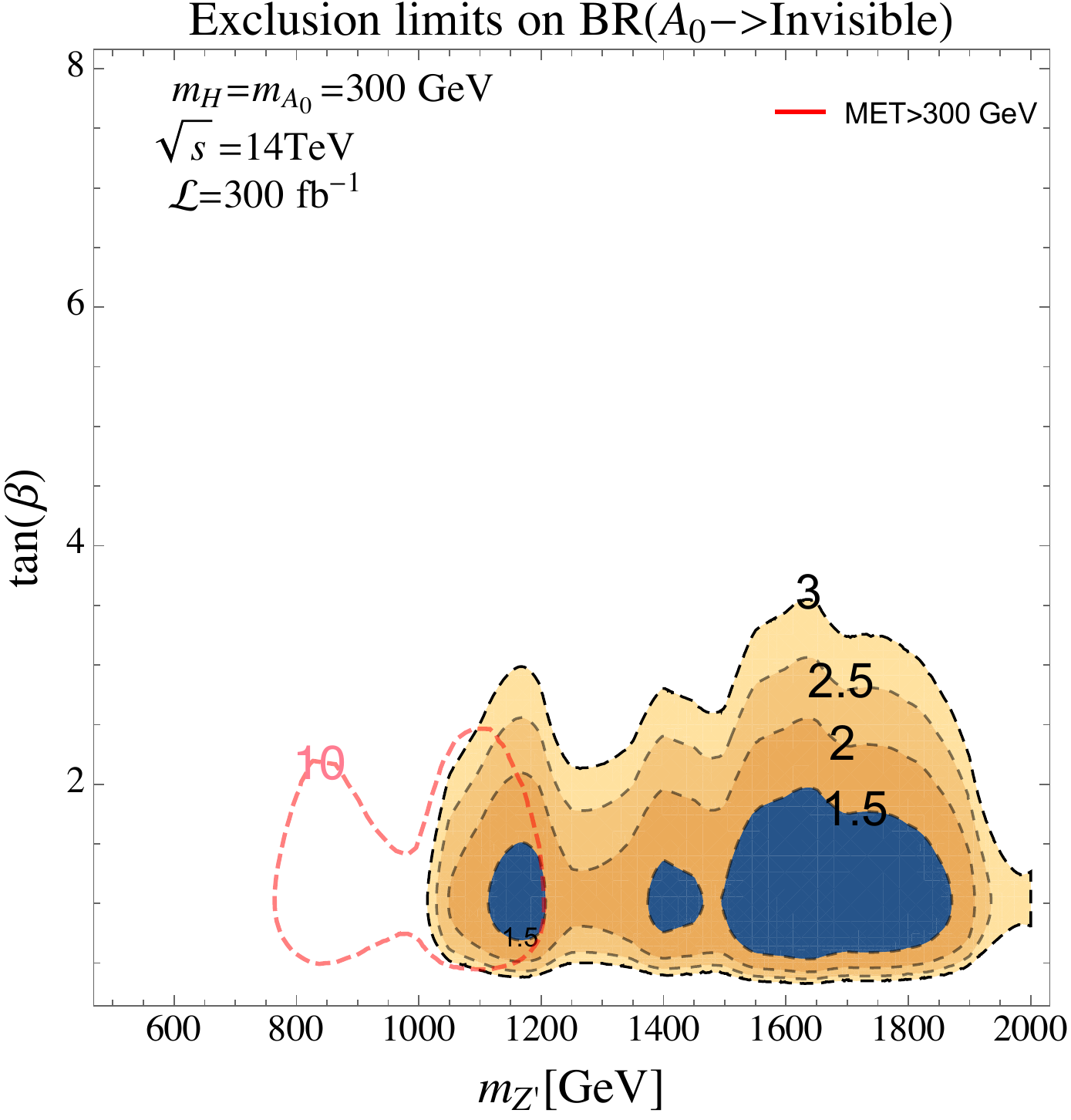}}
\subfigure[]{\includegraphics[scale=0.55]{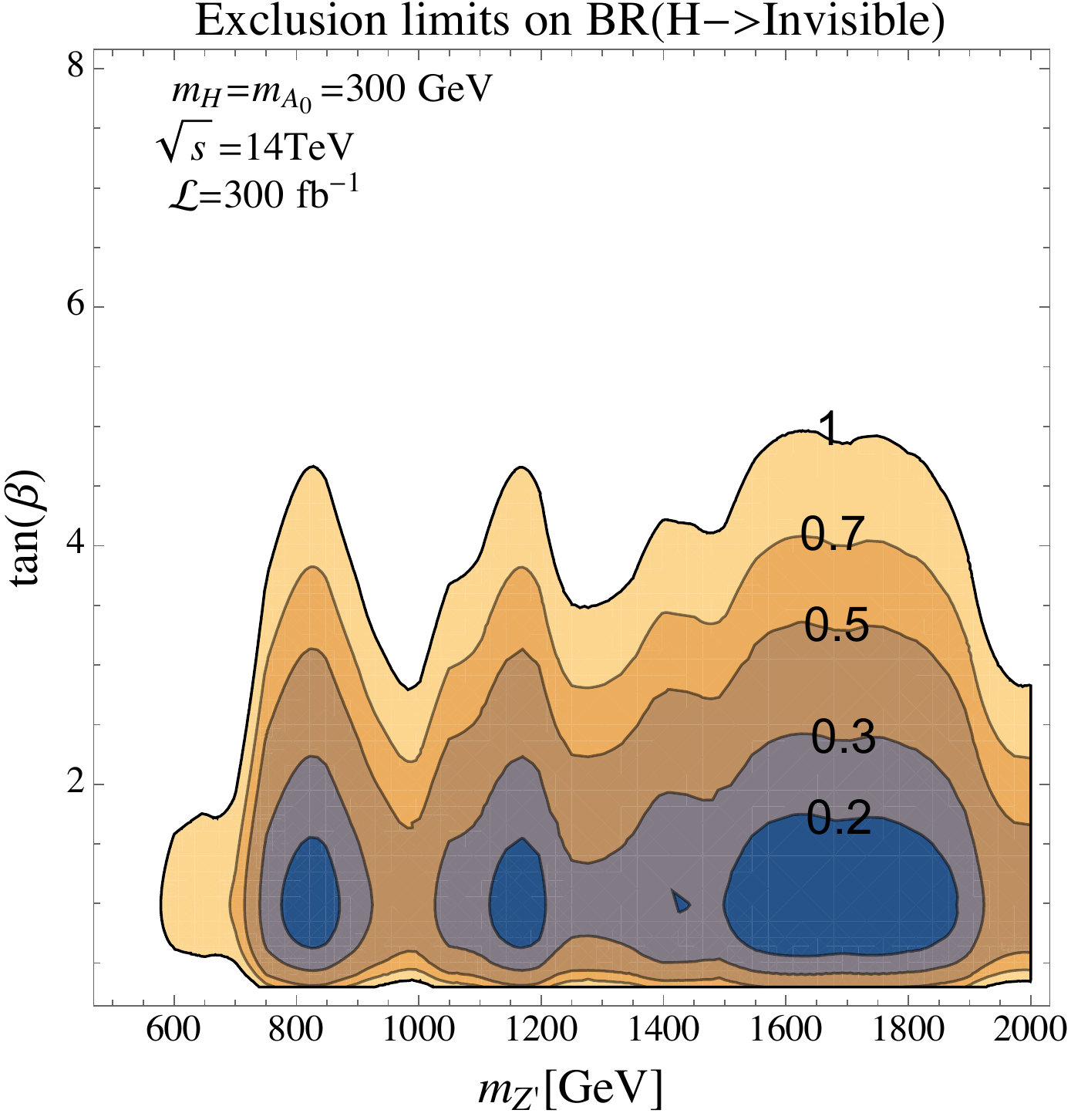}}
 \caption{ 2HDM model: Limit on BR($A_0\rightarrow E_T$) due to mono-Higgs analysis a) and on BR($H\rightarrow E_T$) due to mono-$Z$  b) in the $M_{Z'}-\tan\beta$ plane. c) and d) show projections at $14$ TeV with 300 fb$^{-1}$. The $Z'$ production cross section has been set to saturate current and projected dijet resonance limits respectively, as explained in the text. Contour lines for upper limits greater than 1 in figure (c) are represented as dashed lines. The red curve in (c) represents the exclusion limit obtained from the less stringent cut $\met\geq 300$GeV and closely mimics the limit obtained in~\cite{Berlin:2014cfa} which, however, exploited a different analysis~\cite{TheATLAScollaboration:2013lia}.} 
  \label{fig:2hdm_8T}
\end{figure}

\ba
\rho=1+\epsilon^2\left(\frac{m_{Z'}^2-m_Z^2}{m_Z^2}\right),\\
\epsilon\equiv \frac{(m^0_Z)^2}{m_{Z'}^2-m_Z^2}\frac{g_q\cos \theta_W}{g}\sin^2\beta,
\ea

where $m^0_Z$ is the SM $Z$ boson mass in the absence of mixing. Furthermore, as $u_R$ is charged under the new $U(1)_{Z'}$ gauge symmetry, di-jet resonance searches for the $Z'$ performed at hadron colliders constrain the $Z'$ coupling to the initial state quarks (see Fig.~\ref{fig:dijet1}).  We apply these constraints and take the coupling of the $Z'$ to the initial state quarks ($g_{qqZ'}$) to saturate the combined constraints.  The couplings of $Z'$ to $hA^0$ and $ZH$, which lead to mono-Higgs and mono-$Z$ signals, arise from the covariant derivative of the kinetic term of $H_u$.  

We show in Fig.~\ref{fig:2hdm_xsection} the dependence of mono-Higgs and mono-$Z$ production cross-sections at 8 and 14~TeV on the $Z'$ mass and $\tan\beta$. Both channels have similar dependence on the parameter space because the $Z'A^0h$ and $Z'ZH$ couplings are both inversely proportional to $\tan\beta$, but mono-$Z$ covers a larger parameter space with the same production cross-section. In Figs.~\ref{fig:2hdm_8T} a) and b), we vary the branching ratio of $A^0$ and $H$ to DM and show the mono-Higgs and mono-$Z$ constraints on the $Z'$ mass-$\tan\beta$ plane. The 14~TeV projection, performed with the procedure described in Appendix~\ref{app:b}, is shown in Fig.~\ref{fig:2hdm_8T} c) and d). Here again, we find that the mono-$Z$ channel is able to constrain a larger parameter region compared to the corresponding mono-Higgs channel. Let us again note that whether DM couples to $A^0$ or $H$ largely depends on the UV completion in the dark sector. Hence, both mono-Higgs and mono-$Z$ searches are equally useful for constraining this type of simplified model.



\subsection{Squarks with mono-$Z$}
\begin{figure}[H]
\centering
\includegraphics[scale=0.66]{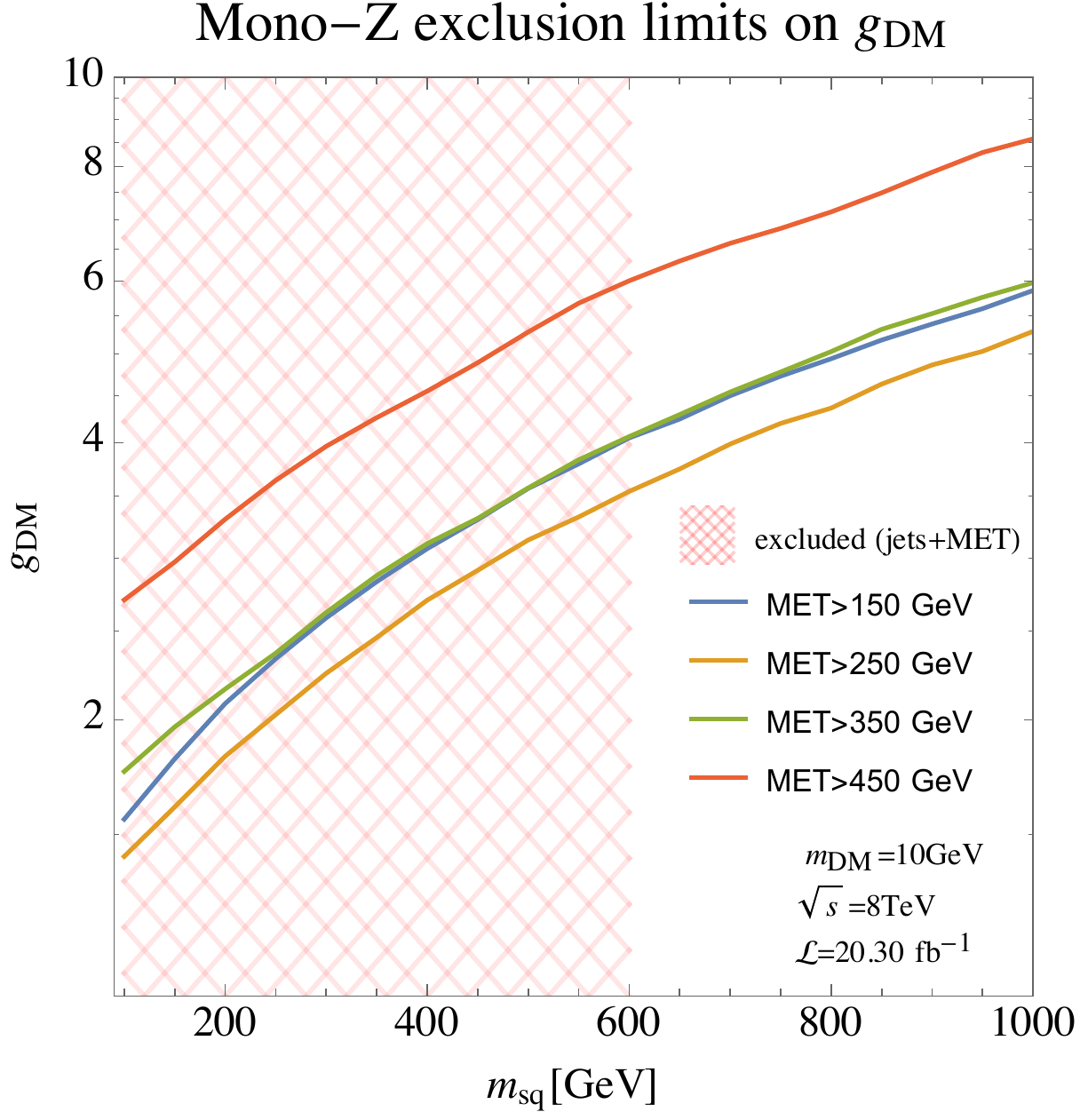}
\includegraphics[scale=0.65]{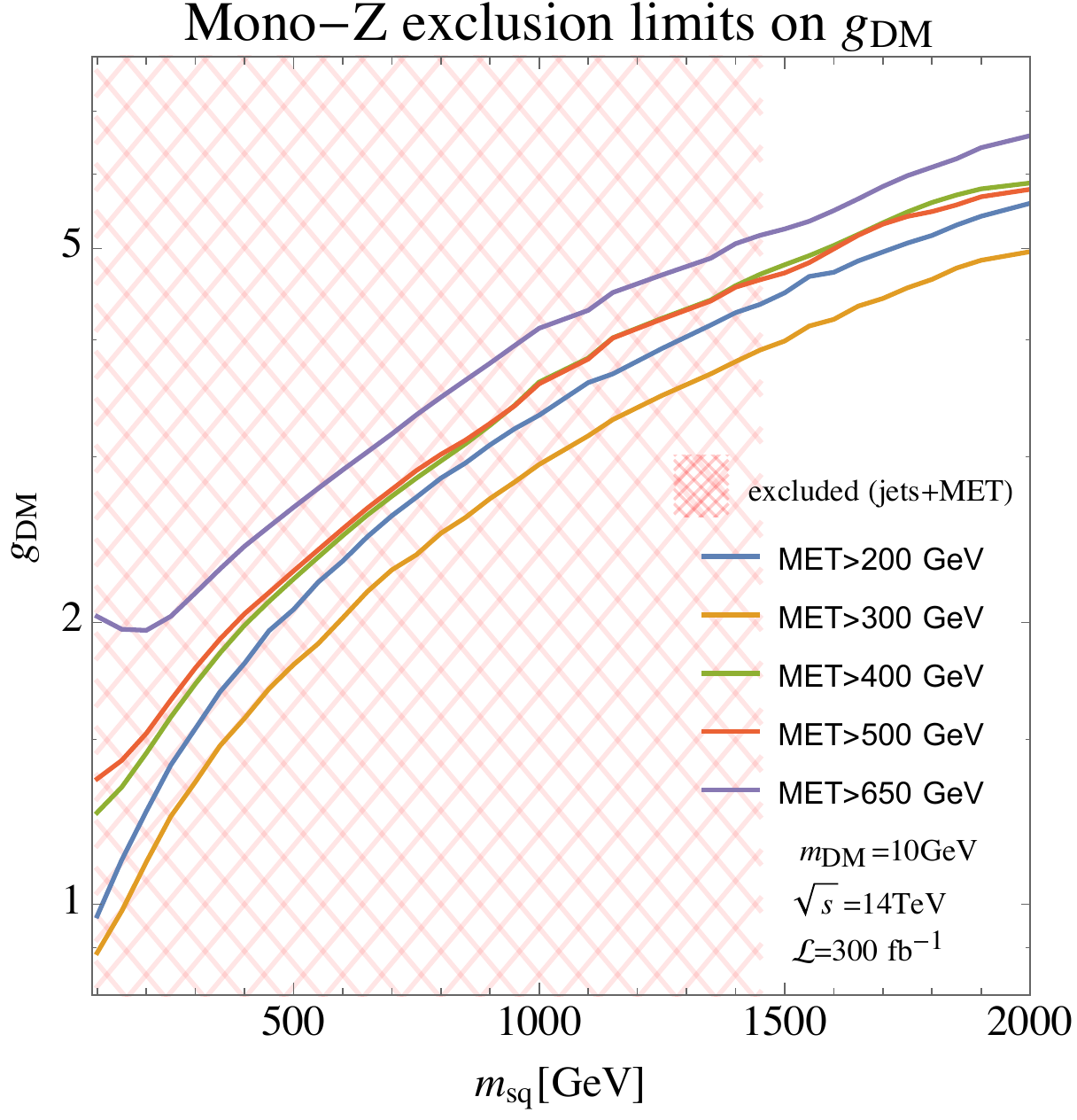}
 \caption{Squarks model: 95\% exclusion limits on the quark-squark-dark matter coupling $g_{DM}$. The mass of the dark matter is fixed to $m_{DM}=10$ GeV.  The left panel shows the 8~TeV constraints and the right panel shows the 14~TeV projection with 300 fb$^{-1}$. The shaded region corresponds to values of the squark mass excluded by multi+jets~+~$\met$ analysis. Projections for jets~+~$\met$ limits are taken from Ref.~\cite{Cohen:2013xda}.  }
 \label{fig:SquarkMonoZ}
\end{figure}

We now consider a scenario which, in SUSY notation, involves a singlino as DM and 8 squarks as mediators:
\be
\mathcal L \supset  g_{DM}\sum_{i=1,2} \left(\widetilde{Q}_L^i \bar{Q}^i_L +  \tilde{u}^i_R \bar{u}^i_R + \tilde{d}^i_R \bar{d}^i_R \right)\chi + \text{mass terms} +h.c. 
\label{eq:la}
\ee
Let us note that for the case where the mixing between left and right-handed squarks is zero, the mono-Higgs production cross-section is highly suppressed by the negligibly small quark masses. Although it is possible to introduce $A$-terms that could enhance the mono-Higgs signal, this would in general lead to severe tuning in the quark Yukawa couplings (see {\em e.g.}~\cite{Hamaguchi:2014pja}). Hence, we opt to leave out this possibility in this work. This is essentially the simplified model proposed in Ref.~\cite{Bell:2012rg} and used by the ATLAS collaboration to present their mono-$Z$ searches at Run I~\cite{Aad:2014vka}. We show in Fig.~\ref{fig:SquarkMonoZ} the constraint on $g_{DM}$ as a function of the mediating squark mass.  We can see that in comparison to the mono-jets and jets~+~$\met$ constraints derived in~\cite{Papucci:2014iwa}, the constraints from mono-$Z$ production are very weak.  14~TeV projections, performed with the procedure described in Appendix~\ref{app:b}, are shown in the right panel of Fig.~\ref{fig:SquarkMonoZ}. They improve the constraints, but are unlikely to be competitive with the di-jet and jets~+~$\met$ constraints.

\begin{figure}[H]
\centering
\includegraphics[scale=0.659]{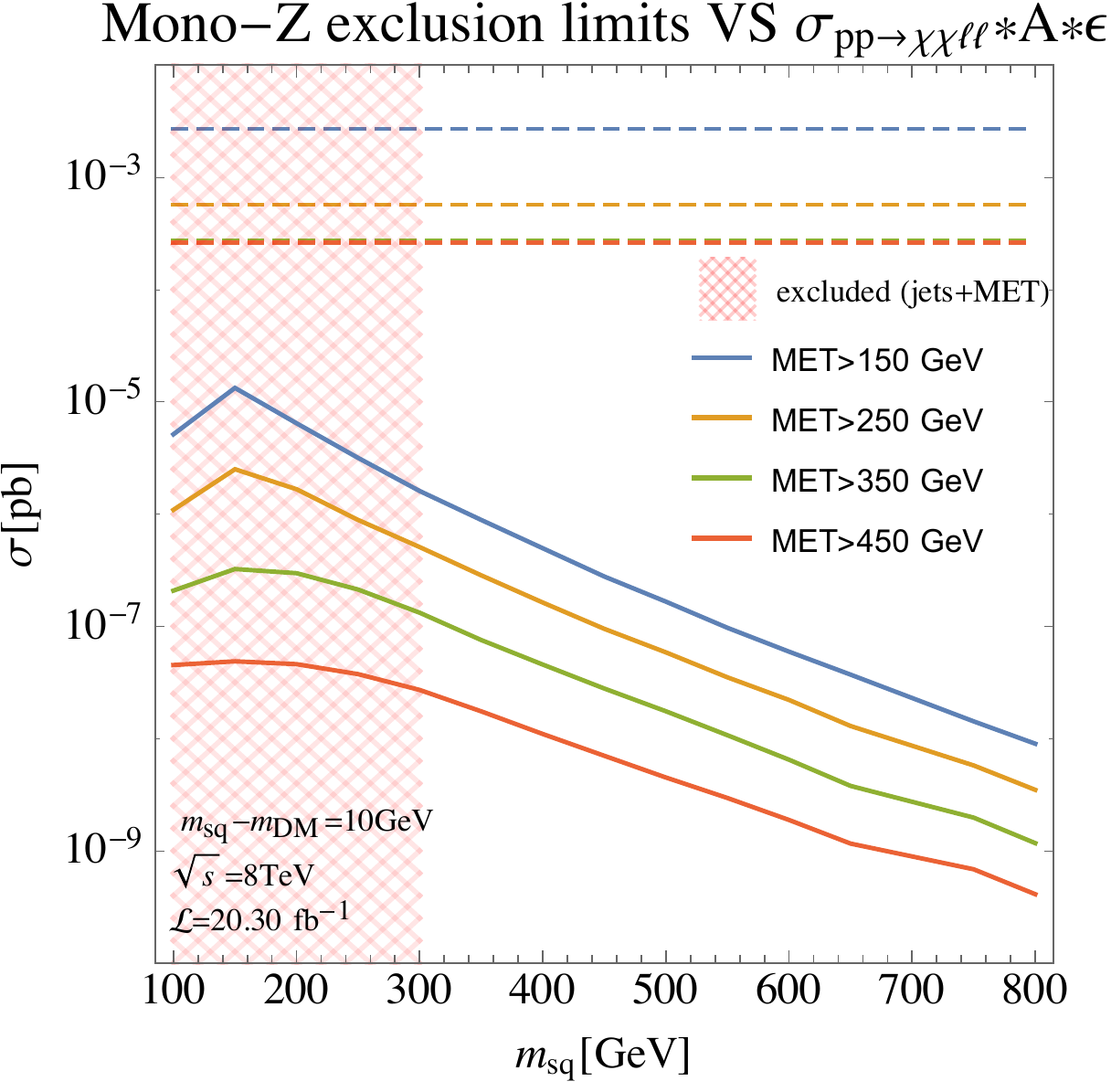}
\includegraphics[scale=0.64]{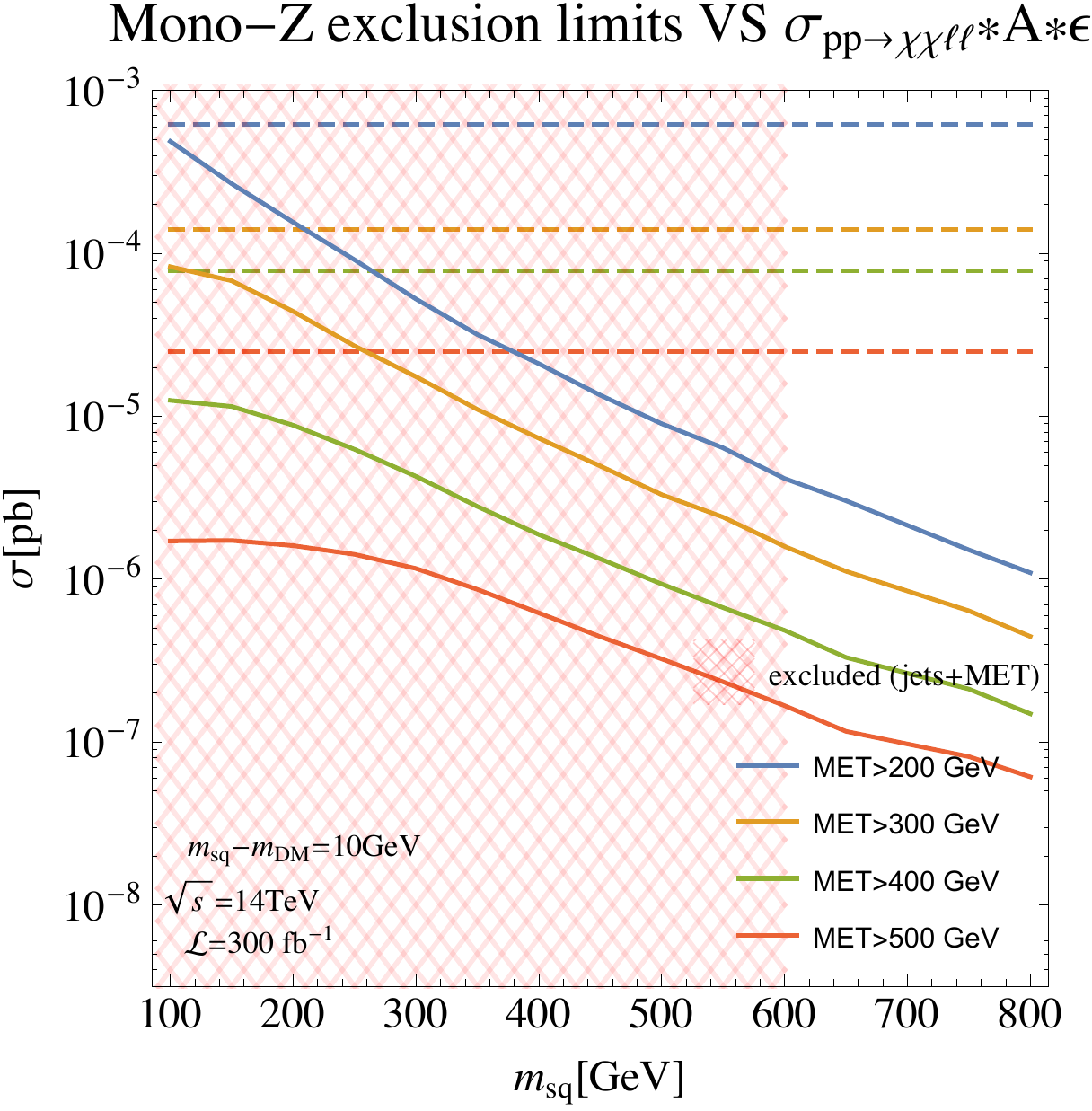}
 \caption{Squarks model: 95\% exclusion limits on the mono-$Z$ cross-section, shown as dashed lines, and 95\% exclusion limits on the cross-section after cuts for DM pair production in association with a $Z$, shown as solid lines. The mass of the dark matter is taken in the compressed region,  $m_{sq} - m_{DM}=10$ GeV, since this enhances the cross-section (see text).  The left panel shows the 8~TeV constraints and the right panel shows the 14~TeV projection with 300 fb$^{-1}$. Projections for jets~+~$\met$ limits are taken from Ref.~\cite{Cohen:2013xda}. The four colors represent the four different $\met$ choices in the mono-$Z$ analysis.} 
 \label{fig:SquarkMonoZCompressed}
\end{figure}

Mono-$Z$ searches could in principle allow to access the compressed case,  $m_{sq} - m_{DM}\ll m_{sq}$, as shown in Fig.~\ref{fig:SquarkMonoZCompressed}.  In this squeezed regime one can take advantage of the gluon-gluon initiated squark pair production, where the squarks then decay into dark matter plus soft jets. Attaching a $Z$ boson to one of the squark lines gives a process consistent with the mono-$Z$ cuts\footnote{This is true only for compressed spectra: a larger mass separation would give rise to hard jet that would not pass the mono-$Z$ cuts on jet $p_T$.} and similar to the monojet topology. Even in this case, where the direct squark limits from the jets~+~$\met$ analysis~\cite{Aad:2014wea} only places a constraint $m_{sq} \gtrsim 300$ GeV, we find that mono-$Z$ searches are much weaker. The 14~TeV projections shown in the right panel improve the limit and are indeed able to exclude compressed spectra up to $m_{sq} \gtrsim 100$ GeV; nevertheless direct searches for squarks will continue to be much more powerful~\cite{Cohen:2013xda}.   On the other hand, as explained in Appendix~\ref{app:b}, our projections do not optimize the cuts to suppress the ratio of background over signal.  Furthermore, we do not have access to the bin correlations: hence we conservatively assumed a $30\%$ uncertainty in each bin. Future studies by the experimental collaborations are likely to improve the limits presented here, though it seems unlikely they will qualitatively change our conclusions.

\begin{figure}[H]
\centering
\includegraphics[scale=0.8]{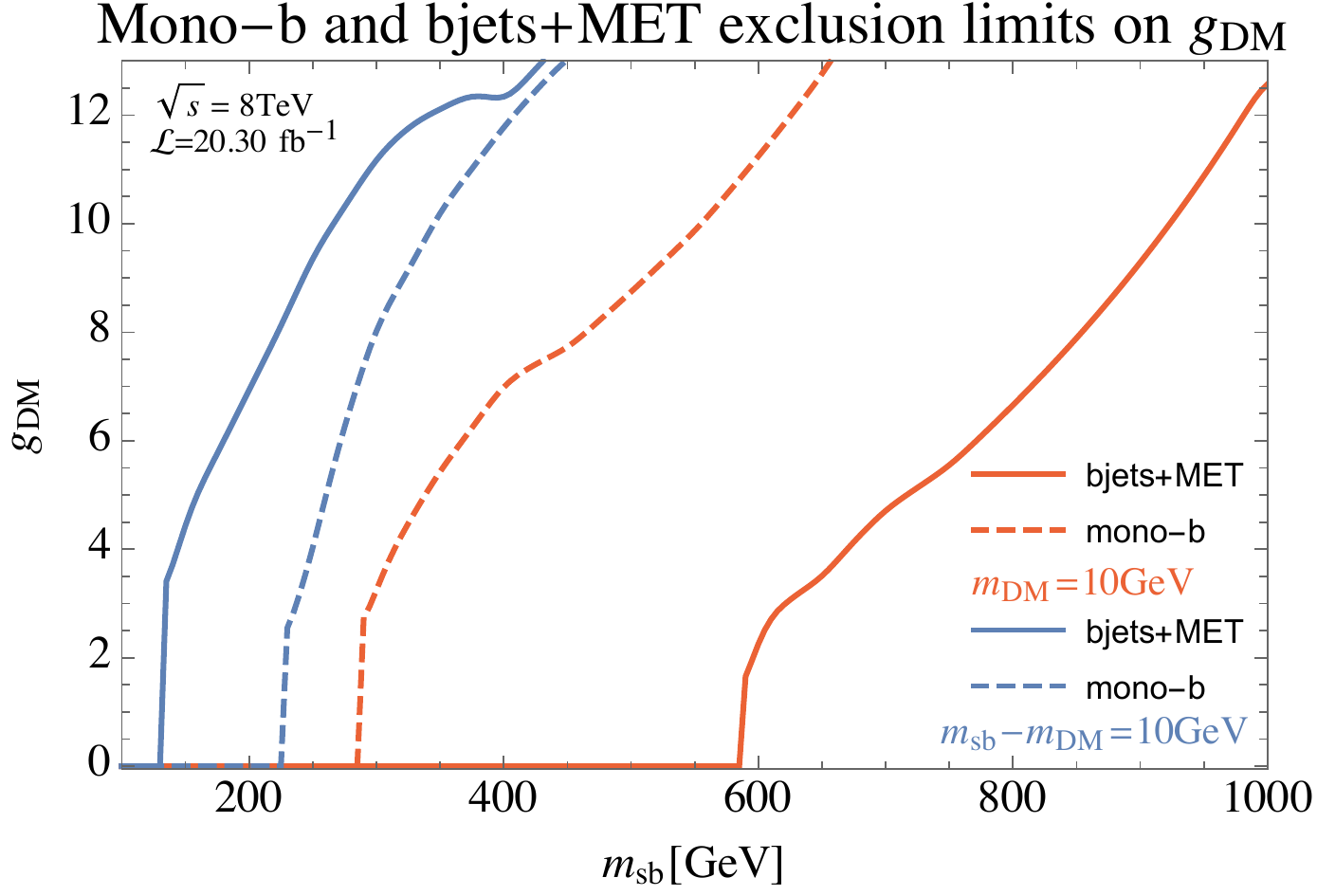}
 \caption{ \begingroup
 \fontsize{10pt}{11pt}\selectfont
    \setstretch{0.5}
Sbottom model:  95\% exclusion limits at 8~TeV from mono-$b$jet and 2-$b$jet~+~$\met$ searches on the sbottom-bottom-DM coupling. The continuous red and blue lines represent bounds from 2b~+~$\met$ searches while dashed lines those from mono-$b$ searches. Different colors correspond to the limiting cases where the mass of the fermonic DM is taken light (red), $m_{DM}=10$ GeV, or in the compressed region (blue),  $m_{sb} - m_{DM}=10$ GeV. At the limit $g_{DM} \to 0$ in the figure, it is implicitly assumed that $g_{DM}$ is small enough such that sbottom pair productions are initiated solely by gluon-gluon processes, but $g_{DM}$ is large enough that to make sbottoms decay promptly to DM.
\endgroup 
} 
 \label{fig:sbottomsearches}
\end{figure}

\subsection{Sbottoms with mono-$b$, mono-$h$ and mono-$Z$}

Similarly to the squark case, we take the Lagrangian as follows  
\be
\mathcal L =  g_{DM} \left(\widetilde{Q}_L^3 \bar{Q}^3_L +  \tilde{b}_R \bar{b}^i_R \right)\chi + \text{mass terms} +  g_h |H_{SM}|^2(|\widetilde{Q}^3_L|^2+|\tilde{b}_R|^2) +h.c ,
\label{eq:sbottomLagrangian}
\ee
where $H_{SM}$ is the SM Higgs doublet.  Notice that we do not  normalize the sbottom coupling with the Higgs boson to the bottom Yukawa coupling.  
We consider first the direct sbottom search constraints in Fig.~\ref{fig:sbottomsearches}, comparing with the mono-$b$ search for the light DM ($m_{DM}=10$ GeV) and compressed region ($m_{sb} - m_{DM}=10$ GeV) cases.  We can see that in the non-compressed region ({\em i.e.} for relatively large mass splitting between the sbottom and neutralino) the traditional sbottom searches dominate the constraints.  On the other hand, in the compressed region, the mono-$b$ search becomes important.  Note that in the non-compressed region, constraints lie around $m_{sb} = 600 $ GeV.

\begin{figure}[H]
\centering
\subfigure[]{\includegraphics[scale=0.6]{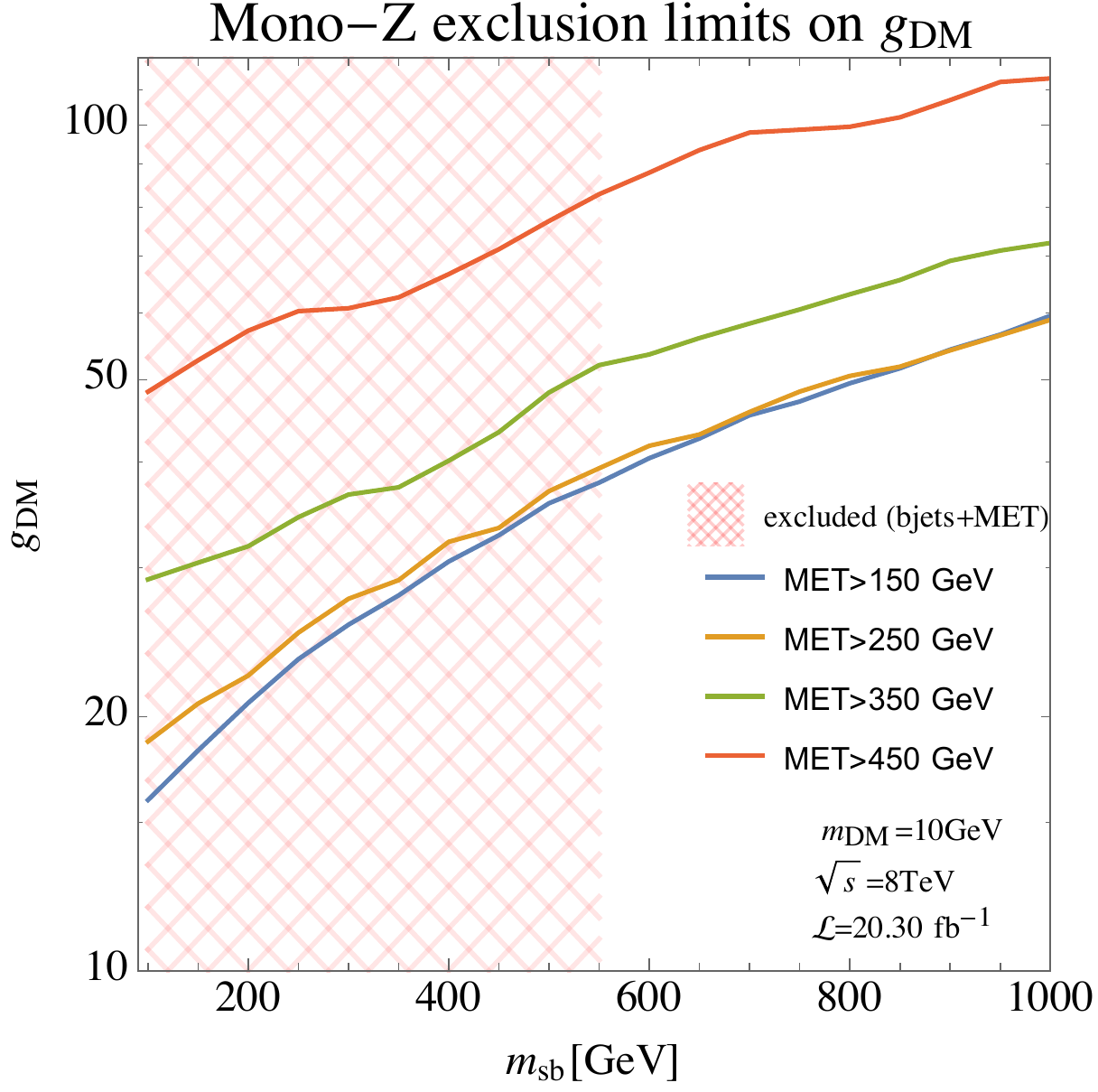}}
\subfigure[]{\includegraphics[scale=0.633]{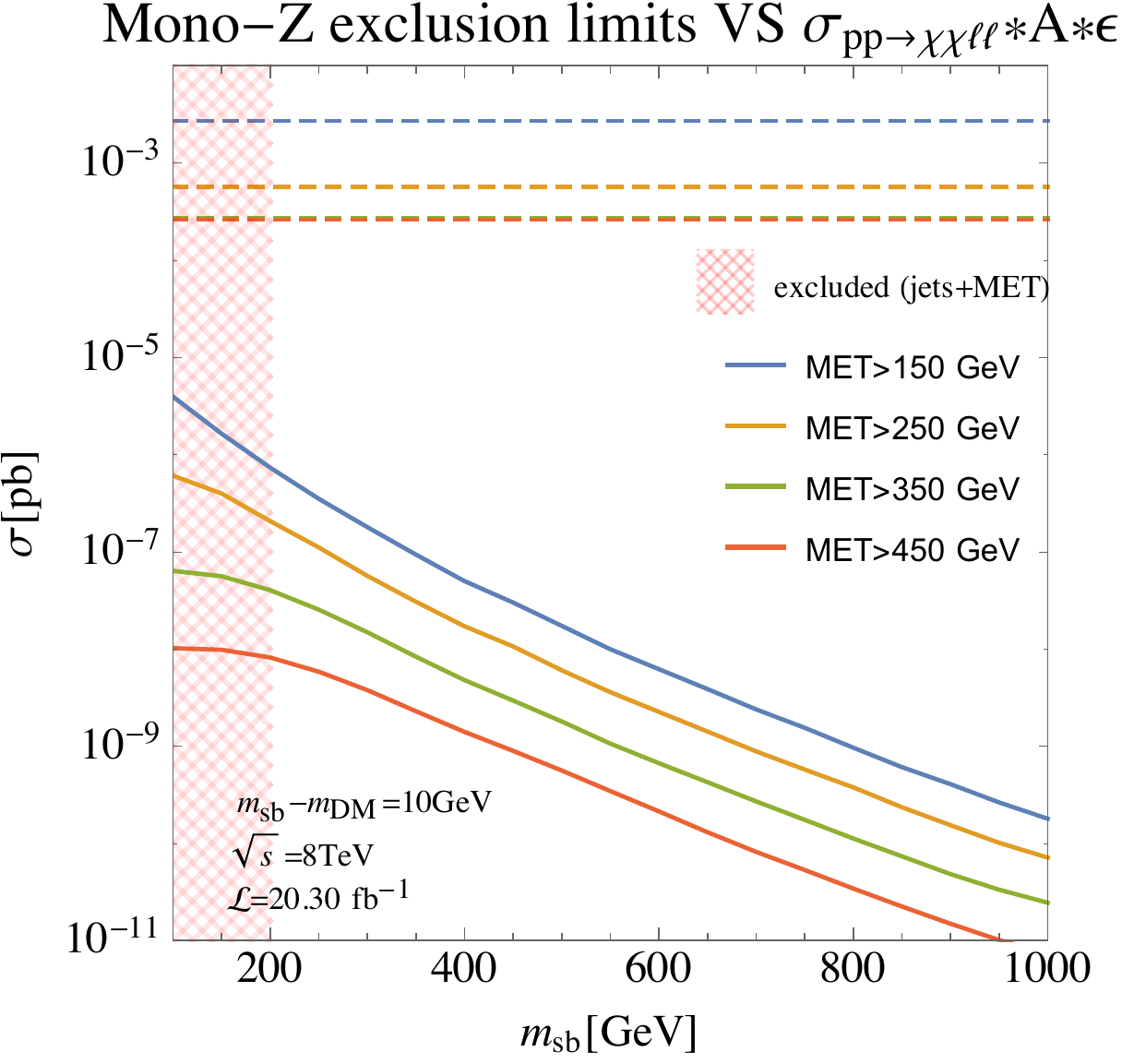}}
\subfigure[]{\includegraphics[scale=0.6]{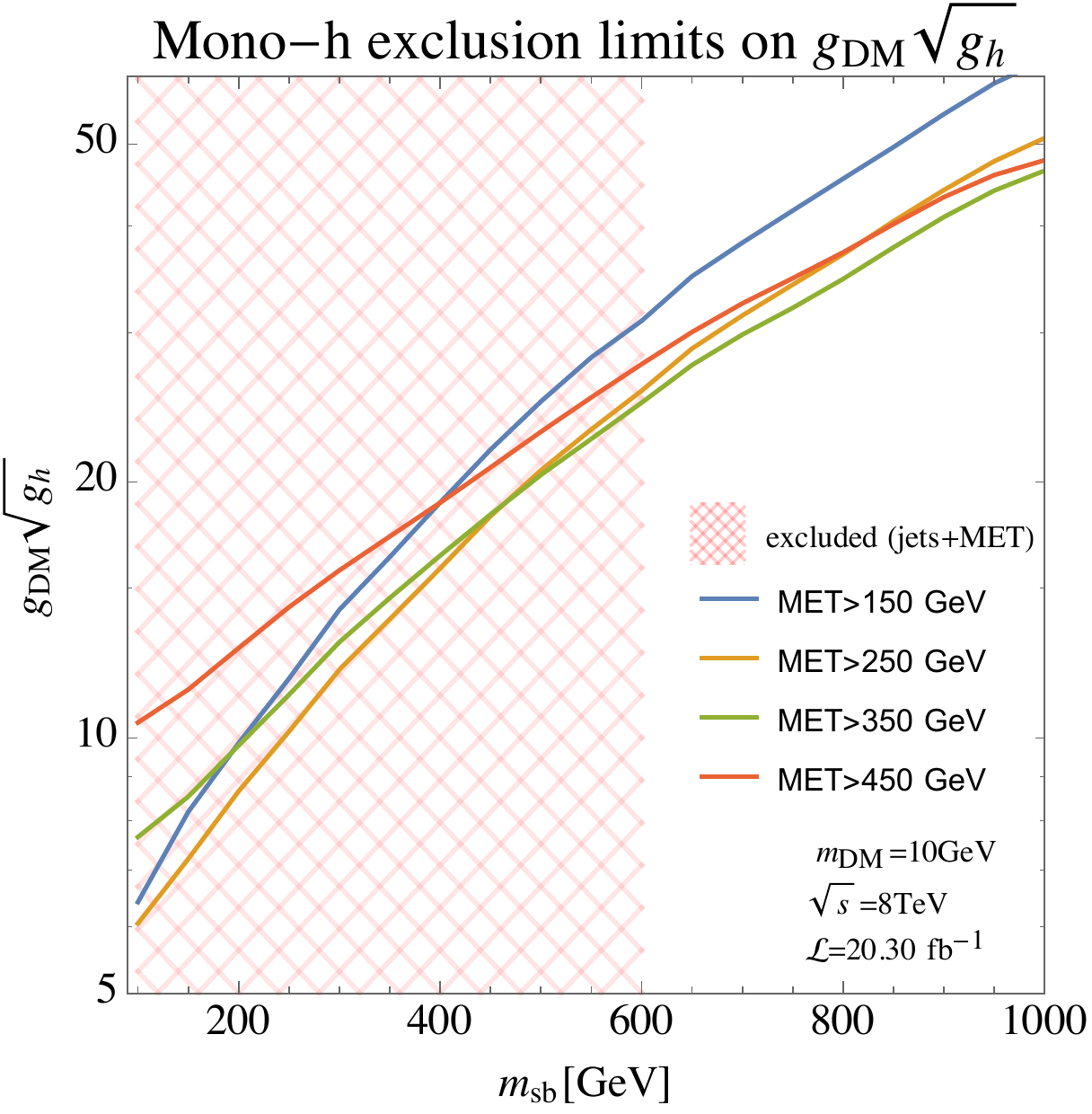}}
\subfigure[]{\includegraphics[scale=0.38]{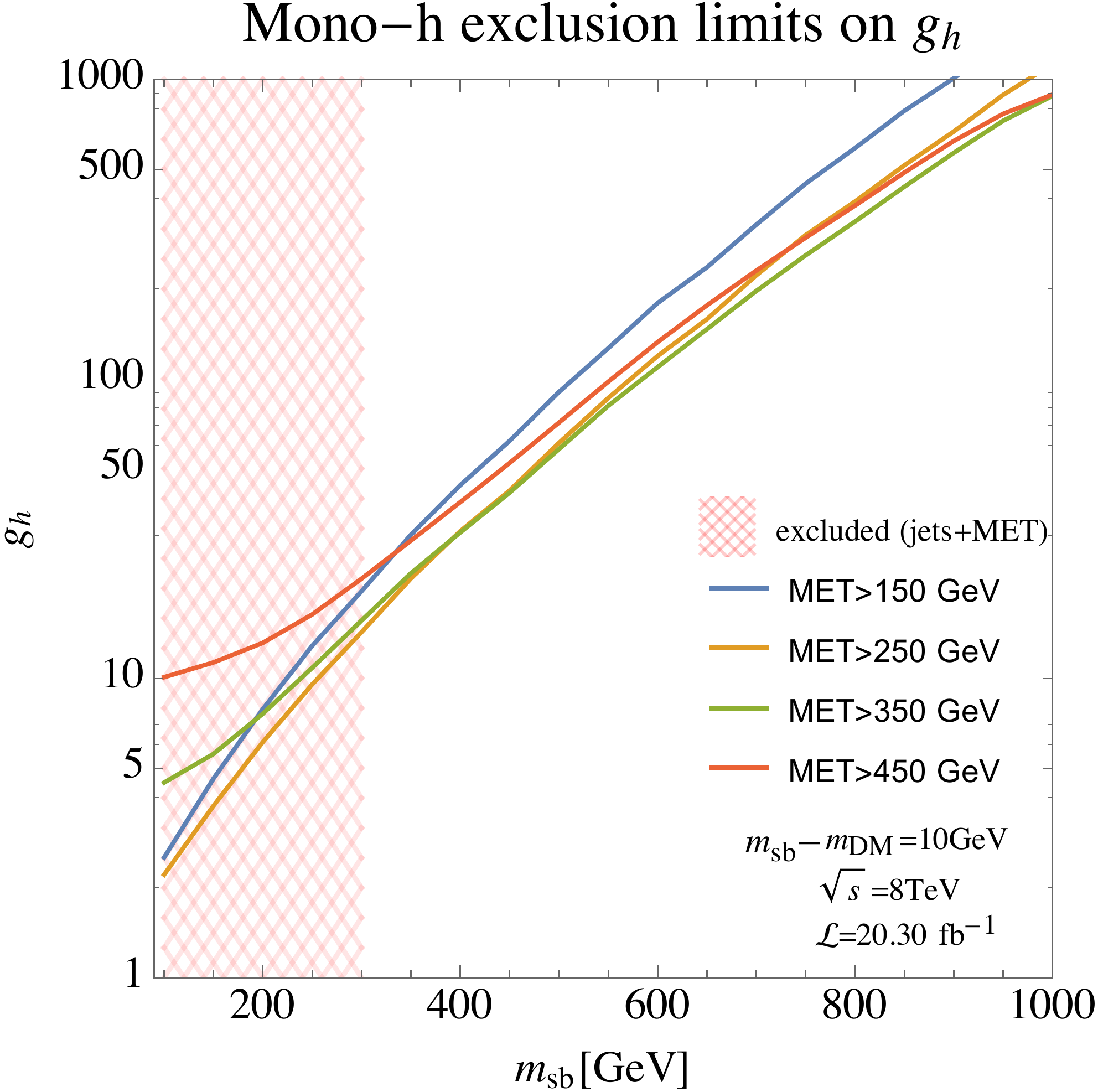}}
 \caption{ \begingroup
Sbottom model: 95\% exclusion limits at 8~TeV from various searches for the sbottom plus fermionic DM model. Different colors correspond to different choices of $\met$ cut in the respective analysis. a) Limits on the sbottom-bottom-DM coupling from a mono-$Z$ search  as a function of the $\widetilde b$ mass. The mass of the dark matter is fixed to $m_{DM}=10$ GeV. b) 95 \% exclusion limits on the cross-section for the mono-$Z$ analysis, shown as dashed lines. 95 \% exclusion limits on the cross-section after cuts for DM in association with a $Z$ decaying into leptons, shown as solid lines. Here $m_{sb} - m_{DM}=10$ GeV and the DM is produced though a sbottom pair. c) Limits from a mono-$h$ search on the product $g_{DM}\sqrt{g_h}$ as a function of the $\widetilde b$ mass, where $g_h$ is the Higgs-sbottom coupling. The mass of the dark matter is fixed to $m_{DM}=10$ GeV.  d) Limits on $g_{h}$ from a mono-$h$ search in the compressed regime $m_{sb} - m_{DM}=10$ GeV.
\endgroup } 
 \label{fig:sbottomsearch}
\end{figure}

Next we compare these results to mono-$Z$ and mono-$h$ constraints in Fig.~\ref{fig:sbottomsearch}.  Again we focus on two extremal cases: light DM and a compressed spectrum, where the process of  gluon-gluon  initiated sbottom pair production increases substantially the cross-section.

\begin{figure}[H]
\centering
\subfigure[]{\includegraphics[scale=0.63]{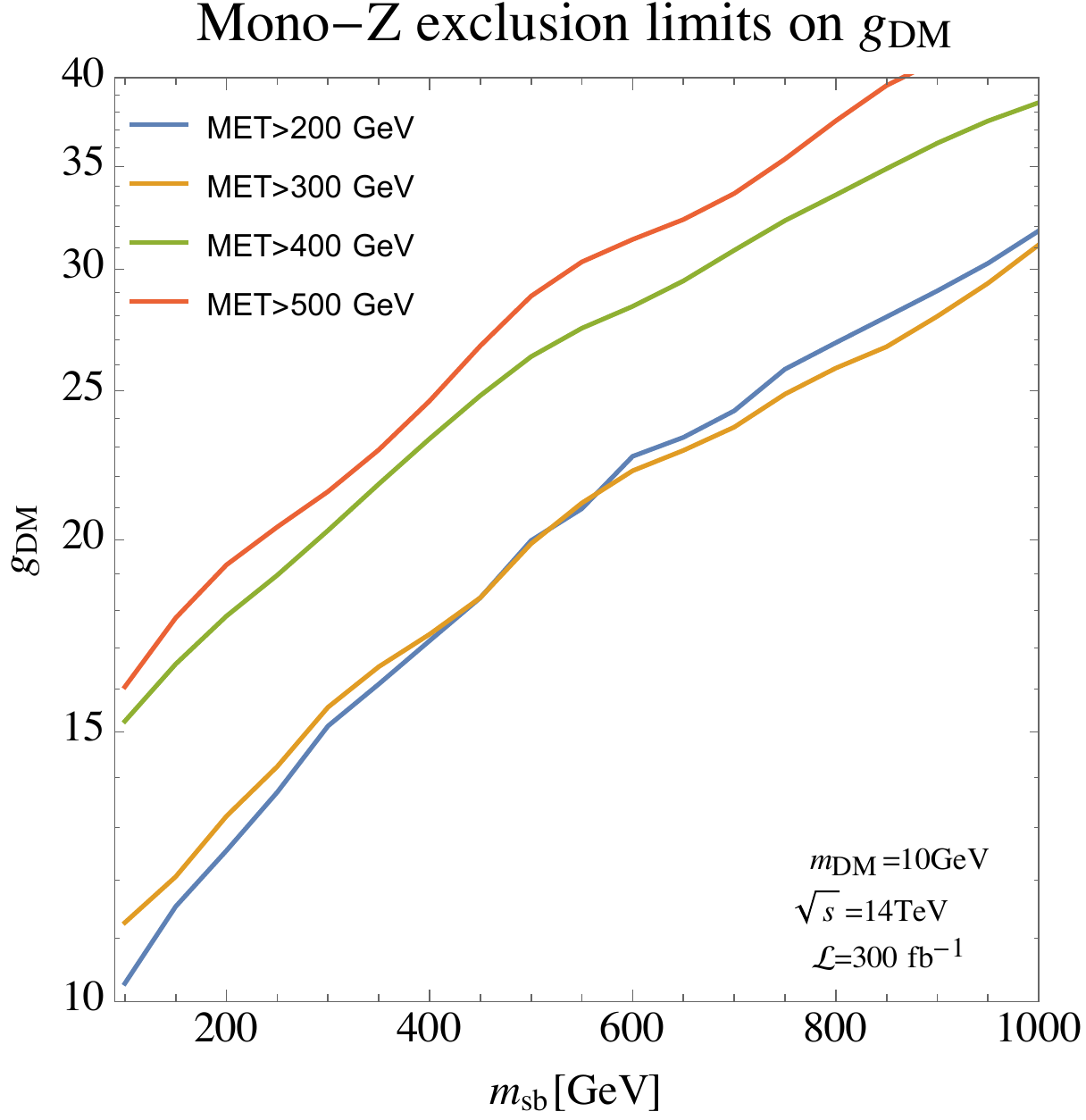}}
\subfigure[]{\includegraphics[scale=0.68]{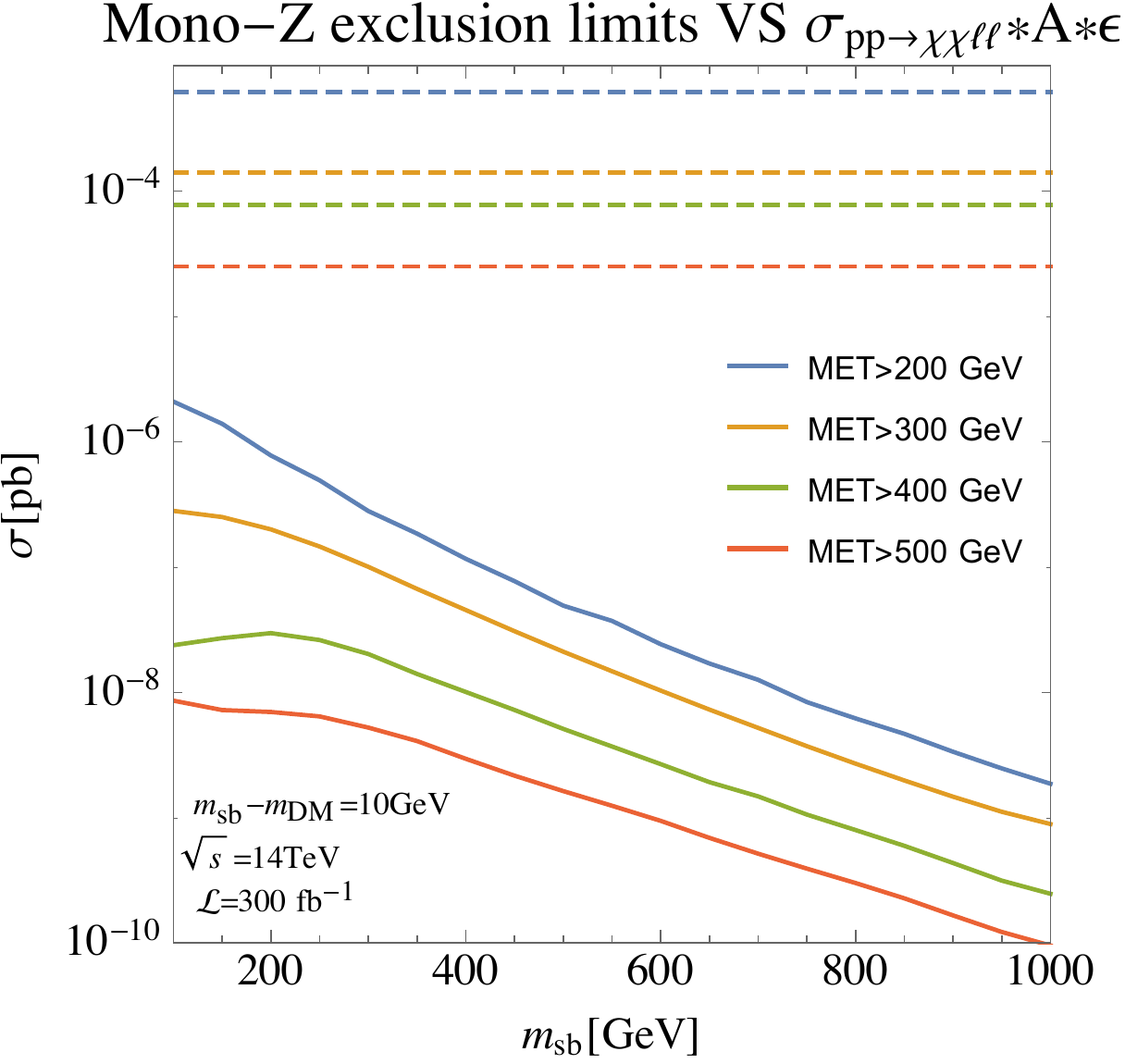}}
\subfigure[]{\includegraphics[scale=0.64]{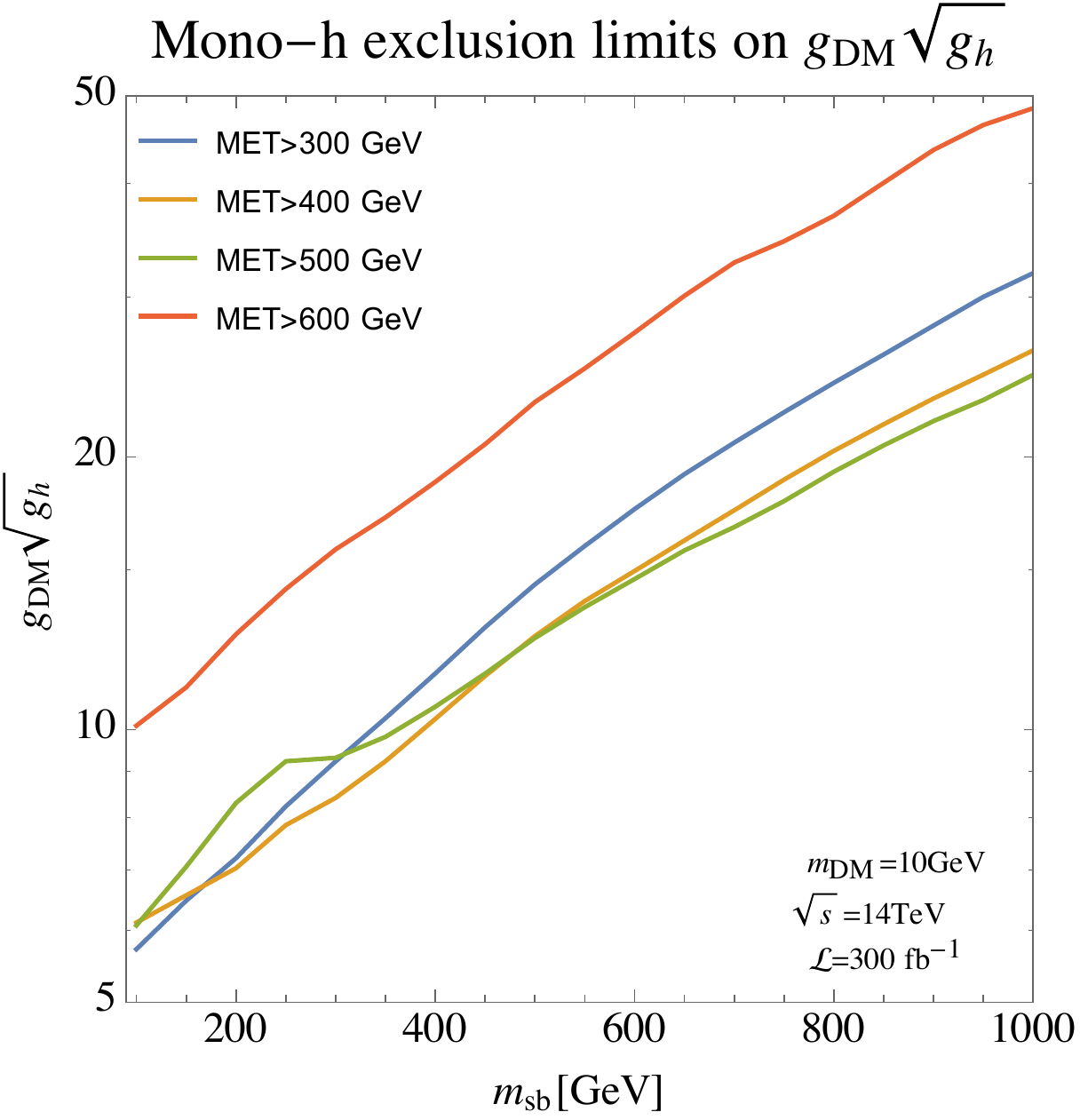}}
\subfigure[]{\includegraphics[scale=0.65]{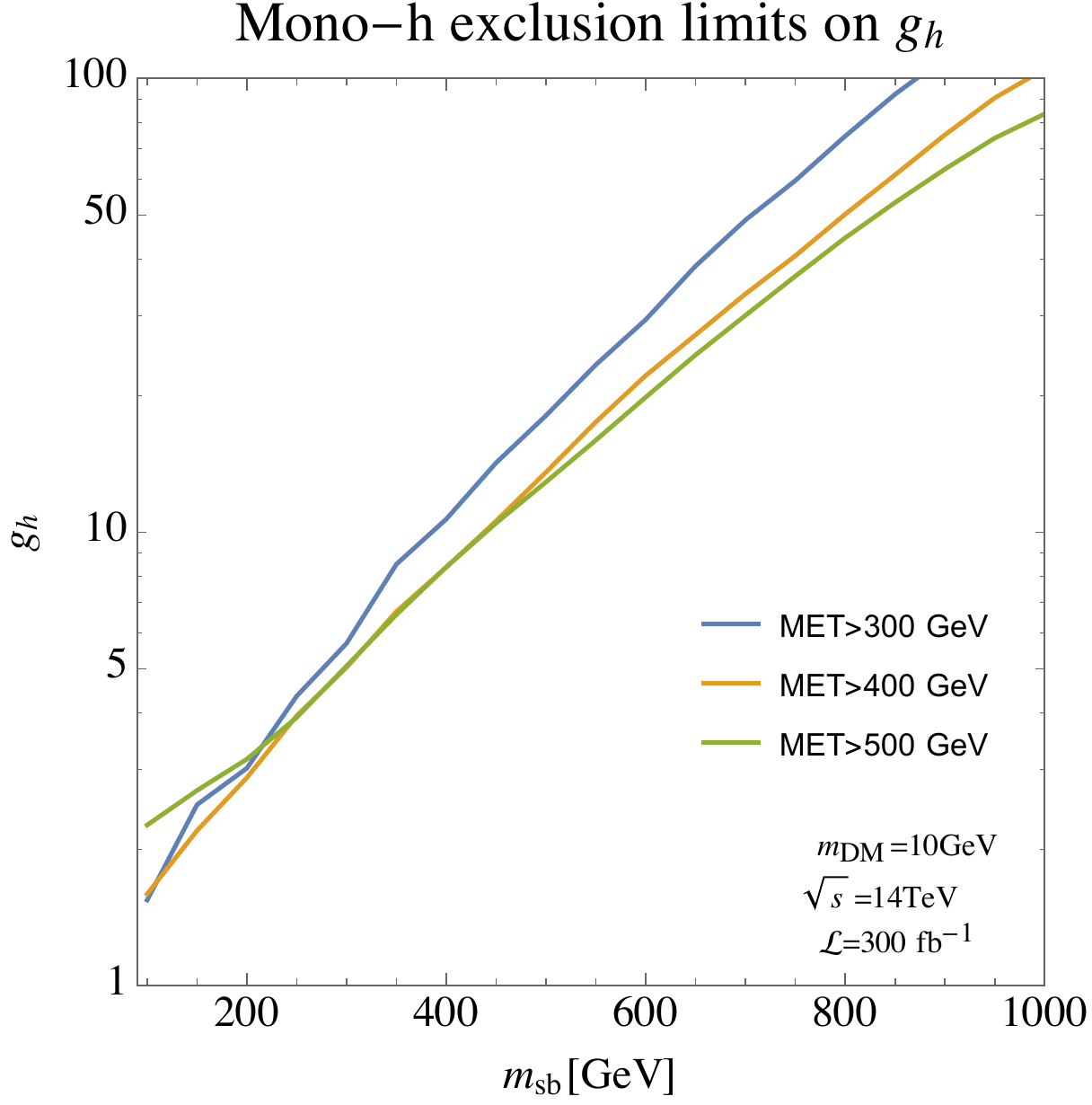}}
 \caption{  
Sbottom model: Projection at 14~TeV with 300 fb$^{-1}$ of the 95\% exclusion limits on the cross-section.  Conventions are as in Fig~\ref{fig:sbottomsearch}. We are not aware of existing mono-$b$ and $b$-jets+$\met$ projections at $14$ TeV.} 
 \label{fig:sbottomprojections}
\end{figure}

It is worth noting that different configurations translate into bounds on different combinations of couplings. For generic $m_{DM}$, the mono-$Z$ search sets a limit on the sbottom-bottom-DM coupling, while mono-$h$ constrains the combination $g_{DM}\sqrt{g_h}$. On the other hand, in the compressed regime the dependence on $g_{DM}$ is lost. We show the projection at 14~TeV in Fig.~\ref{fig:sbottomprojections}, performed with the procedure described in Appendix~\ref{app:b}.
Our results show that the mono-$Z$ analysis is never able to set a limit on perturbative values of the couplings. Stated in a different way, the cross-section rescaling needed to exclude a given point of the parameter space is nowhere close to one, both at LHC8 and LHC14, although the latter slightly improves over the former. This is not surprising, given the results of the previous subsection and the fact that the $Z$ boson does not distinguish between (s)quarks of different generations.

On the other hand, as shown in Fig.~\ref{fig:sbottomsearch} (c,d) the mono-$h$ analysis is instead able to set a limit\footnote{The limit we found makes sense because of our normalization of $g_h$ in Eq.~\ref{eq:sbottomLagrangian}.} on the coupling $g_h$. The bound is further improved at LHC14, as shown in Fig.~\ref{fig:sbottomprojections} (c,d).

\subsection{$s$-channel vector mediator}

Having investigated several models that can be constrained dominantly (at least in certain regions of parameter space) by various mono-$X$ searches, we now step back and consider models with an $s$-channel mediator that have been constrained previously by mono-jet, mono-Higgs and mono-$Z$.  

We first consider the production of Higgs in association with a new massive gauge boson $Z'$ which subsequently decays to DM. This mono-Higgs process occurs via an $s$-channel $Z'$, and has been studied previously in Ref.~\cite{Carpenter:2013xra}. Our purpose here is to compare the constraints obtained there with di-jet and monojet constraints on the $Z'$ mediator, which one expects to be important since the mediating $Z'$ particle has interactions with quarks as well as DM. We write the interaction Lagrangian of this simplified model as
\bea
\mathcal{L} &\supset&  g_{q} Z'_{\mu} \sum_{i=1,2} \left(\bar{Q}_L^i\gamma^\mu  Q^i_L 
+ \bar{u}^i_R \gamma^\mu u^i_R +  \bar{d}^i_R\gamma^\mu d^i_R \right) \nonumber \\
&&+ g_{DM} Z'_\mu \bar{\chi}\gamma^\mu \chi + g_H m_{Z'}hZ'_{\mu}Z'^{\mu}. 
\label{eq:Zplagrangian}
\eea 
Such an interaction of a $Z'$ with quarks and DM can originate from a baryon number gauge symmetry $U(1)_B$, assuming DM is also gauged under $U(1)_B$. We further assume that the $Z'$ obtains its mass $m_{Z'}$ from the spontaneous $U(1)_B$ symmetry breaking due to a new scalar $h_B$ gauged under $U(1)_B$.  DM production associated with a Higgs is made possible by mixing the new scalar with the SM Higgs. $Z-Z'$ mixing is not required to reproduce the mono-$h$ topology, and therefore the model is not constrained by precision electroweak measurements. See Ref.~\cite{Carpenter:2013xra} for a more detailed discussion\footnote{Another simplified model with the same DM production topology has been considered in Ref.~\cite{Carpenter:2013xra}, where the $hZ'$ production occurs via an $s$-channel SM $Z$ boson. In order to observe or constrain such a process at the LHC, however, one requires large $Z-Z'$ mixing, which has already been disfavored by precision electroweak measurements.  We do not consider this simplified model further.}. In this framework:
\bea
g_H = \frac{m_{Z'}\sin\theta}{v_B}, \qquad \tan\theta = \frac{v_B}{v}, \quad \langle h_B\rangle = v_B,
\label{eq:couplingsdef}
\eea
where $v$ is the usual Higgs vev.

\begin{figure}[H]
\centering
\subfigure[]{\includegraphics[scale=0.66]{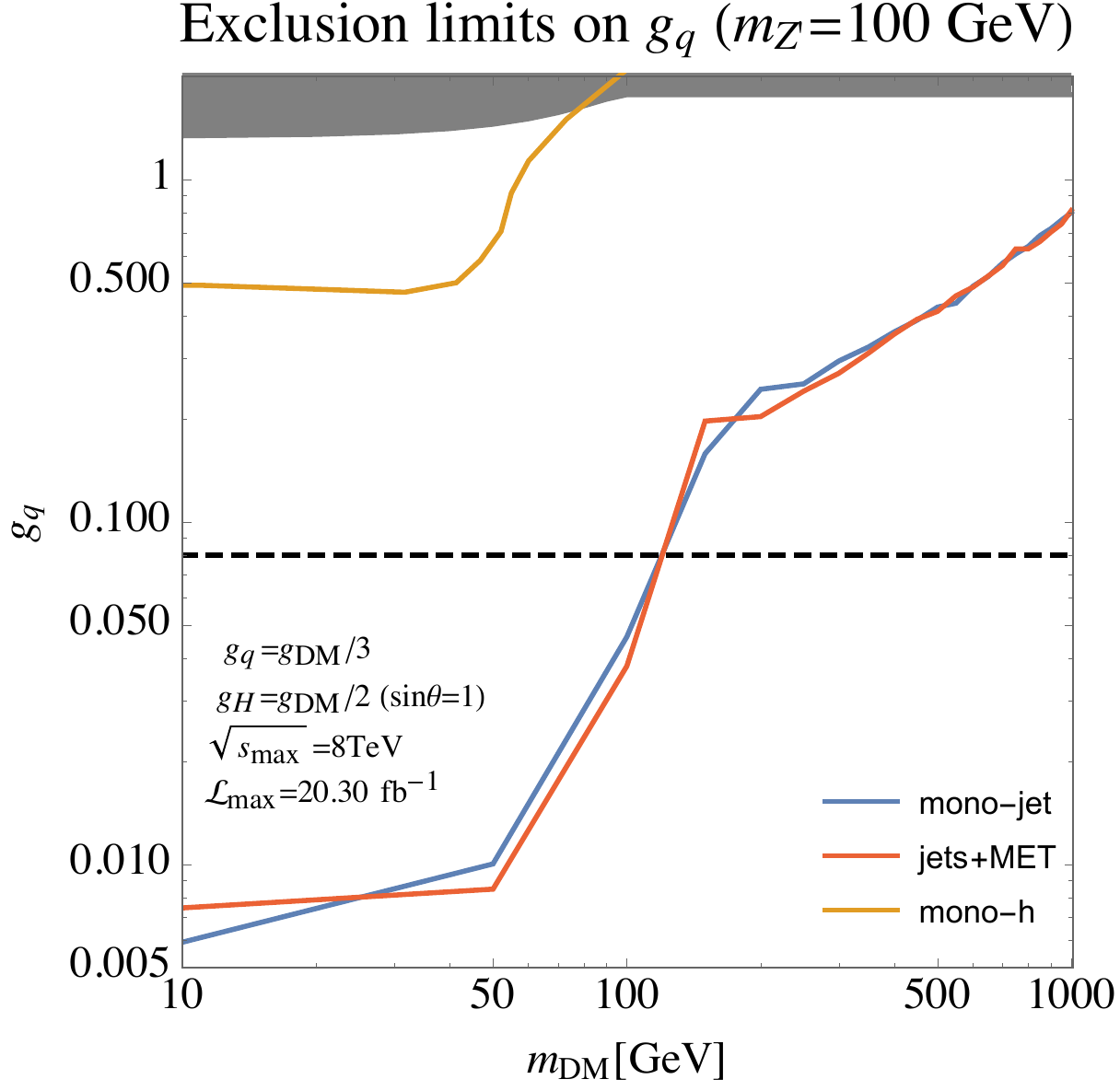}}
\subfigure[]{\includegraphics[scale=0.66]{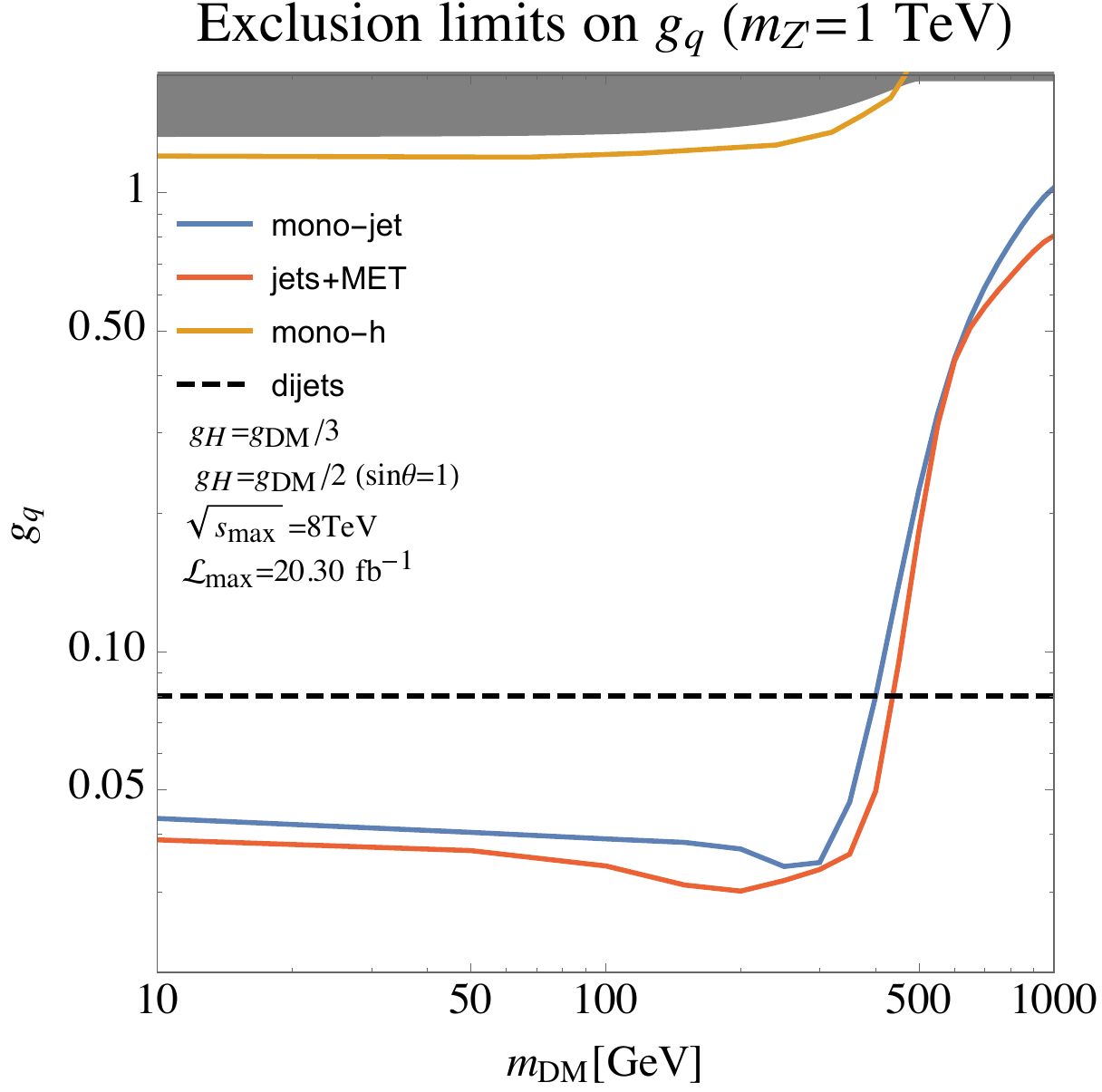}}
\caption{ $s$-channel vector mediator model: comparison of exclusion limits at 95\% C.L. on the quark-Z' coupling $g_{q}$ as a function of the DM mass obtained from monojet, jets~+~$\met$, di-jet resonances and mono-$h$ searches. The latter is taken from Ref.~\cite{Carpenter:2013xra}. We show two benchmark points: $m_{Z'} = 100$ GeV and  $m_{Z'} = 1000$ GeV, $g_q=g_{DM}/3$. $\sqrt{s_{max}}$ and $\mathcal L_{max}$ represent the maximum energy and luminosity among the analyses used.  A detailed list is reported in Table~\ref{tab:search}. The shaded region corresponds to the non-perturbative region defined by the condition on the $Z'$ width $\Gamma_{Z'}=m_{DM}$.} 
\label{Model1}
\end{figure}

 We compare constraints from mono-Higgs to those obtained from di-jet searches for the mediator, di-jet~+~$\met$, monojet, and mono-Higgs in Fig.~\ref{Model1}. Different searches constrain different combinations of $g_q$, $g_{DM}$ and $g_H$.  To perform a meaningful comparison and make contact with the analysis already performed in Ref.~\cite{Carpenter:2013xra} we properly rescale one of their benchmarks and translate all the bounds to the quark and Higgs couplings:
\be
 3 g_q = g_{DM} = g_B, \qquad  g_H = 3g_q/2\,,
\label{eq:bechmark}
 \ee
where $g_B$ is the $Z'$ gauge coupling, while the coupling to the Higgs boson $g_H$ has been taken at the formal limit of the perturbative regime consistent with Eq.~\ref{eq:couplingsdef} in order to maximize the constraining power of mono-Higgs analysis. 
The only limits not present in the literature are those coming from jets~+~$\met$. These were obtained with a full recasting along the lines of Ref.~\cite{Papucci:2014iwa}, using the minimal $Z'$ width resulting from the couplings in Eq.~\ref{eq:Zplagrangian}. Details are provided in Appendix~\ref{app:a}. The limits from di-jets are taken directly from the literature~\cite{Khachatryan:2016ecr,CMS:2016jog}, taking into account the factor of $1/2$ difference in the normalization of the coupling.

For the choice of parameters in Eq.~\ref{eq:bechmark}, jets~+~$\met$ and monojet appear comparable and much more constraining than mono-Higgs searches. For heavier $Z'$ masses ({\em e.g.} $M_{Z'}\simeq 1.5$ TeV), jets~+~$\met$ is less constraining, while the di-jet bound plays the dominant role.

\subsection{$s$-channel scalar mediator}

We next replace the vector $s$-channel mediator in the previous scenario with a scalar mediator in order to realize the DM production topology in the second row of Table~\ref{tab:ab}. This is possible by introducing a singlet $S$ that acts as a portal between DM and the SM Higgs:
\bea
 \mathcal{L} \supset - y S\bar{\chi}\chi + \frac12 m_{hS}^2 hS.
\eea

 Specifically, we consider the following Lagrangian:
\bea
&&\mathcal L = \mathcal{L}_{SM} +i\bar{\chi}\slashed{\partial}\chi + \frac12 (\partial_\mu S)^2 - \frac12 m_S^2 S^2 -\eta (H_{SM}^\dagger H_{SM}) S -\lambda  (H_{SM}^\dagger H_{SM}) S^2   - y S\bar{\chi}\chi,
\eea
where $H_{SM}$ is the SM Higgs doublet.   The SM Higgs sector is, as usual:
\bea
 &&  \mathcal{L}_{SM} \supset  \frac12 m_h^2  (H_{SM}^\dagger H_{SM}) - \frac{m_h^2}{2 v^2} (H_{SM}^\dagger H_{SM})^2 + \sum_i \left( y^i_u H_{SM}^\dagger \bar{Q}_L^i u^i_R + y^i_d H_{SM} \bar{Q}_L^i d^i_R \right).
\eea
This model was also considered in Ref.~\cite{Carpenter:2013xra}, where they found that neither mono-$h$ nor mono-$Z$ is strongly constraining.  Here we consider whether a monojet search can be constraining on the parameter space of this model.
We use the parameterization of a singlet mixed with the Higgs boson, defining 
\bea
&&H_{SM} = \frac{1}{\sqrt2}\begin{pmatrix} 0\\ v+h \end{pmatrix}\\
&&h = \cos_\theta h' + \sin_\theta S'\\
&&S = -\sin_\theta h' + \cos_\theta S'\\
&& \tan2\theta = \frac{2\eta v}{m_S^2+\lambda v^2-m_h^2}.
\eea
We obtain the Lagrangian in terms of the mass eigenstates $h'$ and $S'$. After the field redefinition, the new scalar $S'$ couples to all quarks with strength $\frac{m_q}{v}\sin_\theta$. In addition, all the Higgs couplings will be rescaled by a factor of $\cos_\theta$. These shifts are taken into account in our analyses and plots. 

\begin{figure}[H]
\centering
\includegraphics[scale=0.7]{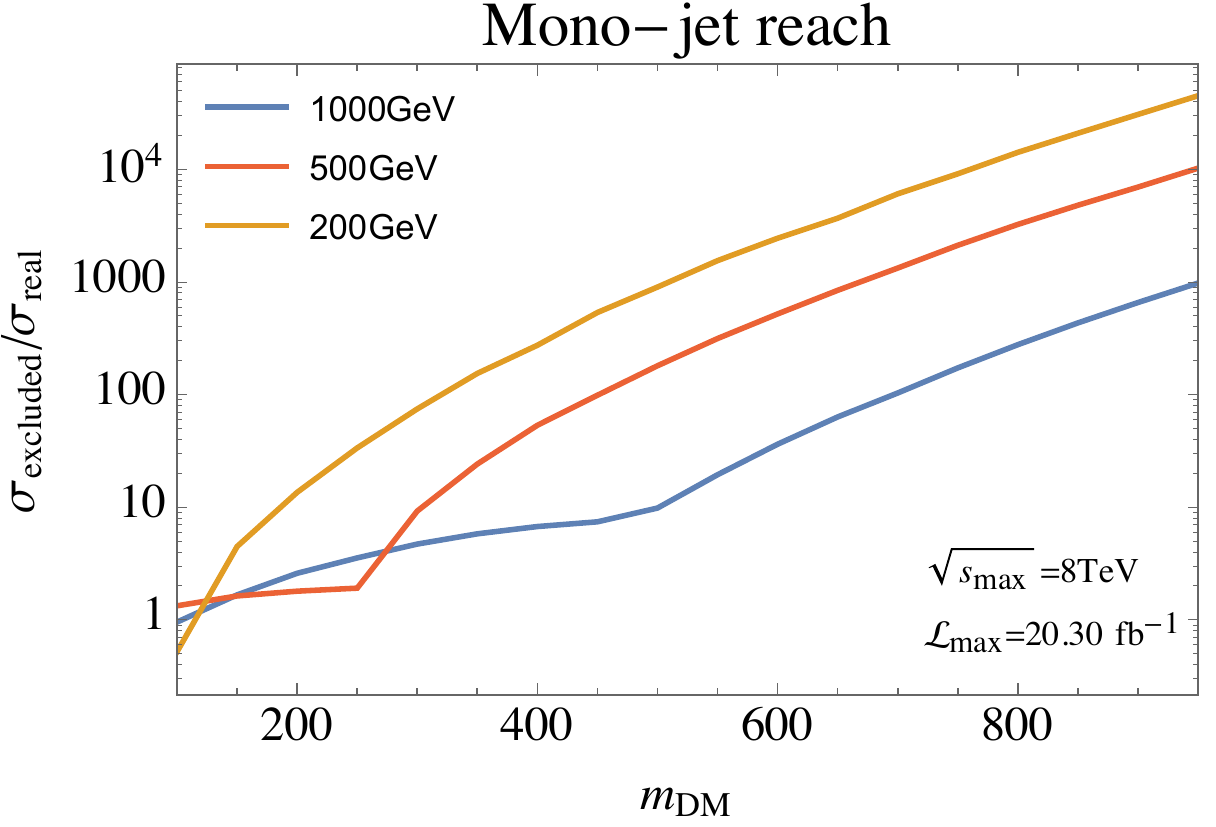}
\caption{$s$-channel scalar mediator model: rescaling needed in the mono-jet~+~$\met$ cross-section in order for LHC8 to be sensitive. Different curves correspond to different values of the singlet mass $S$. $\sqrt{s_{max}}$ and $\mathcal L_{max}$ represent the maximum energy and luminosity amongst the analyses used.  A detailed list is reported in Table~\ref{tab:search}.} 
\label{fig:Model2}
\end{figure}

We find that the constraints from the monojet search on such model are also generally very weak, as shown in Fig.~\ref{fig:Model2}.

\subsection{Inelastic squarks}


Until this point our simplified models have demanded that the dark matter not only be charge neutral, but also a singlet.
In this (last) section we consider a scenario where the dark matter is not directly coupled to a squark-like particle, but is instead produced though an additional intermediate state. Although it is possible to build such a model using only singlet dark matter, one would engineer a rather complicated construction in order to produce sizable mono-$h$ and mono-$Z$ signals\footnote{Such a model would consist of a squark-like particle and two neutral states $\chi,\chi'$. Sizable cross-section for mono-$h$ and mono-$Z$ are obtained through $Z-Z'$ and $h-S$ mixing. Here $Z'$ and $S$ are two additional vector and scalar fields.}. 
Thus, for the sake of simplicity, here we abandon the singlet requirement in favor of a more elegant and simple model. 

We study a model consisting of colored scalar mediators (the eight light flavor squarks) and two electroweak fermion doublets (Higgisnos, $\widetilde{H}_{1,2}$) acting as the mediators. The Higgsinos have a Dirac-like mass $\mu$-term, and their neutral components mix with a singlino $\chi$ (DM) via the SM Higgs to form mass eigenstates $\chi_i$, with $i=1-3$. Squarks couple to the singlino and $\widetilde{H}$'s. The Lagrangian is:
\ba
{\cal L} &\supset& - \frac{m_S}{2}\chi\chi - m_{D} \widetilde{H}_{1}\widetilde{H}_{2}-y_1 \chi\widetilde{H}_{1} H_{SM} -y_2 \overline{\widetilde{H}_2}\bar{\chi}H_{SM} \nonumber \\
 & &+g_{DM} \sum_{i=1,2}\left(\widetilde{Q}_L^i \bar{Q}^i_L+\tilde{q}^i_R \bar{q}^i_R \right)\chi  +g_{\tilde{H}}\left(\overline{\widetilde{Q}_L^i}q_R^i\widetilde{H}_2 +\bar{q}_R^i\widetilde{Q}_L^i\widetilde{H}_1\right) + h.c.
\ea

This model (and its pure electroweak subsector), being a generalization of a sector of the MSSM where the SUSY relations between gauge and Yukawa couplings have been relaxed, has been considered in the literature for many applications~\cite{An:2013xka,Papucci:2014iwa,Garny:2014waa,Mahbubani:2005pt,Cohen:2011ec}.

Here, we consider the production of $\chi \widetilde{H}$ through squarks in the $t$-channel at the LHC. $\widetilde{H}$ then decays into $Z$ ($H$) and $\chi$,  giving a mono-$Z$ (mono-$h$) signature. In order for the mono-$h$ and mono-$Z$ channels to compete with other direct searches we focus on the region of parameter space where the squarks predominantly decay to Higgsinos. In particular for our benchmark point we fix the decay branching ratios of the squarks to be  ${\rm Br}(\tilde{q}\to q +\chi_{1}^{\pm}):{\rm Br}(\tilde{q}\to q+\chi_{2,3}):{\rm Br}(\tilde{q}\to q+\chi_{1})\simeq 6:3:1$ (this is achieved for example by choosing the ratio of the couplings $g_{\widetilde{H}}/g_{DM}$ to be $\sqrt{5}$). Furthermore, we require that the neutrali Higgsinos to have equal branching rations for the decays into a $H$ or a $Z$ and the DM particle. While the full parameter space will not be explored in this paper, we identify three mass spectra as our benchmark scenarios, corresponding to non-compressed mass spectrum, compressed $\widetilde{H}$-$\chi_1$ mass spectrum, and compressed $\tilde{q}$-$\widetilde{H}$ mass spectrum.  

\begin{figure}[H]
\centering
\subfigure[]{\includegraphics[scale=0.6]{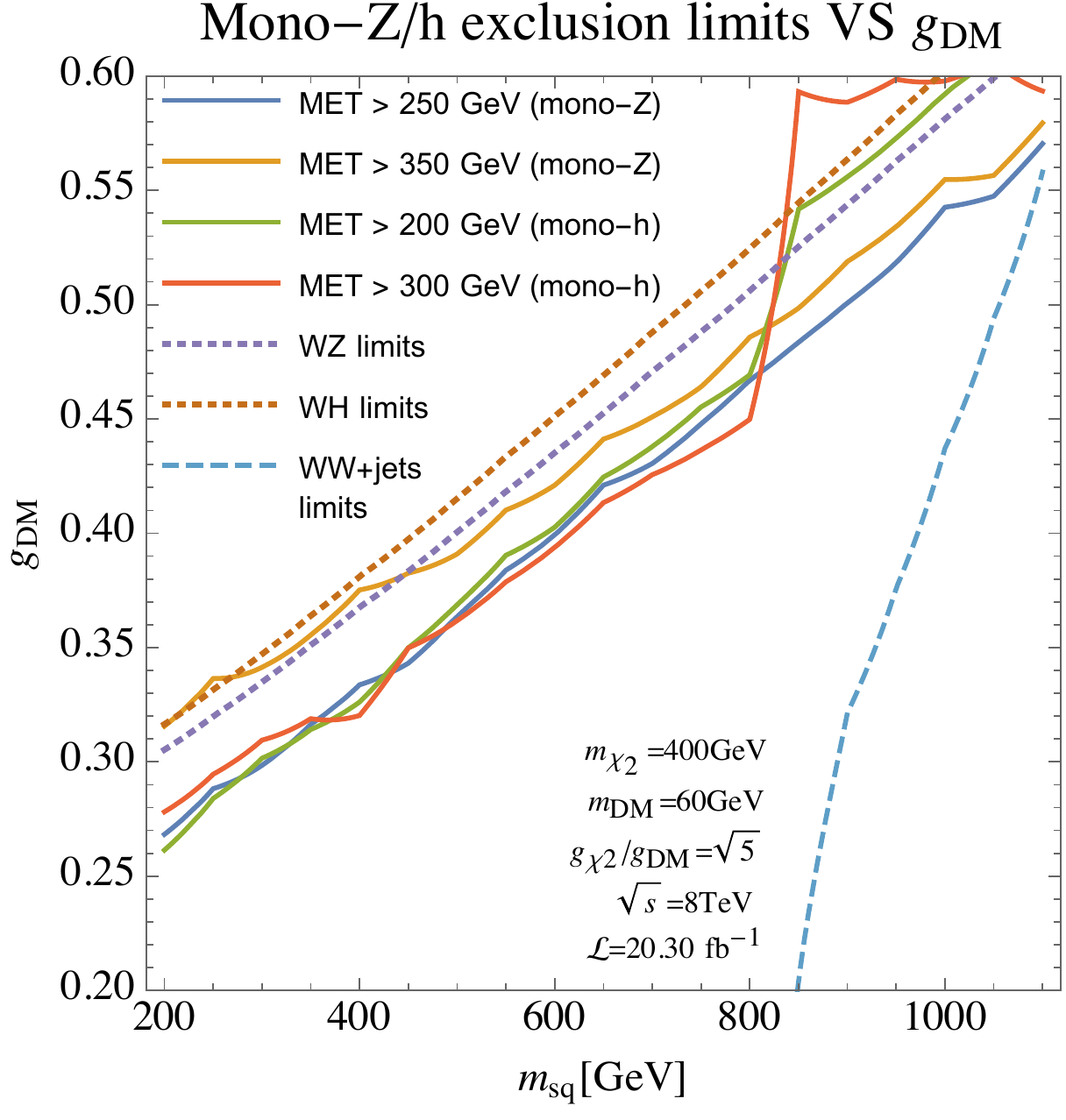}}
\subfigure[]{\includegraphics[scale=0.6]{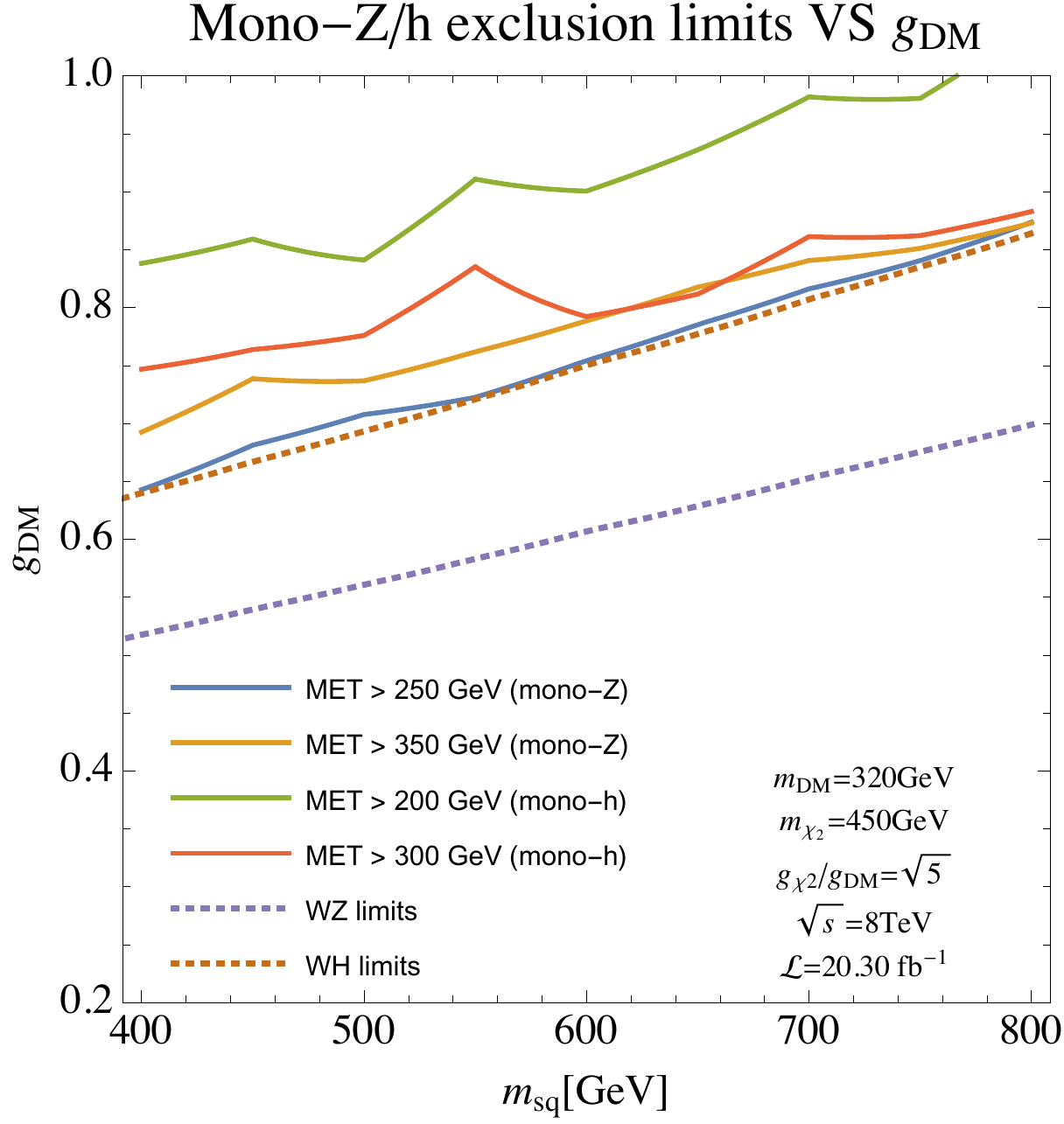}}
\subfigure[]{\includegraphics[scale=0.61]{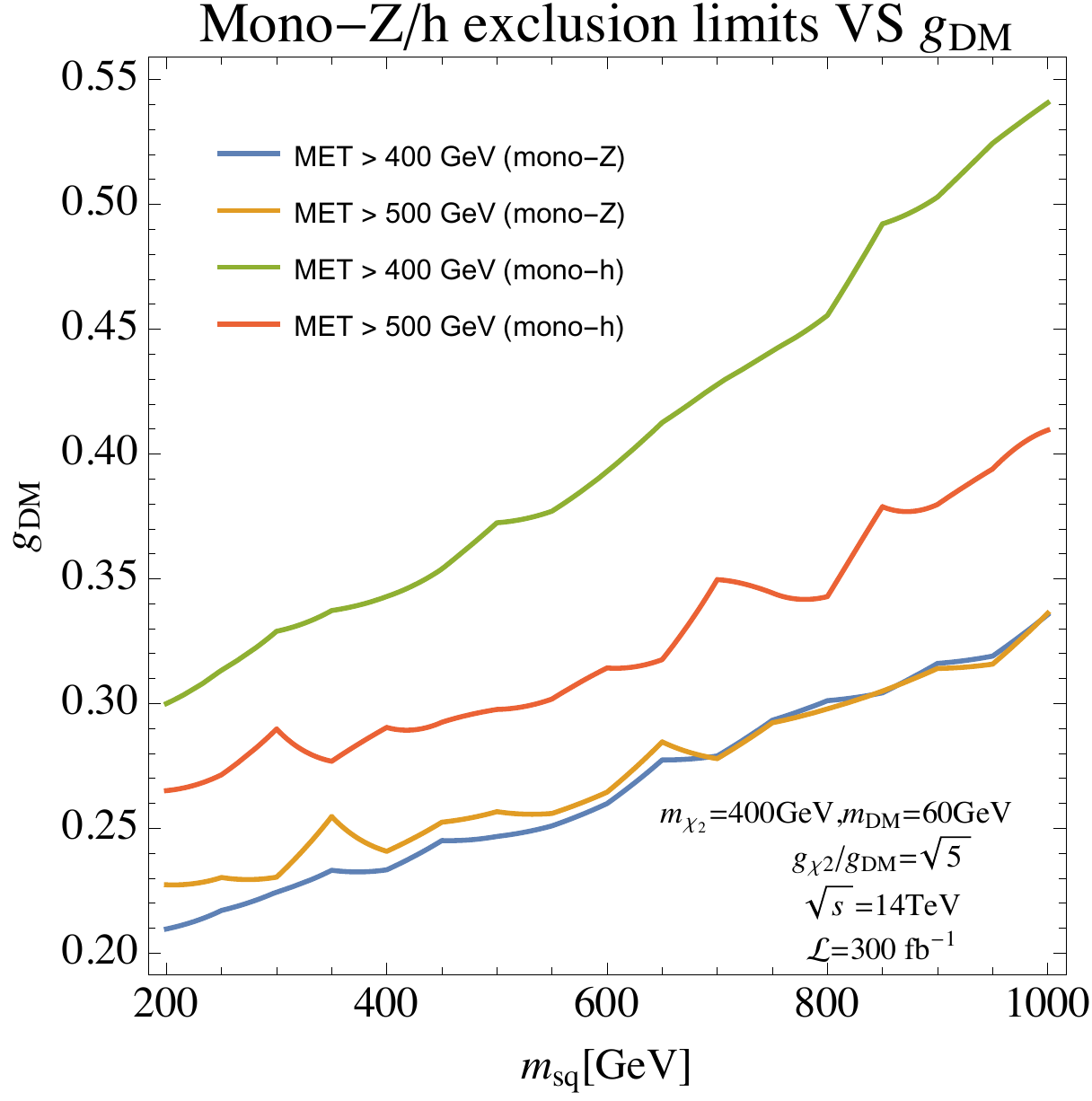}}
\subfigure[]{\includegraphics[scale=0.6]{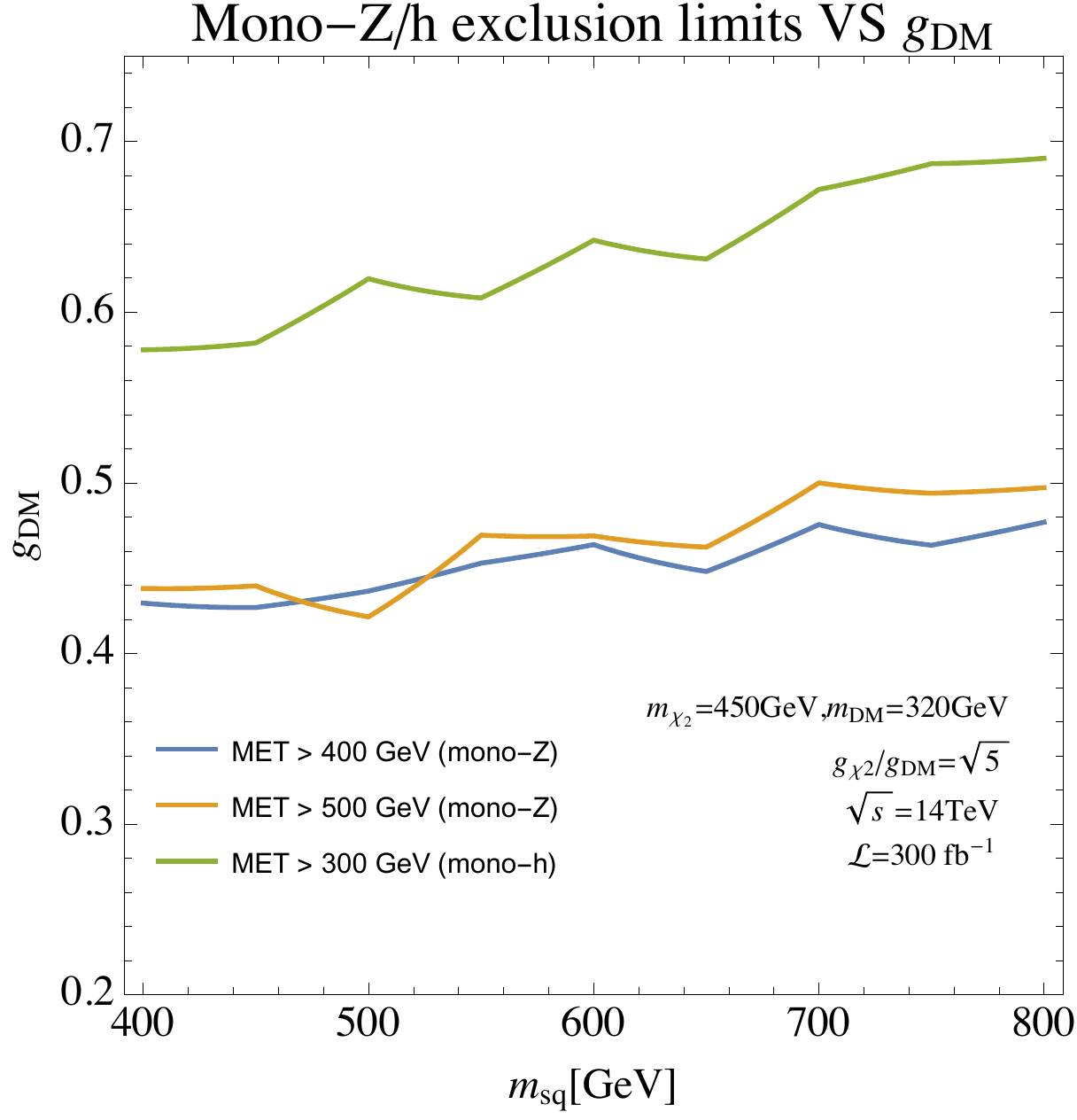}}
 \caption{Inelastic squark model: mono-$Z$ and mono-Higgs, shown as solid lines with different colors corresponding to different $\met$ cuts, as well as 8~TeV $WZ$, $WWj$ and $WH$~+~$\met$, 95\% exclusion limits on the inelastic squark model, shown as dashed lines, with (a) DM mass 60 GeV, Higgsino mass 400 GeV, and (b) DM mass 320 GeV, Higgsino mass 450 GeV. The $WZ/WH/WWj$~+~$\met$ limits are estimated using efficiency tables and cross-section upper limits given in~\cite{Aad:2014vma,Khachatryan:2014mma,Aad:2015mia}. Panels (c) and (d) show the 14~TeV projections at 300 ${\rm fb}^{-1}$.} 
 \label{fig:in_sq}
\end{figure}

For the scenario with non-compressed mass spectrum, the mass of $\chi_{2,3}$ and $\chi_1$ are 400 GeV and 60 GeV respectively. Note that the current LHC constraints on the electroweak production of electroweak-inos are irrelevant for this choice of parameters. In addition to $\chi \widetilde{H}$ production, the process $pp \to \chi_{2,3}\chi_{2,3} \to \chi_1\chi_1 Z V(\equiv Z/h) \to \chi_1\chi_1 \nu \bar{\nu} V$ also contributes to the mono-$Z$ ($h$) production. The reach of mono-$Z$ and mono-$h$ at 8~TeV are shown in Fig.~\ref{fig:in_sq} (a) on the $g_{DM}$-$m_{sq}$ plane. On the same plot we show the constraints on $g_{DM}$ from the $WZ$~+~$\met$, $WH$~+~$\met$, and the $WW$~+~jets~+~$\met$ searches~\cite{Aad:2014vma,Khachatryan:2014mma,Aad:2015mia}. The first two constraints arise from the processes $pp \to \chi_{2,3}\chi_1^{\pm} \to \chi_1 \chi_1 W^{\pm} V$ via $t$-channel squarks, where $\chi_1^{\pm}$ is the charged Higgsino. The $WW$~+~jets~+~$\met$ search corresponds to constraints from the direct squark decay to the W boson ($\tilde{q}\to j \chi_1^{\pm}\to j\chi_1 W^{\pm}$).  This search tags the leptonic decay mode of the $W$ boson, and is more constraining than the standard jets~+~$\met$ searches. It can be observed that the mono-$Z$/$h$ search imposes weaker constraints than the $WW$~+~jets~+~$\met$ search.  Fig.~\ref{fig:in_sq} (b) shows LHC constraints ($WV$~+~$\met$ searches) on the scenario with compressed $\widetilde{H}$-$\chi_1$ mass spectrum, with masses $\chi_{2,3}$ and $\chi_1$ set to 450 GeV and 320 GeV respectively. Overall the limits on $g_{DM}$ are expectedly weakened in this compressed mass region. Even so, the $WV$~+~$\met$ searches are the more powerful probe of this parameter region compared to mono-$Z$/$h$. As the $WV$~+~$\met$ constraints alone are sufficient to overcome mono-$Z$/$h$, the $WW$~+~jets~+~$\met$ constraint is not shown in the plot. While mono-$Z$/$h$ limits are expected to improve at 14~TeV as shown in Fig.~\ref{fig:in_sq} (c) and Fig.~\ref{fig:in_sq} (d), the 8~TeV $WZ/WH/WWj$~+~$\met$ searches still outperform mono-$Z$/$h$.

\begin{figure}[t]
\centering
\subfigure[]{\includegraphics[scale=0.6]{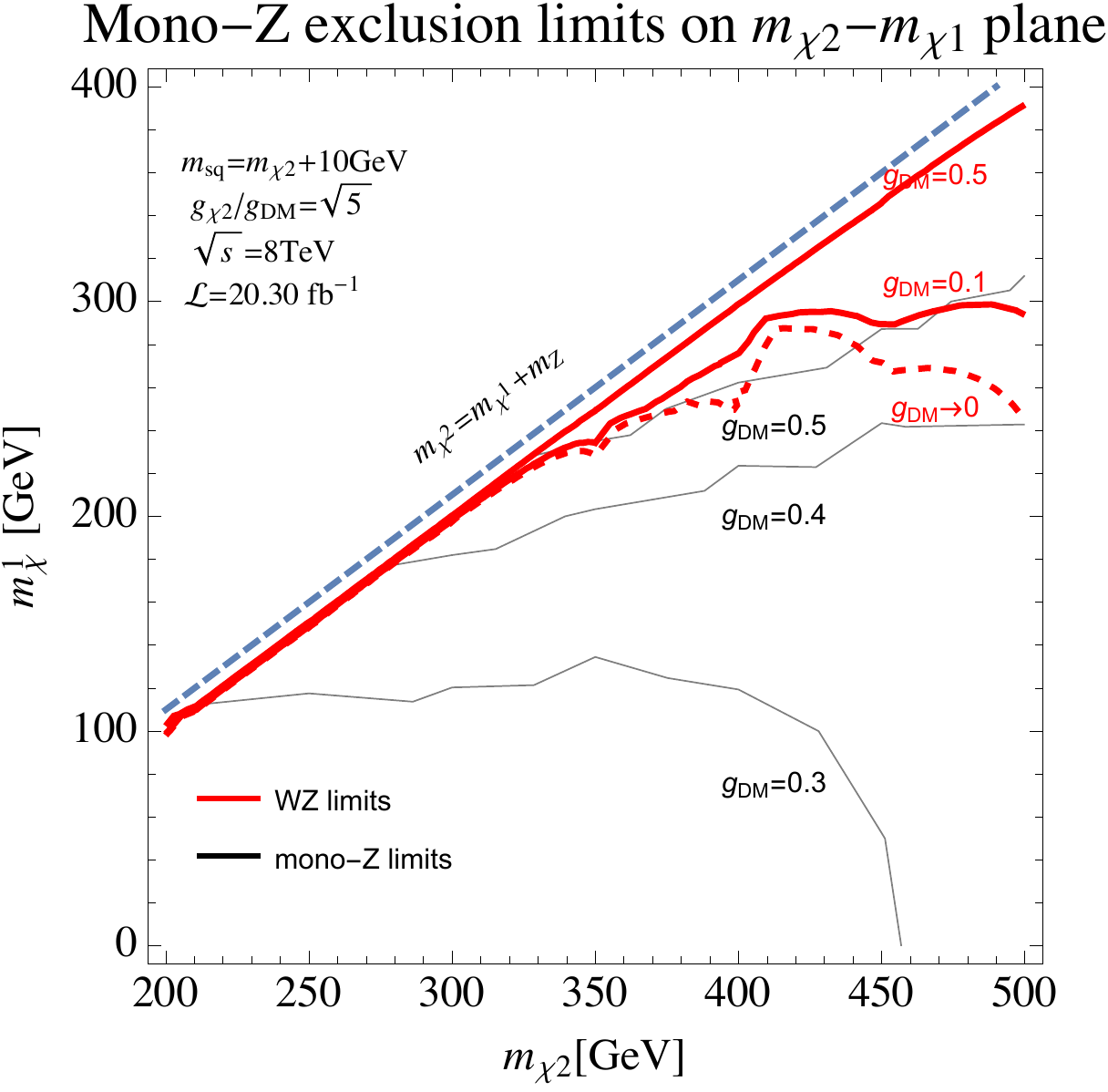}}
\subfigure[]{\includegraphics[scale=0.6]{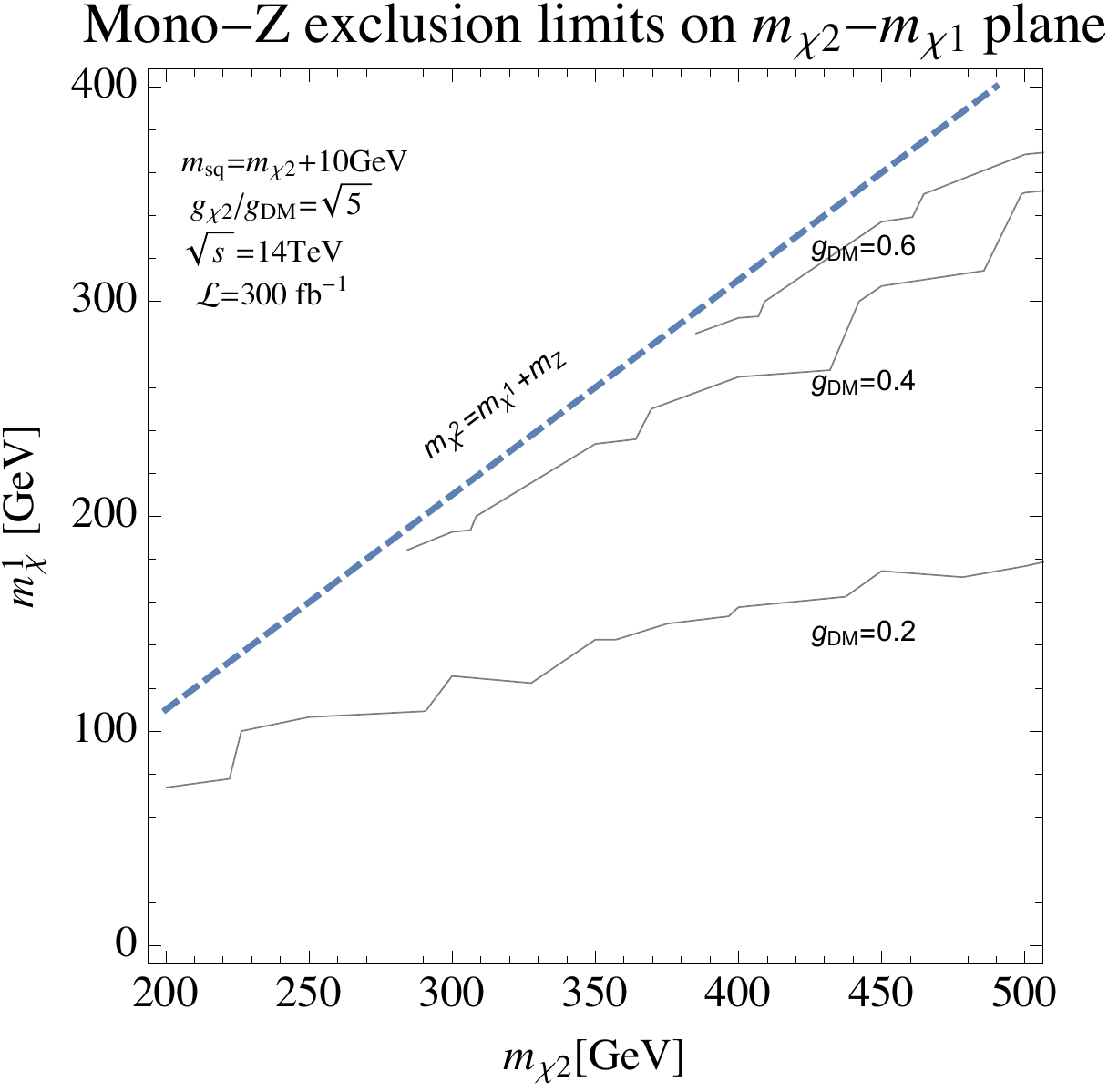}}
\caption{Inelastic squark model: mono-$Z$ limits (black lines) in the compressed mass region (squarks are 10 GeV heavier than $\chi_2$) at 8~TeV (a). The red dashed and solid lines represent limits from the electroweakino search in the WZ~+~$\met$ final states~\cite{Aad:2014vma}. The electroweakino search is dominant over mono-$Z$ in all parameter space investigated. The 14~TeV projections are shown in panel (b).} 
\label{fig:in_sq_monoz}
\end{figure}

\begin{figure}[t]
\centering
\includegraphics[scale=0.7]{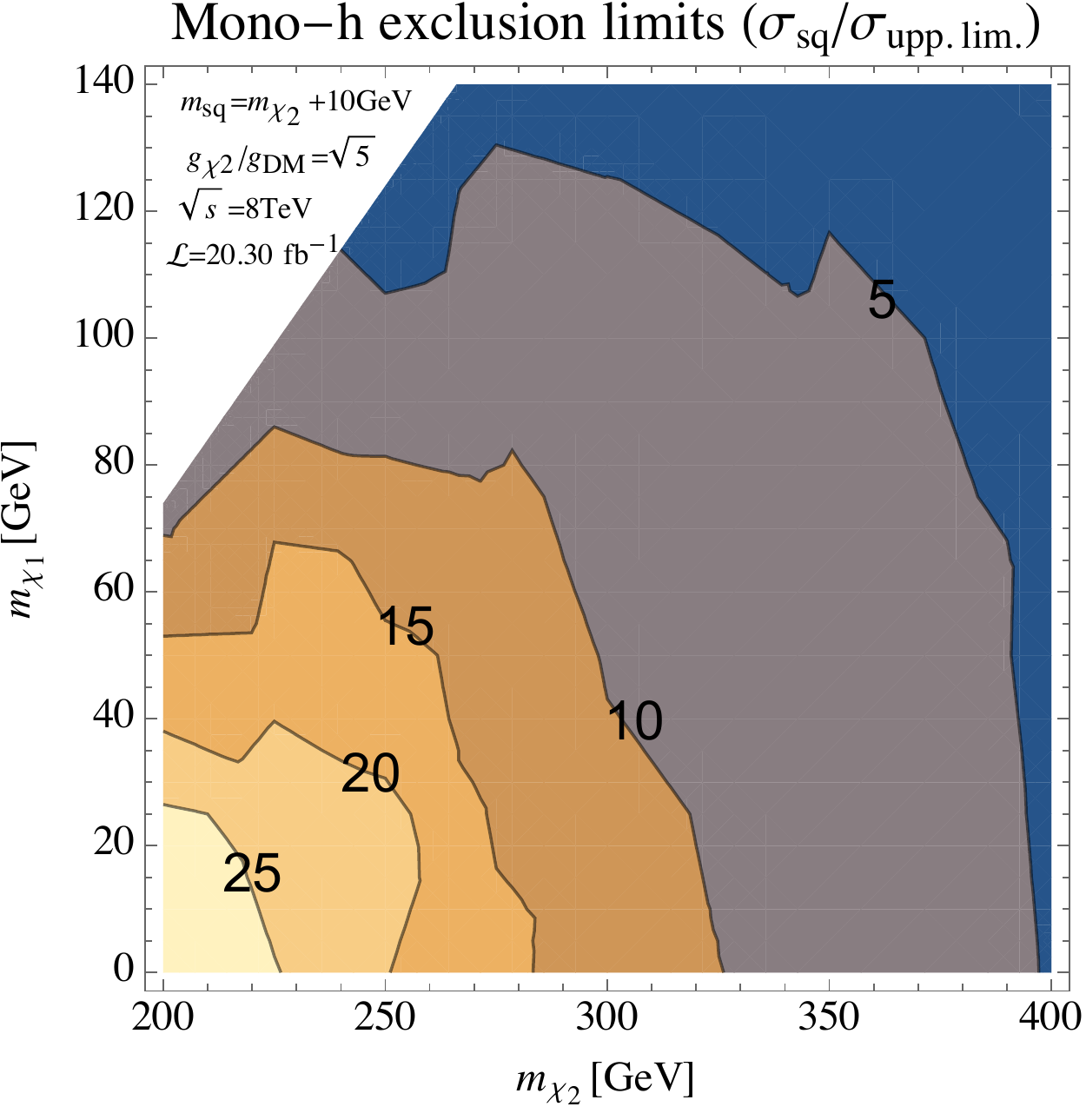}
\caption{Inelastic squark model: mono-$h$ limits in the compressed mass region (squarks are 10 GeV heavier than $\chi_2$). The contour lines represent the values of the squark production cross-section divided by the cross-section upper limits from the electroweakino search in the WH~+~$\met$ final states~\cite{Khachatryan:2014mma}. The electroweakino search is dominant over mono-$h$ in all parameter space investigated.} 
\label{fig:in_sq_monoh}
\end{figure}

Another scenario of interest lies in the compressed $\tilde{q}$-$\widetilde{H}$ mass region. We take squarks to be 10 GeV heavier than $\widetilde{H}$, and vary the masses of $\widetilde{H}$ and $\chi$. Soft jets from the squark decay can escape detection, and the cascade decay of squarks contribute sizably to the $WV$~+~$\met$ channel. However, as can be observed in Fig.~\ref{fig:in_sq_monoz}, the $WZ$~+~$\met$ channel is more constraining than mono-$Z$, taking into account constraints from the QCD squark production and the process $pp \to \chi_{2,3}\chi_1^{\pm} \to \chi \chi W^{\pm} V$ ($g_{DM}\neq 0$). In Fig.~\ref{fig:in_sq_monoh}, one observes that the $WH$~+~$\met$ constraint from QCD squark production is dominant over mono-$h$ regardless of the values of $g_{DM}$.

In summary, we do not find parameter space where mono-$Z$/$h$ is dominant over direct searches for the inelastic squark model.

\section{Conclusions}

It is essential to broadly explore DM simplified models at the LHC, elucidating how well the mono-$X$ and direct searches constrain each simplified model. In this paper, we proposed a set of simplified models covering mono-$X$ DM production topologies thoroughly, and we provided details of possible UV completions that realize the simplified model DM production topologies. Each model which produces a mono-$X$ signature through mediator decay to DM universally predicts other signatures, such as when the mediator decays back to the initial state particles that produced it ({\em e.g.} to a pair of jets).  Generally, the direct search for the mediator through visible states such as di-jets and diboson will generate stronger constraints than the mono-$X$ constraints from DM decays, even when the DM coupling to the mediating particle is at the perturbative limit.  However, each mono-$X$ search has a model, or region of parameter space, where the mono-$X$ signature dominates.  This is summarized in Table~\ref{tab:mono}. 

While mono-$X$ signatures are not generic searches for DM, as they are typically not the dominant channel, they are a useful tool in the hunt for physics beyond the SM.  




\section*{Acknowledgments}

SPL is supported by JSPS Research Fellowships for Young Scientists and the Program for Leading Graduate Schools, MEXT, Japan. MP and KZ are supported by the DoE under contract DE-AC02-05CH11231. AV is supported by the Swiss National Science Foundation under grant no. PP00P2-163670.

\appendix

\section{Experimental analyses and simulation details}
\label{app:a}
In this Appendix, we give descriptions of experimental analyses and simulation details of our study. For reference,  we list all relevant collider searches utilized in our analysis in Table~\ref{tab:searches}.

In the case of monojet (mono-$b$), (b-)jets~+~$\met$, and diboson signatures we made use of the cross-section limits on simplified models provided by experimental collaborations\footnote{This method neglects finite width effects, as extensively discussed in~\cite{Papucci:2014iwa}.}.
For mono-$Z$ and mono-$h$ analyses we generate events and implement the cuts using the Madgraph~\cite{Alwall:2011uj}, Pythia~\cite{Sjostrand:2007gs} and Delphes~\cite{deFavereau:2013fsa} pipeline. Our set of simplified models is implemented with the FeynRules package~\cite{Alloul:2013bka}.
For all the other searches (mono-jet and jets~+~$\met$) we also performed a full simulation, following a somewhat different procedure: first we simulated events with MadGraph. The we showered using Pythia, which were then passed through Atom~\cite{Atom}. The procedure follows closely the one described in Ref.~\cite{Papucci:2014iwa} and we refer to it for all the details. All the simulated events used the minimal width resulting from the couplings of the simplified model.

 Upper limits on mono-$X$ cross-sections are either taken from the experimental collaborations' reports, or extracted following the  $CL_S$ prescription~\cite{Cowan:2010js,Mistlberger:2012rs}. We summarize all LHC searches used in this work in Table~\ref{tab:search}.

\begin{figure}[t]
\centering
\includegraphics[scale=0.6]{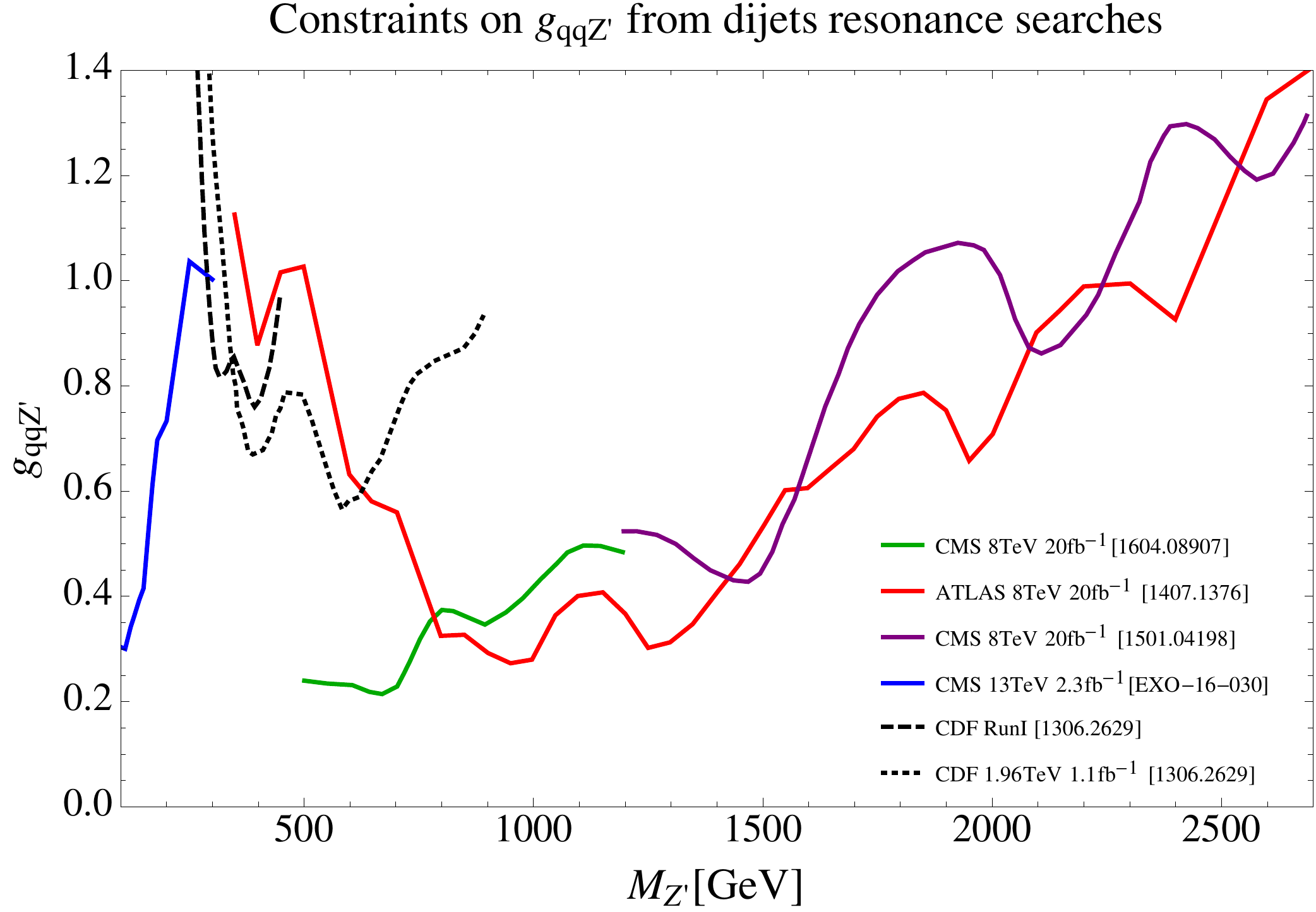}
\caption{Upper limits on the $u_R$-$Z'$ coupling at 95\% from di-jet resonance searches, taken from~\cite{Berlin:2014cfa,Aad:2014aqa,Khachatryan:2016ecr}, where different data sets are used to set the upper limits. For~\cite{Aad:2014aqa,Khachatryan:2016ecr}, we have rescaled the coupling upper limits as presented in~\cite{Dobrescu:2013coa} by recalculating the di-jet production cross-section of the relevant processes to reflect the assumption in our model, which has $Z'$ coupled only to $u_R$.} 
\label{fig:dijet1}
\end{figure}

 We report di-jet bounds on the $u_R$-$Z'$ coupling at 95\% from three different sources~\cite{Berlin:2014cfa,Aad:2014aqa,Khachatryan:2016ecr}, which use different data sets and have somewhat different results, as shown in Fig.~\ref{fig:dijet1}. The first (second) only provides bounds for $m_{Z'}\geq 300$ GeV ($m_{Z'}\geq 150$ GeV). It should also be noted that in Fig.~\ref{fig:dijet1}, $Z'$ presumably decays into jets with branching ratio 100 \%. In our models, $Z'$ can also decay into DM with a certain branching ratio, meaning that the $u_R$-$Z'$ coupling given in Fig.~\ref{fig:dijet1} has to be rescaled when $m_{Z'} > 2m_{DM} $. In our analysis, we calculate the partial width generated by the decay into DM and rescale the saturated di-jet constraints accordingly to take this into account.

\begin{table}[t] 
\centering 
\begin{tabular}{|l|l|l|l|l|}
\hline
Signature & Channel &Signal regions &search for  & refs 
\\
\hline
\multirow{4}{*}{jet(s)~$(+ \met)$} & $2j$    			&di-jet resonance &$Z'$    		& \cite{Berlin:2014cfa,Aad:2014aqa,Khachatryan:2016ecr}\\
						& $2j + \met$ 		& &$\tilde{q}$   		& \cite{Aad:2014wea, Chatrchyan:2014lfa,Chatrchyan:2013mys, ATLAS:2012sma,Chatrchyan:2012lia,Chatrchyan:2012wa}\\
						&$1j + \met$ 		& &monojet 		&\cite{Aad:2014nra,Aad:2015zva,Khachatryan:2014rra,ATLAS:2012ky,Chatrchyan:2012me}\\
\hline
\multirow{3}{*}{b-jet(s)~$(+ \met)$} & $H(\to 2b)+ \met$   	&$\met > 150, 200,$ &mono-$h$   	& \cite{Aad:2015dva}\\
					            & $2b + \met$ & $300, 400~\GEV$& sbottom & \cite{Aad:2013ija}\\
					            & $1b + \met$ && mono-$b$ &  \cite{Aad:2014nra}\\
\hline
\multirow{3}{*}{lepton(s)~$(+j+ \met)$}  & $Z (\to l l)W(\to 2j) + \met$&$\met > 150, 250,$&$\chi_1^{\pm}\chi^0_2$ & \cite{Aad:2014vma} \\
							& $Z (\to l l) + \met$&$350, 450~\GEV$ & mono-$Z$ &\cite{Aad:2014vka} \\
		&$W(\to l\nu)W(2j)+j$'s$+\met$&&$\tilde{q}$&\cite{Aad:2015mia} \\
							\hline
combined & $H+W+\met$&&$\chi_1^{\pm}\chi^0_2$ &\cite{Khachatryan:2014mma}\\
\hline
\end{tabular} 
\caption{LHC searches used in this work.}
\label{tab:search} 
\end{table}

\section{14~TeV Projections}
\label{app:b}
The 14~TeV projected signal and background events are generated using the same pipeline. The total integrated luminosity is taken to be 300 ${\rm fb}^{-1}$. The dominant SM background for mono-$Z$ is the diboson process $pp\to ZZ\to l^+l^-\nu \bar{\nu}$. In order to project the mono-$Z$ reach at 14~TeV, we tweak the 8~TeV event selection criterion by increasing the $\met$ thresholds (200, 300, 400, 500, 650 and 800 GeV) to maintain approximately the same number of background events for the leading SM background contribution.  Other event selection criteria are kept to be the same as in the 8~TeV analysis. For mono-$h$, the $Z$~+~jets, $t\bar{t}$ and diboson backgrounds are found to be important. Four SRs are defined according to the $\met$ thresholds at 14~TeV: 300, 400, 500 and 600 GeV respectively. Similar to the mono-$Z$ projections, other event selection criteria are kept to be consistent with the 8~TeV analysis.
This prescription was validated by repeating it at 13~TeV and comparing it with corresponding 2016 Run II analyses when these were available and found to yield good agreement. 

The expected cross-section times branching ratio upper limit for each signal region is calculated using the $CL_S$ prescription. A systematic uncertainty of 30\% is assumed in our estimate. In Tables~\ref{tab:monoz14} and~\ref{tab:monoh14} we summarize the current status and prospects of mono-$Z$ and mono-$h$ searches. 

\begin{table}[t]
 \centering
\begin{tabular}{|c|c|rrrr|}
\hline
\multirow{3}{*}{8~TeV (mono-$Z$)} & $\met$ cut [GeV] & $>150$ & $>250$ & $ >350$ & $>450$ \\
\cline{2-6} 
& SM BG after cuts & 52 & 7.2 & 1.4 &0.4 \\
\cline{2-6} 
& obs. limit [fb] & 1.5 & 0.32 & 0.15 & 0.15\\
\hline
\end{tabular}
\begin{tabular}{|c|c|rrrrrr|}
\hline
\multirow{3}{*}{14~TeV (mono-$Z$)} & $\met$ cut [GeV] &$>200$ &$>300$ &$>400$ &$>500$ &$>650$ &$>800$ \\
\cline{2-8} 
&  SM BG after cuts &311.9&66.7&33.4 &6.2 &1.0 &0.2 \\
\cline{2-8} 
& exp. limit [fb] &0.62&0.14& 0.078 & 0.025 & 0.0099 & 0.0099\\
\hline
\end{tabular}
 \caption{Signal regions, SM background events after applying cuts and cross-section times branching ratio upper limits at 95 \% C.L. for the mono-$Z$ search. The 8~TeV results (background and observed cross-section times branching ratio upper limits) are taken from~\cite{Aad:2014vka}. The expected cross-section times branching ratio upper limits for the 14~TeV projections are estimated assuming a systematic uncertainty of 30\%. The total integrated luminosity is 300 ${\rm fb}^{-1}$.}
  \label{tab:monoz14}
 \end{table}

\begin{table}[t]
 \centering
\begin{tabular}{|c|c|rrrr|}
\hline
\multirow{3}{*}{8~TeV (mono-$h$)} & $\met$ cut [GeV] & $>150$ & $>200$ & $ >300$ & $>400$ \\
\cline{2-6} 
& SM BG after cuts & 148 & 62 & 9.4 &1.7 \\
\cline{2-6} 
& obs. limit [fb] & 3.7 & 1.3 & 0.45 & 0.20\\
\hline
\hline
\multirow{3}{*}{14~TeV (mono-$h$)} & $\met$ cut [GeV] & $>300$ & $>400$ & $ >500$ & $>600$ \\
\cline{2-6} 
& SM BG after cuts & 402.9 & 79.4 & 19.4 &7.6 \\
\cline{2-6} 
& exp. limit [fb] & 0.80 & 0.17 & 0.048 & 0.027\\
\hline
\end{tabular}
 \caption{Signal regions, SM background events and cross-section times branching ratio upper limits at 95 \% C.L. for the mono-$h$ search. The 8~TeV results (background and observed cross-section times branching ratio upper limits) are taken from~\cite{Aad:2015dva}. The expected cross-section times branching ratio upper limits for the 14~TeV projections are estimated assuming a systematic uncertainty of 30\%. The total integrated luminosity is 300 ${\rm fb}^{-1}$.}
  \label{tab:monoh14}
 \end{table}

\bibliography{monox}
\bibliographystyle{apsrev4-1}
\end{document}